# Comprehensive Overview of Bottom-Up Proteomics using Mass Spectrometry

This manuscript ([permalink](#)) was automatically generated from [jessegmeyerlab/proteomics-tutorial@03cc708](#) on November 13, 2023.

## Authors

**Yuming Jiang**  [0000-0001-7444-3849](#) ·  [jymbcrc](#) ·  [yumingjiang94](#)  Department of Computational Biomedicine, Cedars Sinai Medical Center

**Devasahayam Arokia Balaya Rex†**  [0000-0002-9556-3150](#) ·  [ArokiaRex](#) ·  [rexp___](#) Center for Systems Biology and Molecular Medicine, Yenepoya Research Centre, Yenepoya (Deemed to be University), Mangalore 575018, India

**Dina Schuster†**  [0000-0001-6611-8237](#) ·  [dschust-r](#) ·  [dina_sch](#)  Department of Biology, Institute of Molecular Systems Biology, ETH Zurich, Zurich 8093, Switzerland; Department of Biology, Institute of Molecular Biology and Biophysics, ETH Zurich, Zurich 8093, Switzerland; Laboratory of Biomolecular Research, Division of Biology and Chemistry, Paul Scherrer Institute, Villigen 5232, Switzerland

**Benjamin A. Neely\***  [0000-0001-6120-7695](#) ·  [neely](#) ·  [neely615](#)  Chemical Sciences Division, National Institute of Standards and Technology, NIST Charleston · Funded by NIST

**Germán L. Rosano\***  [0000-0002-8313-6813](#) ·  [ger225](#) ·  [GermanRosano](#)  Mass Spectrometry Unit, Institute of Molecular and Cellular Biology of Rosario, Rosario, Argentina · Funded by Grant PICT 2019-02971 (Agencia I+D+i)

**Norbert Volkmar\***  [0000-0003-0766-5606](#) ·  [norbertvolkmar](#) ·  [NorbertVolkmar](#) Department of Biology, Institute of Molecular Systems Biology, ETH Zurich, Zurich 8093, Switzerland

**Amanda Momenzadeh**  [0000-0002-8614-0690](#) ·  [amandamomenzadeh](#)  Department of Computational Biomedicine, Cedars Sinai Medical Center, Los Angeles, California, USA

**Trenton M. Peters-Clarke**  [0000-0002-9153-2525](#) ·  [petersclarke](#) ·  [trentmpc](#) Department of Pharmaceutical Chemistry, University of California-San Francisco

**Susan B. Egbert**  [0000-0001-5458-1099](#) ·  [lichenlady94](#) ·  [lichenlady94](#)  Department of Chemistry, University of Manitoba, Winnipeg, Cananda

**Simion Kreimer**  [0000-0001-6627-3771](#) ·  [KreimerSimion](#)  Smidt Heart Institute, Cedars Sinai Medical Center; Advanced Clinical Biosystems Research Institute, Cedars Sinai Medical Center

**Emma H. Doud**  0000-0003-0049-0073 ·  edoud1 ·  fireinlab  Center for Proteome Analysis, Indiana University School of Medicine, Indianapolis, Indiana, USA

**Oliver M. Crook**  0000-0001-5669-8506 ·  ococrook ·  OllyMCrook  Oxford Protein Informatics Group, Department of Statistics, University of Oxford, Oxford OX1 3LB, United Kingdom

**Amit Kumar Yadav**  0000-0002-9445-8156 ·  aky ·  theoneamit  Translational Health Science and Technology Institute · Funded by Grant BT/PR16456/BID/7/624/2016 (Department of Biotechnology, India); Grant Translational Research Program (TRP) at THSTI funded by DBT

**Muralidharan Vanuopadath**  0000-0002-9364-917X ·  vanuopadathmurali ·  V_MuraleeDhar  School of Biotechnology, Amrita Vishwa Vidyapeetham, Kollam-690 525, Kerala, India · Funded by Department of Health Research, Indian Council of Medical Research, Government of India (File No.R.12014/31/2022-HR)

**Martín L. Mayta**  0000-0002-7986-4551 ·  martinmayta ·  MartinMayta2  School of Medicine and Health Sciences, Center for Health Sciences Research, Universidad Adventista del Plata, Libertador San Martín 3103, Argentina; Molecular Biology Department, School of Pharmacy and Biochemistry, Universidad Nacional de Rosario, Rosario 2000, Argentina

**Anna G. Duboff**  0009-0002-7316-3831 ·  agduboff  Department of Chemistry, University of Washington · Funded by Summer Research Acceleration Fellowship, Department of Chemistry, University of Washington

**Nicholas M. Riley**  0000-0002-1536-2966 ·  rileynm ·  riley_nm1  Department of Chemistry, University of Washington · Funded by National Institutes of Health Grant R00 GM147304

**Robert L. Moritz**  0000-0002-3216-9447 ·  r_l_moritz  Institute for Systems biology, Seattle, WA, USA, 98109 · Funded by National Institutes of Health Grants R01GM087221, R24GM127667, U19AG023122, S10OD026936; National Science Foundation Award 1920268

**Jesse G. Meyer** ✉  0000-0003-2753-3926 ·  jessegmeyerlab ·  j_my_sci  Department of Computational Biomedicine, Cedars Sinai Medical Center · Funded by National Institutes of Health Grant R21 AG074234; National Institutes of Health Grant R35 GM142502

† These authors contributed equally as co-second author.
* These authors contributed equally as co-third author.
✉ — Correspondence to Jesse.Meyer@cshs.org

## Abstract

Proteomics is the large scale study of protein structure and function from biological systems through protein identification and quantification. "Shotgun proteomics" or "bottom-up proteomics" is the prevailing strategy, in which proteins are hydrolyzed into peptides that are analyzed by mass spectrometry. Proteomics studies can be applied to diverse studies ranging from simple protein identification to studies of proteoforms, protein-protein interactions, protein structural alterations, absolute and relative protein quantification, post-translational modifications, and protein stability. To enable this range of different experiments, there are diverse strategies for proteome analysis. The nuances of how proteomic workflows differ may be challenging to understand for new practitioners. Here, we provide a comprehensive overview of different proteomics methods to aid the novice and experienced researcher. We cover from biochemistry basics and protein extraction to biological interpretation and orthogonal validation. We expect this work to serve as a basic resource for new practitioners in the field of shotgun or bottom-up proteomics.

## Introduction

Proteomics is the large-scale study of protein structure and function. Proteins are translated from messenger RNA (mRNA) transcripts that are transcribed from the complementary DNA-based genome. Although the genome encodes potential cellular functions and states, the study of proteins in all their forms is necessary to truly understand biology.

Currently, proteomics can be performed with various methods. Mass spectrometry has emerged within the past few decades as the premier tool for comprehensive proteome analysis. The ability of mass spectrometry (MS) to detect charged chemicals enables the identification of peptide sequences and modifications for diverse biological investigations. Alternative (commercial) methods based on affinity interactions of antibodies or DNA aptamers have been developed, namely Olink and SomaScan. There are also nascent methods that are either recently commercialized or still under development and not yet applicable to whole proteomes, such as motif scanning using antibodies, variants of N-terminal degradation, and nanopores [1,2,3,4]. Another approach uses parallel immobilization of peptides with total internal reflection microscopy and sequential Edman degradation [5]. However, by far the most common method for proteomics is based on mass spectrometry coupled to liquid chromatography (LC).

Modern proteomics had its roots in the early 1980s with the analysis of peptides by mass spectrometry and low efficiency ion sources, but started ramping up around the year 1990 with the introduction of soft ionization methods that enabled, for the first time, efficient transfer of large biomolecules into the gas phase without destroying them [6,7]. Shortly afterward, the first computer algorithm for matching peptides to a database was introduced [8]. Another major milestone that allowed identification of over 1,000 proteins were improvements to chromatography upstream of MS anlaysis [9]. As the volume of data exploded, methods for statistical analysis transitioned from the wild west of ad hoc empirical analysis to modern informatics based on statistical models [10] and false discovery rate [11].

Two strategies of mass spectrometry-based proteomics differ fundamentally by whether proteins are analyzed as a whole chain or cleaved into peptides before analysis: "top-down" versus "bottom-up". Bottom-up proteomics (also refered to as shotgun proteomics) is defined by the intentional hydrolysis of proteins into peptide pieces using enzymes called proteases [12]. Therefore, bottom-up proteomics does not actually measure proteins, but instead infers protein presence and abundance from identified peptides [10]. Sometimes, proteins are inferred from only one peptide sequence representing a small fraction of the total protein sequence predicted from the genome. In contrast, top-down proteomics attempts to measure intact proteins [13,14,15,16]. The potential benefit of top-down proteomics is the ability to measure the many varied proteoforms [14,17,18]. However, due to myriad analytical challenges, the depth of protein coverage that is achievable by top-down proteomics is considerably less than that of bottom-up proteomics [19].

In this tutorial we focus on the bottom-up proteomics workflow. The most common version of this workflow is generally comprised of the following steps. First, proteins in a biological sample must be extracted. Usually this is achieved by mechanically lysing cells or tissue while denaturing and solubilizing the proteins and disrupting DNA to minimize interference in analysis procedures. Next, proteins are hydrolyzed into peptides, most often using the protease trypsin, which generates peptides with basic C-terminal amino acids (arginine and lysine) to aid in fragment ion series production during tandem mass spectrometry (MS/MS). Peptides can also be generated by chemical reactions that induce residue specific hydrolysis, such as cyanogen bromide that cleaves after methionine. Peptides from proteome hydrolysis must be purified; this is often accomplished with reversed-phase liquid chromatography (RPLC) cartridges or tips to remove interfering molecules in the sample such as salts and buffers. The peptides are then almost always separated by reversed-phase LC before they are ionized and introduced into a mass spectrometer, although recent reports also describe LC-free proteomics by direct infusion [20,21,22]. The mass spectrometer then collects precursor and fragment ion data from those peptides. Peptides must be identified from the tandem mass spectra, protein groups are inferred from a proteome database, and then quantitative values are assigned. Changes in protein abundances across conditions are determined with statistical tests, and results must be interpreted in the context of the relevant biology. Data interpretation is the rate limiting step; data collected in less than one week can take months or years to understand.

There are many variations to this workflow. The diversity of experimental goals that are achievable with proteomics technology drives an expansive array of workflows. Every choice is important as every choice will affect the results, from instrument procurement to choice of data processing software and everything in between. In this tutorial, we detail all the required steps to serve as a comprehensive overview for new proteomics practitioners.

There are 17 sections in total:

# 1. Biochemistry Basics

## Proteins

Proteins are large biomolecules or biopolymers made up of a backbone of amino acids which are linked by peptide bonds. They perform various functions in living organisms ranging from structural roles to functional involvement in cellular signaling and the catalysis of chemical reactions (enzymes). Proteins are made up of 20 different amino acids (not counting pyrrolysine, hydroxyproline, and selenocysteine, which only occur in specific organisms) and their sequence is encoded in their corresponding genes. The human genome encodes approximately 19,778 of the predicted canonical proteins coded in the human genome (see www.neXtProt.org) [23]. Each protein is present at a different abundance depending on the cell type or bodily fluid. Previous studies have shown that the concentration range of proteins can span at least seven orders of magnitude to up to 20,000,000 copies per cell, and that their distribution is tissue-specific [24,25]. Protein abundances can span more than ten orders of magnitude in human blood, while a few proteins make up most of the protein by weight in these fluids, making blood and plasma proteomics one of the most challenging matrices for mass spectrometry to analyze. Due to genetic variation, alternative splicing, and co- and post-translational modifications (PTMs), multiple different proteoforms can be produced from a single gene (**Figure 1**) [14,26].

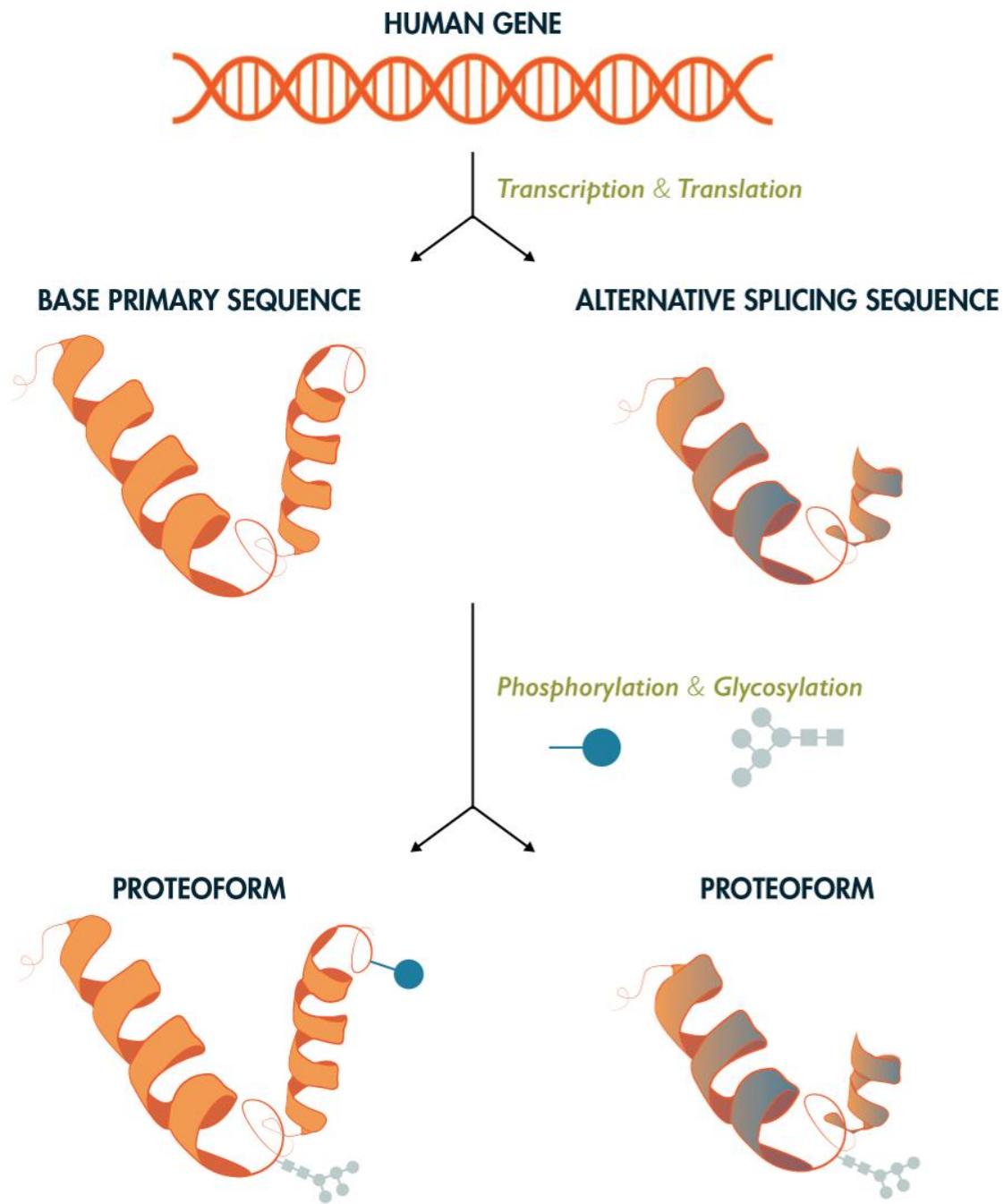

Figure 1: **Proteome Complexity.** *Each gene may be expressed in the form of multiple protein products, or proteoforms, through alternative splicing and incorporation of post-translational modifications. As such, there are many more unique proteoforms than genes. While there exist 20,000 - 23,000 coding genes in the human genome, upwards of 1,000,000 unique human proteoforms may exist. The study of the structure, function, and spatial and temporal regulation of these proteins is the subject of mass spectrometry-based proteomics*

## PTMs

After protein biosynthesis, enzymatic and nonenzymatic processes change the protein sequence through proteolysis or covalent chemical modification of amino acid side chains. Post-translational modifications (PTMs) are important biological regulators contributing to the diversity and function of the cellular proteome. Proteins can be post-translationally modified through enzymatic and non-enzymatic reactions *in vivo* and *in vitro* [27]. PTMs can be reversible or irreversible, and they change protein function in multiple ways, for example by altering substrate–enzyme interactions, subcellular localization or protein-protein interactions [28,29].

More than 400 biological PTMs have been discovered in both prokaryotic and eukaryotic cells. There are many more chemical artifact PTMs that occur during sample preparation, such as carbamylation. These modifications are crucial in controlling protein functions and signal transduction pathways [30]. The most commonly studied and biologically relevant post-translational modifications include phosphorylation (Ser, Thr, Tyr, His), glycosylation (Arg, Asp, Cys, Ser, Thr, Tyr, Trp), disulfide bonds (Cys-Cys), ubiquitination (Lys, Cys, Ser, Thr, N-term), succinylation (Lys), methylation (Arg, Lys, His, Glu, Asn, Cys), oxidation (especially Met, Trp, His, Cys), acetylation (Lys, N-term), and lipidations [31].

Protein PTMs can alter its function, activity, structure, spatiotemporal status and interaction with proteins or small molecules. PTMs alter signal transduction pathways and gene expression control [32] regulation of apoptosis [33,34] by phosphorylation. Ubiquitination generally regulates protein degradation [35], SUMOylation regulates chromatin structure, DNA repair, transcription, and cell-cycle progression [36,37], and palmitoylation regulates the maintenance of the structural organization of exosome-like extracellular vesicle membranes by [38]. Glycosylation is a ubiquitous modification that regulates various T cell functions, such as cellular migration, T cell receptor signaling, cell survival, and apoptosis [39,40]. Deregulation of PTMs is linked to cellular stress and diseases [41].

Several non-MS methods exist to study PTMs, including in vitro PTM reaction tests with colorimetric assays, radioactive isotope-labeled substrates, western blot with PTM-specific antibodies and superbinders, and peptide and protein arrays [42,43,44]. While effective, these approaches have many limitations, such as inefficiency and difficulty in producing pan-specific antibodies. MS-based proteomics approaches are currently the predominant tool for identifying and quantifying changes in PTMs.

## Protein Structure

Almost all proteins (except for intrinsically disordered proteins[45]) fold into three-dimensional (3D) structures either by themselves or assisted by molecular chaperones [46]. There are four levels relevant to the folding of any protein:
   - Primary structure: The protein's linear amino acid sequence, with amino acids connected through peptide bonds.
   - Secondary structure: The amino acid chain's folding: α-helix, β-sheet or turn.
   - Tertiary structure: The three-dimensional structure of the protein.
   - Quaternary structure: The structure of several protein molecules/subunits in one complex.

Of recent note, the development of AlphaFold, has enabled the high-accuracy three-dimensional structural prediction of all human proteins and for proteins of many other species, enabling a more thorough study of protein folding and is used to predict the relationship between fold and function [47,48].

# 2. Types of Experiments

A wide range of questions are addressable with proteomics technology, which translates to a wide range of variations of proteomics workflows. In some workflows, the identification of proteins in a given sample is desired. For other experiments, the quantification of as many proteins as possible is essential for the success of the study. Therefore, proteomic experiments can be both qualitative and quantitative. The following sections give an overview of several common proteomics experiments.

## Protein abundance changes

A common experiment is a discovery-based, unbiased mapping of proteins along with detection of changes in their abundance across sample groups. This is achieved using methods such as label free quantification (LFQ) or tandem mass tags (TMT), which are described in more detail in subsequent sections. In these experiments, data should be collected from at least three biological replicates of each condition to estimate the variance of measuring each protein. Depending on the experiment design, different statistical tests are used to calculate changes in measured protein abundances between groups. If there are only two groups, the quantities might be compared with a t-test or with a Wilcoxon signed-rank test. The latter is a non-parametric version of the t-test. If there are more than two sample groups, then Analysis of Variance (ANOVA) is used instead. With either testing scheme, the p-values from the first set of tests must be corrected for multiple tests. A common method for p-value correction is the Benjamini-Hochberg method [49]. These types of experiments have revealed wide ranges of proteomic remodeling from various biological systems.

## PTMs

Proteins may become decorated with various chemical modifications during or after translation [31], or through proteolytic cleavage such as N-terminal methionine removal [50]. Several proteomics methods are available to detect and quantify each specific type of modification. See also the section on Protein/Peptide Enrichment and Depletion. For a good online resource listing potential modifications, sites of attachment, and their mass differences, the website www.unimod.org is an excellent curated and freely accessible database.

### Phosphoproteomics

Phosphoproteomics is the study of protein phosphorylation, wherein a phosphate group is covalently attached to a protein side-chain (most commonly serine, threonine, or tyrosine). Although western blotting can measure one phosphorylation site at a time, mass spectrometry-based proteomics can measure thousands of sites from a sample at the same time. After proteolysis of the proteome, phosphopeptides need to be enriched to be detected by mass spectrometry. Various methods of enrichment have been developed [51,52,53,54]. See the Peptide/Protein Enrichment and Depletion section for more details. A key challenge of phosphoproteomics is sensitivity. It is important to ensure that there is sufficient amount of protein before completing a phosphoproteomics project as typical enrichment workflows may extract only ~1% of the proteome. Many phosphoproteomics workflows start with at least 1 mg of total protein per sample. Newer, more sensitive instrumentation is enabling detection of protein phosphosites from much less material, down to the nanogram-level of peptide loading on the the LC-MS system. Despite advancement in phosphoproteomics technology, the following challenges still exist: limited sample amounts, highly complex samples, and wide dynamic range [55]. Additionally, phosphoproteomic analysis is often time-consuming and requires the use of expensive equipment such as enrichment kits.

## Glycoproteomics

Glycosylation is gaining interest due to its ubiquity and emerging functional roles. Protein glycosylation sites can be N-linked (asparagine-linked) or O-linked (serine/threonine-linked). Understanding the function of protein glycosylation will help us understand numerous biological processes since this is a universal protein modification across all domains of life, especially at the cell surface [56,57,58,59].

Studies of phosphorylation and glycosylation share several experimental pipeline steps including sample preparation. Protein clean-up approaches for glycoproteomics may differ from other proteomics experiments because glycopeptides are more hydrophilic than most peptides. Some approaches mentioned in the literature include: filter-aided sample preparation (FASP), suspension traps (S-traps), and protein aggregation capture (PAC) [56,60,61,62,63,64,65]. Multiple proteases may be used to increase the sequence coverage and detect more modification sites, such as: trypsin, chymotrypsin, Pepsin, WaLP/MaLP [66], GluC, AspN, Pronase, Proteinase K, OgpA, StcEz, BT4244, AM0627, AM1514, AM0608, Pic, ZmpC, CpaA, IMPa, PNGase F, Endo F, Endo H, and OglyZOR [56]. Mass spectrometry has improved over the past decade, and now many strategies are available for glycoprotein structure elucidation and glycosylation site quantification [56]. See also the section on "AminoxyTMT Isobaric Mass Tags" as an example quantitative glycoproteomics method.

# Structural techniques

Several proteomics methods have been developed to reveal protein structure information for simple and complex systems.

## Cross-linking mass spectrometry (XL-MS)

XL-MS is an emerging technology in the field of proteomics. It can be used to determine changes in protein-protein interactions and/or protein structure. XL-MS covalently locks interacting proteins together to preserve interactions and proximity during MS analysis. XL-MS is different from traditional MS in that it requires the identification of chimeric MS/MS spectra from cross-linked peptides [67,68].

The common steps in a XL-MS workflow are as follows [69]:
1. Generate a system with protein-protein interactions of interest (*in vitro* or *in vivo* [70])
2. Add a cross-linking reagent to covalently connect adjacent protein regions (such as disuccinimidyl sulfoxide, DSSO) [68]
3. Proteolysis to produce peptides
4. MS/MS data collection
5. Identify cross-linked peptide pairs using special software (i.e. pLink [71], Kojak [72,73], xQuest [74], XlinkX [75])
6. Generate cross-link maps for structural modeling and visualization [76,77] (optional: 7. Use detected cross-links for protein-protein docking [78])

## Hydrogen deuterium exchange mass spectrometry (HDX-MS)

HDX-MS works by detecting changes in peptide mass due to exchange of amide hydrogens of the protein backbone with deuterium from D2O [79]. The exchange rate depends on the protein solvent accessible surface area, dynamics, and the properties of the amino acid sequence [79,80,81,82]. Although using D2O to make deuterium-labeled samples is simple, HDX-MS requires several controls to ensure that experimental conditions capture the dynamics of interest [79,83,84,85]. If the peptide dissociation process is tuned appropriately, residue-level quantification of changes in solvent accessibility are possible within a measured peptide [86]. HDX can produce precise protein structure measurements with high reproducibility. Masson *et al.* gave recommendations on how to prep samples, conduct data analysis, and present findings in a detailed stepwise manner [79].

## Radical Footprinting

This technique uses hydroxyl radical footprinting and MS to elucidate protein structures, assembly, and interactions within a large macromolecule [87,88]. In addition to proteomics applications, various approaches to make hydroxide radicals have also been applied for footprinting studies in nucleic acid/ligand interactions [89,90,91]. This chapter is very useful in learning more about this topic: [92].

There are several methods of producing radicals for protein footprinting:
1. Fenton and Fenton-like Chemistry [87,93,94]
2. Electron-Pulse Radiolysis [87,95]
3. High-Voltage Electrical Discharge [87,96]
4. Synchrotron X-ray Radiolysis of Water [87,97]
5. Plasma Formation of OH Radicals [87,98]
6. Photolysis of Hydrogen Peroxide [87,99]

## Fast photochemical oxidation of proteins (FPOP) [100]

FPOP is an example of a radical footprinting method. In FPOP, a laser-based hydroxyl radical protein footprinting MS method that relies on the irreversible labeling of solvent-exposed amino acid side chains by hydroxyl radicals in order to understand structure of proteins. A laser produces 248 nm light that causes hydrogen peroxide to break into a pair of hydroxyl radicals [99,101]. The flow rate of solution through the capillary and laser frequency are adjusted such that each protein molecule is irradiated only once. After they are irradiated, the sample is collected in a tube that contins catalase and free methionine in the buffer, quenching the H2O2 and hydroxyl radicals and preventing secondary modification of residues that become exposed due to unfolding after the initial labeling. Control samples are made by running the sample through the flow system without any irradiation. Another experimental control involves the addition of a radical scavenger to tune the extent of protein oxidation [102,103]. FPOP has wide application for proteins including measurements of fast protein folding and transient dynamics.

## Protein Painting [104,105]

Protein painting uses "molecular paints" to noncovalently coat the solvent-accessible surface of proteins. These paint molecules will coat the protein surfaces but will not have access to the hydrophobic cores or protein-protein interface regions that solvents cannot access. If the "paint" covers free amines of lysine side chains, the painted parts will not be subjected to trypsin cleavage, while the unpainted areas will. After proteolysis, the peptides samples will be subjected to MS. A lack of proteolysis in a region is interpreted as solvent accessibility, which gives rough structural information about complex protein mixtures or even a whole proteome.

## LiP-MS (Limited Proteolysis Mass Spectrometry) [106,107,107,108]

Limited proteolysis coupled to mass spectrometry (LiP-MS) is a method that tracks structural changes in complex proteomes in response to a variety of perturbations or stimuli. The underlying tenet of LiP-MS is that a stimuli-induced change in native protein structure (i.e. protein-protein interaction, introduction of a PTM, ligand/substrate binding, or changes in osmolarity or ambient temperature) can be detected by a change in accessibility of a broad-specificity protease (i.e. proteinase K) to the region(s) of the protein where the structural change occurs. For example, small molecule binding may render a disordered region protected from non-specific proteolysis by directly blocking access of the protease to the cleavage site. LiP-MS can therefore provide a somewhat unbiased view of structural changes at the proteome scale. Importantly, LiP-MS necessitates cell lysates or individual proteins be maintained in their native state prior to or during perturbation and protease treatment. LiP-MS can also be applied to membrane suspensions, to facilitate the study of membrane proteins without the need for purification or detergents [109].
For additional information about LiP-MS, please refer to the following article: [110]

# Protein stability and small molecule binding

## Cellular Thermal Shift Assay (CETSA) [111,112]

CETSA involves subjecting a protein sample to a thermal shift assay (TSA), in which the protein is exposed to a range of temperatures, and the resulting changes in protein stability by quantifying protein remaining in the soluble fraction. This is done in live cells immediately before lysis, or in non-denaturing lysates. The original paper reported this method using immunoaffinity approaches for detecting changes in soluble protein. The assay is capable of detecting shifts in the thermal equilibrium of cellular proteins in response to a variety of perturbations, but most commonly in response to in vitro drug treatments.

## Thermal proteome profiling (TPP) [113,114,115,116]

Thermal proteome profiling (TPP) follows the same principle as CETSA, but has been extended to use an unbiased mass spectrometry readout of many proteins. By measuring changes in thermal stability of thousands of proteins, binding to an unknown or unexpected protein can be discovered. During a typical TPP experiment, a protein sample is first treated with a drug of interest to stabilize protein-ligand interactions. The sample is then divided into multiple aliquots, which are subjected to different temperatures to induce thermal denaturation. The resulting drug-induced changes in protein stability curves are detected using mass spectrometry. By comparing protein stability curves across the temperatures between treatment conditions, TPP can provide insight into the proteins that bind a ligand.

# Protein-protein interactions (PPIs)

## Affinity purification coupled to mass spectrometry (AP-MS) [117,118,119]

AP-MS is an approach that involves purification of a target protein or protein complex using a specific antibody followed by mass spectrometry analysis to identify the interacting proteins. In a typical AP-MS experiment, a protein or protein complex of interest is first tagged with a specific epitope or affinity tag, such as a FLAG or HA tag, which is used to selectively capture the target protein using an antibody. The protein complex is then purified from the sample using a series of wash steps, and the interacting proteins are identified using mass spectrometry. The success of AP-MS experiments depends on many factors, including the quality of the antibody or tag used for purification, the specificity and efficiency of the resin used for capture, and the sensitivity and resolution of the mass spectrometer. In addition, careful experimental design and data analysis are critical for accurately identifying and interpreting protein-protein interactions. AP-MS has been used to study a wide range of biological processes, including signal transduction pathways, protein complex dynamics, and protein post-translational modifications. AP-MS has been performed on a whole proteome scale as part of the BioPlex project [120,121,122]. Despite its widespread use, AP-MS has some limitations, including the potential for non-specific interactions, the difficulty in interpreting complex data sets, and the possibility of missing important interacting partners due to constraints in sensitivity or specificity. However, with continued advances in technology and data analysis methods, AP-MS is likely to remain a valuable tool for studying protein-protein interactions.

## APEX peroxidase [123,124]

APEX-MS is a labeling technique that utilizes a peroxidase genetically fused to a protein of interest. When biotin-phenol is transiently added in the presence of hydrogen peroxide, nearby proteins are covalently biotinylated [125]. APEX thereby enables the discovery of interacting proteins in living cells. One of the major advantages of APEX is its ability to label proteins in their native environment, allowing for the identification of interactions that occur under physiological conditions. Despite its advantages, APEX has some limitations, including the potential for non-specific labeling, the difficulty in distinguishing between direct and indirect interactions, and the possibility of missing interactions that occur at low abundance or in regions of the cell that are not effectively labeled.

## Proximity-dependent biotin identification (BioID) [126,127,128,129]

BioID is a proximity labeling technique that allows for the identification of protein-protein interactions. BioID involves the genetic tagging of a protein of interest with a promiscuous biotin ligase in live cells, which then biotinylates proteins in close proximity to the protein of interest. One of the advantages of BioID is its ability to label proteins in their native environment, allowing for the identification of interactions that occur under physiological conditions. BioID has been used to identify a wide range of protein interactions, including receptor-ligand interactions, signaling complexes, and protein localization. BioID is a slower reaction than APEX and therefore may pick up more transient interactions. BioID has the same limitations as APEX. For more information on BioID, please refer to [130].

# 3. Protein Extraction

Protein extraction from the sample of interest is the initial phase of any mass spectrometry-based proteomics experiment. Thought should be given to any planned downstream assays, specific needs of proteolysis (LiP-MS, PTM enrichments, enzymatic reactions, glycan purification or hydrogen-deuterium exchange experiments), long-term project goals (reproducibility, multiple sample types, low abundance samples), as well as to the initial experimental question (coverage of a specific protein, subcellular proteomics, global proteomics, protein-protein interactions or affinity enrichment of specific classes of modifications). The 2009 version of Methods in Enzymology: guide to Protein Purification [131] serves as a deep dive into how molecular biologists and biochemists traditionally carried out protein extraction. The Protein Protocols handbook [132] and the excellent review by Linn [133] are good sources of general proteomics protocols. Another excellent resource is the "Proteins and Proteomics: A Laboratory Manual" by Richard J. Simpson [134,134]. This manual is 926 pages packed full of bench tested protocols and procedures for carrying out protein centric studies. Any change in extraction conditions should be expected to create potential changes in downstream results. Be sure to plan and optimize the protein extraction step first and use a protocol that works for your needs. To reproduce the results of another study, one should begin with the same extraction protocols.

# Buffer choice

## General proteomics

A common question to proteomics core facilities is, "What is the best buffer for protein extraction?" Unfortunately, there is no one correct answer. For global proteomics experiments where maximizing the number of protein or peptide identifications is a goal, a buffer of neutral pH (50-100 mM phosphate buffered saline (PBS), tris(hydroxymethyl)aminomethane (Tris), 4-(2-hydroxyethyl)-1-piperazineethanesulfonic acid (HEPES), ammonium bicarbonate, triethanolamine bicarbonate; pH 7.5-8.5) is used in conjunction with a chaotrope or surfactant to denature and solubilize proteins (e.g., 8 M urea, 6 M guanidine, 5% sodium dodecyl sulfate (SDS)) [135,136]. Often other salts like 50-150 mM sodium chloride (NaCl) are also added. Although there are a range of buffers that can be used to provide the correct working pH and ionic strength, not all buffers are compatible with downstream workflows and some buffers can induce modifications of proteins (e.g., ammonium bicarbonate promotes methionine oxidation and we generally suggest Tris-HCl instead to minimize oxidation. A great online resource to help calculate buffer compositions and pH values is the website by Robert Beynon at http://phbuffers.org. Complete and quick denaturation of proteins in the sample is required to limit changes to protein status by endogenous proteases, kinases, phosphatases, and other enzymes. If intact protein separations are planned (based on size or isoelectric point), choose a denaturant compatible with those methods, such as SDS [137]. Compatibility with the protease (typically trypsin) and peptide cleanup steps must be considered. Of note, detergents can be incompatible with LC-MS workflows as they can cause ion suppression and column clogging. It is therefore advisable to remove detergents using detergent-removal kits or precipitation techniques (i.e. deoxycholate precipitates at low pH and can easily be removed by filtration or centrifugation). For more information, see section "Removal of buffer/interfering small molecules". Alternatively, mass-specrtrometry-compatible detergents may be used (i.e. n-dodecyl-beta-maltoside). Urea must be diluted to 2 M or less for trypsin and chymotrypsin digestions, while guanidine and SDS should be removed either through protein precipitation, through filter-assisted sample preparation (FASP), or similar solid phase digestion techniques. Some buffers can potentially introduce modifications to proteins such as carbamylation from urea at high temperatures [138].

## Protein-protein interactions

Denaturing conditions will efficiently extract proteins, but will denature proteins and therefore disrupt most protein-protein interactions. If you are working on an immune- or affinity purification of a specific protein and expect to analyze enzymatic activity, structural features, and/or protein-protein interactions, a non-denaturing lysis buffer should be utilized [139,140]. Check the calculated isoelectric point (pI) and hydrophobicity (e.g., try the Expasy.org resource ProtParam) for a good idea of starting pH/conductivity, but a stability screen may be needed. In general, a good starting point for the buffer will still be close to neutral pH with 50-250 mM NaCl, but specific proteins may require pH as low as 2 or as high as 9 for stable extraction. A low percent of mass spectrometry compatible detergent may also be used, such as n-dodecyl-beta-maltoside. Newer mass spectrometry-compatible detergents are also useful for protein extraction and ease of downstream processing – including Rapigest® (Waters), N-octyl-β-glucopyranoside, MS-compatible degradable surfactant (MaSDeS) [141], Azo [142], PPS silent surfactant [143], sodium laurate [144], and sodium deoxycholate [145]. Avoid using tween-20, triton-X, NP-40, and polyethylene glycols (PEGs) as these compounds are challenging to remove after digestion [146].

## Optional additives

For non-denaturing buffer conditions, which preserve tertiary and quaternary protein structures, additional additives may not be necessary for successful extraction and to prevent proteolysis or PTMs throughout the extraction process. Protease, phosphatase and deubiquitinase inhibitors are optional additives in less denaturing conditions or in experiments focused on specific PTMs. For a broad range of inhibitors, a premixed tablet can be added to the lysis buffer, such as Roche cOmplete Mini Protease Inhibitor Cocktail tablets. Protease inhibitors may impact desired proteolysis from the added protease, and will need to be diluted or removed prior to protease addition. To improve extraction of DNA- or RNA-binding proteins, adding a small amount of nuclease or benzonase is useful for degradation of any bound nucleic acids and results in a more consistent digestion [147].

# Mechanical or Sonic Disruption

## Cell lysis

One typical lysis buffer is 8 M urea in 100 mM Tris, pH 8.5; the pH is based on optimum trypsin activity [148] Small mammalian cell pellets and exosomes will lyse almost instantly upon addition denaturing buffer. If non-denaturing conditions are desired, osmotic swelling and subsequent shearing or sonication can be applied [149]. Efficiency of extraction and degradation of nucleic acids can be improved using various sonication methods: 1) probe sonicator with ice; 2) water bath sonicator with ice or cooling; 3) bioruptor® sonication device 4) Adaptive focused acoustics (AFA®) [150]. Key to these additional lysis techniques is to keep the temperature of the sample from rising significantly which can cause proteins to aggregate or degrade. Some cell types may require additional force for effective lysis (see below). For cells with cell walls (i.e. bacteria or yeast), lysozyme is often added in the lysis buffer. Any added protein will be present in downstream results, however, so excessive addition of lysozyme is to be avoided unless tagged protein purification will occur.

## Tissue/other lysis

Although small pieces of soft tissue can often be successfully extracted with the probe and sonication methods described above, larger/harder tissues as well as plants/yeast/fungi are better extracted with some form of additional mechanical force. If proteins are to be extracted from a large amount of sample, such as soil, feces, or other diffuse input, one option is to use a dedicated blender and filter the sample, followed by centrifugation. If samples are smaller, such as tissue, tumors, etc., cryo-homogenization is recommended. The simplest form of this is grinding the sample with liquid nitrogen and a mortar and pestle. Tools such as bead beaters (i.e. FastPrep-24®) are also used, where the sample is placed in a tube with appropriately sized glass or ceramics beads and shaken rapidly. Cryo-mills are chambers where liquid nitrogen is applied around a vessel and large bead or beads. Cryo-fractionators homogenize samples in special bags that are frozen in liquid nitrogen and smashed with various degrees of force [151]. In addition, rapid bead beating mills such as the Bertin Precellys Evolution are both economical, effective and detergent compatible for many types of proteomics experiments at a scale of 96 samples per batch. After homogenization, samples can be sonicated by one of the methods above to fragment DNA and increase solubilization of proteins.

# Measuring the efficiency of protein extraction

Following protein extraction, samples should be centrifuged (10-14,000 g for 10-30 min depending on sample type) to remove debris and any unlysed material prior to determining protein concentration. The amount of remaining insoluble material should be noted throughout an experiment as a large change may indicate protein extraction issues. Protein concentration can be calculated using a number of assays or tools [152,153]; generally absorbance measurements are facile, fast and affordable, such as Bradford or BCA assays. Protein can also be estimated by tryptophan fluorescence, which has the benefit of not consuming sample [154]. A nanodrop UV spectrophotometer may be used to measure absorbance at UV280. Consistency in this method is important as each method will have inherent bias and error [155,156]. Extraction buffer components will need to be compatible with any assay chosen; alternatively, buffer may be removed (see below) prior to protein concentration calculation.

# Reduction and alkylation

Typically, disulfide bonds in proteins are reduced and alkylated prior to proteolysis in order to disrupt structures and simplify peptide analysis. This allows better access to all residues during proteolysis and removes the crosslinked peptides created by S-S inter peptide linkages. There are a variety of reagent options for these steps. For reduction, the typical agents used are 5-15 mM concentration of tris(2-carboxyethyl)phosphine hydrochloride (TCEP-HCl), dithiothreitol (DTT), or 2-beta-mercaptoethanol (2BME). TCEP-HCl is an efficient reducing agent, but it also significantly lowers sample pH, which can be abated by increasing sample buffer concentration or resuspending TCEP-HCl in an appropriate buffer system (i.e. 1M HEPES pH 7.5). Following the reducing step, a slightly higher 10-20mM concentration of alkylating agent such as chloroacetamide/iodoacetamide or n-ethyl maleimide is used to cap the free thiols [157,158,159]. In order to monitor which cysteine residues are linked or modified in a protein, it is also possible to alkylate free cysteine residues with one reagent, reduce di-sulfide bonds (or other cysteine modifications) and alkylate with a different reagent [160,161,162]. Alkylation reactions are generally carried out in the dark at room temperature to avoid excessive off-target alkylation of other amino acids.

# Removal of buffer/interfering small molecules

If extraction must take place in a buffer which is incompatible for efficient proteolysis (check the guidelines for the protease of choice), then protein cleanup should occur prior to digestion. This is generally performed through precipitation of proteins. The most common types are 1) acetone, 2) trichloroacetic acid (TCA), and 3) methanol/chloroform/water [163,164]. Proteins are generally insoluble in most pure organic solvents, so cold ethanol or methanol are sometimes used. Pellets should be washed with organic solvent for complete removal especially of detergents. Alternatively, solid phase based digestion methods such as S-trap [165], FASP [166,167], SP3 [168,169] and on column/bead such as protein aggregation capture (PAC) [170] can allow for proteins to be applied to a solid phase and buffers removed prior to proteolysis [171]. Specialty detergent removal columns exist (Pierce/Thermo Fisher Scientific) but add expense and time-consuming steps to the process. Relatively low concentrations of specific detergents, such as 1% deoxycholate (DOC), or chaotropes (i.e. 1M urea) are compatible with proteolysis by trypsin/Lys-C. Often proteolysis-compatible concentrations of these detergents and chaotropes are achieved by diluting the sample in appropriate buffer (i.e. 100 mM ammonium bicarbonate, pH 8.5) after cell or tissue lysis in a higher concentration. DOC can then be easily removed by precipitation or phase separation [172] following digestion by acidification of the sample to pH 2-3. Any small-molecule removal protocol should be tested for efficiency prior to implementing in a workflow with many samples as avoiding detergent (or polymer) contamination in the LC/MS is very important.

# Protein quantification

After proteins are isolated from the sample matrix, they are often quantified. Protein quantification is important to assess the yield of an extraction procedure, and to adjust the scale of the downstream processing steps to match the amount of protein. For example, when purifying peptides, the amount of sorbent should match the amount of material to be bound. Presently, there is a wide variety of techniques to quantitate the amount of protein present in a given sample. These methods can be broadly divided into three types as follows:

## Colorimetry-based methods:

The method includes different assays like Coomassie Blue G-250 dye binding (the Bradford assay), the Folin-Lowry assay, the bicinchoninic acid (BCA) assay and the biuret assay [173]. The most commonly used method is the BCA assay. In the BCA method the peptide bonds of the protein reduce cupric ions [Cu2+] to cuprous ions [Cu+] at a rate which is proportional to the amount of protein present in the sample. Subsequently, the BCA reagent binds to the cuprous ions, leading to the formation of a complex which absorbs 562 nm wavelength light. This permits a direct correlation between sample protein concentration and absorbance [174,175]. The Bradford assay is another method for protein quantification also based on colorimetry principle. It relies on the interaction between the Coomassie brilliant blue dye and the protein based on hydrophobic and electrostatic interactions. Dye binding shifts the absorption maxima from 470 nm to 595 nm [176,177]. Similarly, the Folin- Lowry method is a two-step colorimetric assay. Step one is the biuret reaction wherein complexes of copper with the nitrogen in the protein molecule are formed. In the second step, the complexed tyrosine and tryptophan amino acids react with Folin–Ciocalteu phenol reagent generating an intense, blue-green color absorbing light at 650–750 nm [178].

Another simple but less reliable protein quantification method of UV-Vis Absorbance at 280 nm estimates the protein concentration by measuring the absorption of the aromatic residues; tyrosine, and tryptophan, at 280 nm [179].

## Fluorescence-based methods:

Colorimetric assays are inexpensive and require common lab equipment, but colorimetric detection is less sensitive than fluorescence. Total protein in proteomic samples can be quantified using intrinsic fluorescence of tryptophan based on the assumption that approximately 1% of all amino acids in the proteome are tryptophan [180].

NanoOrange is an assay for the quantitative measurement of proteins in solution using the NanoOrange reagent, a merocyanine dye that produces a large increase in fluorescence quantum yield when it interacts with detergent-coated proteins. Fluorescence is measured using 485-nm excitation and 590-nm emission wavelengths. The NanoOrange assay can be performed using fluorescence microplate readers, fluorometers, and laser scanners that are standard in the laboratory [153].

3-(4-carboxybenzoyl)quinoline-2-carboxaldehyde (CBQCA) is a sensitive fluorogenic reagent for amine detection, which can be used for analyzing proteins in solution. As the number of accessible amines in a protein is modulated by its concentration, CBQCA has a greater sensitivity and dynamic range when measuring protein concentration [181].

# Protein Extraction Summary

Learning the fundamentals and mechanisms of how and why sample preparation steps are performed is vital because it enables flexibility to perform proteomics from a wide range of samples. For bottom-up proteomics, the overreaching goal is efficient and consistent extraction and digestion. A range of mechanical and non-mechanical extraction protocols have been developed and the choice of technique is generally dictated by sample type or assay requirements (i.e. native versus non-native extraction). Extraction can be aided by the addition of detergents and/or chaotropes to the sample, but care should be taken that these additives do not interfere with the sample digestion step or downstream mass-spectrometry analysis.

# 4. Proteolysis

Proteolysis is the defining step that differentiates bottom-up or shotgun proteomics from top-down proteomics. Hydrolysis of proteins is extremely important because it defines the population of potentially identifiable peptides. Generally, peptides between a length of 7-35 amino acids are considered useful for mass spectrometry analysis. Peptides that are too long are difficult to identify by tandem mass spectrometry or may be lost during sample preparation due to irreversible binding with solid-phase extraction sorbents. Peptides that are too short are also not useful because they may match to many proteins during protein inference. There are many choices of enzymes and chemicals that hydrolyze proteins into peptides. This section summarizes potential choices and their strengths and weaknesses.

Trypsin is the most common choice of protease for proteome hydrolysis [182]. Trypsin is favorable because of its specificity, availability, efficiency and low cost. Trypsin cleaves at the C-terminus of basic amino acids, Arg and Lys, if not immediately followed by proline. Many of the peptides generated from trypsin are short in length (less than ~ 20 amino acids), which is ideal for chromatographic separation, MS-based peptide fragmentation and identification by database search. The main drawback of trypsin is that majority (56%) of the tryptic peptides are ≤ 6 amino acids, and hence using trypsin alone limits the observable proteome [183,184,185]. This limits the number of identifiable protein isoforms and post-translational modifications.

Although trypsin is the most common protease used for proteomics, in theory it can only cover a fraction of the proteome predicted from the genome [186]. This is due to production of peptides that are too short to be unique, for example due to R and K immediately next to each other. Peptides below a certain length are likely to occur many times in the whole proteome, meaning that even if we identify them we cannot know their protein of origin. In protein regions devoid of R/K, trypsin may also result in very long peptides that are then lost due to irreversible binding to the solid phase extraction device, or that become difficult to identify due to complicated fragmentation patterns. Thus, parts of the true proteome sequences that are present are lost after trypsin digestion due to both production of very long and very short peptides.

Many alternative proteases are available with different specificities that complement trypsin to reveal different protein sequences [183,187], which can help distinguish protein isoforms [188] (Figure 2). The enzyme choice mostly depends on the application. In general, for a mere protein identification, trypsin is often chosen due to the aforementioned reasons. However, alternative enzymes can facilitate *de novo* assembly when the genomic data information is limited in the public database repositories [189,190,191,192,193]. Use of multiple proteases for proteome digestion also can improve the sensitivity and accuracy of protein quantification [194]. Moreover, by providing an increased peptide diversity, the use of multiple proteases can expand sequence coverage and increase the probability of finding peptides which are unique to single proteins [66,186,195]. A multi-protease approach can also improve the identification of N-Termini and signal peptides for small proteins [196]. Overall, integrating multiple-protease data can increase the number of proteins identified [197,198], increase the identified post-translational modifications [66,195,199] and decrease the ambiguity of the inferred protein groups [195].

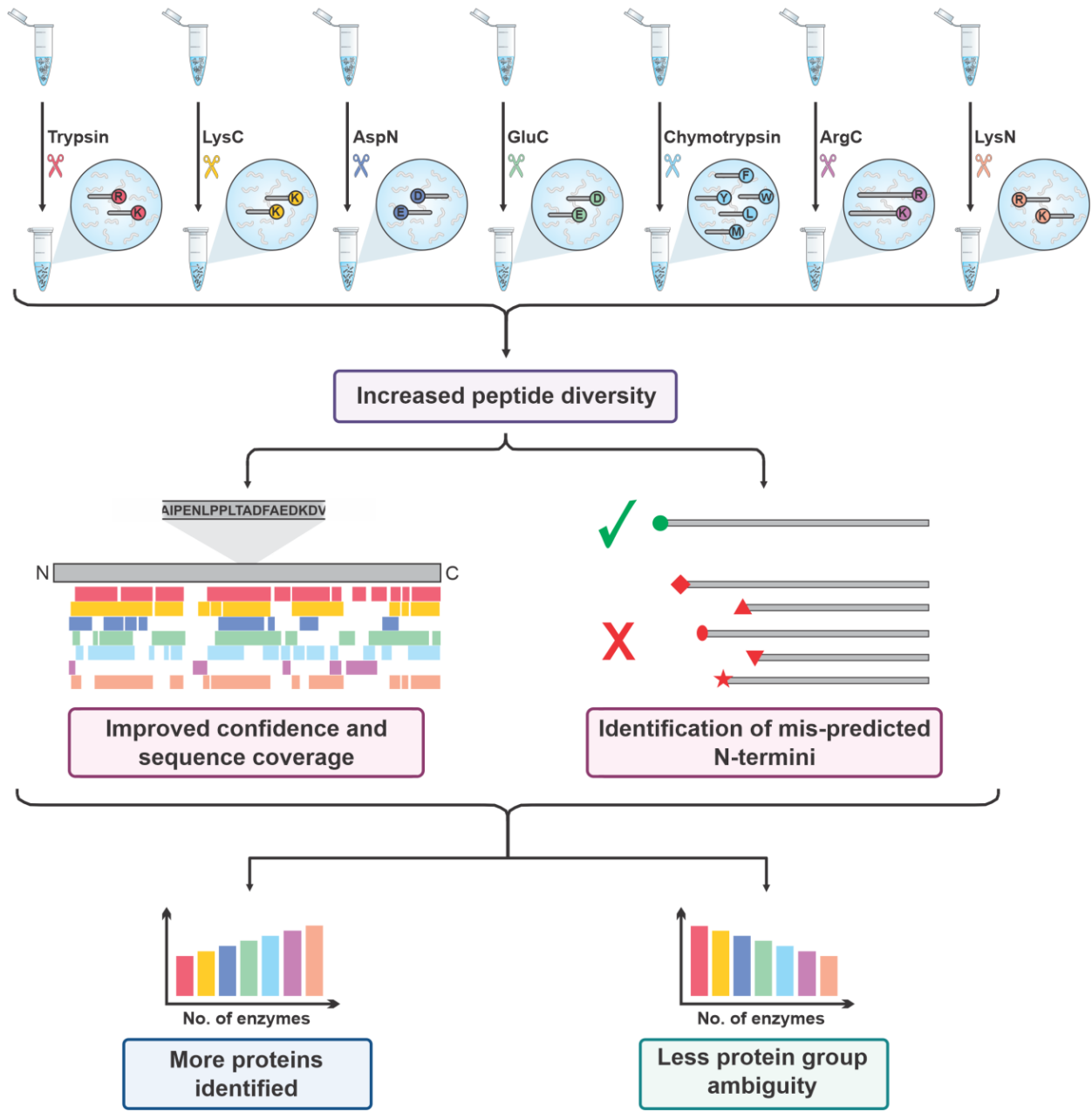

*Figure 2: **Multiple protease proteolysis improves protein inference** The use of other proteases beyond Trypsin such as Lysyl endopeptidase (Lys-C), Peptidyl-Asp metallopeptidase (Asp-N), Glutamyl peptidase I, (Glu-C), Chymotrypsin, Clostripain (Arg-C) or Peptidyl-Lys metalloendopeptidase (Lys-N) can generate a greater diversity of peptides. This improves protein sequence coverage and allows for the correct identification of their N-termini. Increasing the number of complimentary enzymes used will increase the number of proteins identified by single peptides and decreases the ambiguity of the assignment of protein groups. Therefore, this will allow more protein isoforms and post-translational modifications to be identified than using Trypsin alone.*

Lysyl endopeptidase (Lys-C) obtained from *Lysobacter enzymogenesis* is a serine protease involved in cleaving carboxyl terminus of Lys [184,200]. Like trypsin, the optimum pH range required for its activity is from 7 to 9. A major advantage of Lys-C is its resistance to denaturing agents, including 8 M urea - a chaotrope commonly used to denature proteins *prior* to digestion [188]. Trypsin is less efficient at cleaving Lys than Arg, which could limit the quality of quantitation from tryptic peptides. Hence, to achieve complete protein digestion with minimal missed cleavages, Lys-C is often used simultaneously with trypsin digestion [201].

Alpha-lytic protease (aLP) is another protease secreted by the soil bacterial *Lysobacter enzymogenesis* [202]. Wild-type aLP (WaLP) and an active site mutant of aLP, M190A (MaLP), have been used to expand proteome coverage [66]. Based on observed peptide sequences from yeast proteome digestion, WaLP showed a specificity for small aliphatic amino acids like alanine, valine, and glycine, but also threonine and serine. MaLP showed specificity for slightly larger amino acids like methionine, phenylalanine, and surprisingly, a preference for leucine over isoleucine. The specificity of WaLP for threonine enabled the first method for mapping endogenous human SUMO sites [37].

Glutamyl peptidase I, commonly known as Glu-C or V8 protease, is a serine protease obtained from *Staphyloccous aureus* [203]. Glu-C cleaves at the C-terminus of glutamate, but also after aspartate [203,204].

Peptidyl-Asp metallopeptidase, commonly known as Asp-N, is a metalloprotease obtained from *Pseudomonas fragi* [205]. Asp-N catalyzes the hydrolysis of peptide bonds at the N-terminal of aspartate residues. The optimum activity of this enzyme occurs at a pH range between 4 and 9. As with any metalloprotease, chelators like EDTA should be avoided for digestion buffers when using Asp-N. Studies also suggest that Asp-N cleaves at the amino terminus of glutamate when a detergent is present in the proteolysis buffer [205]. Asp-N often leaves many missed cleavages [188].

Chymotrypsin or chymotrypsinogen A is a serine protease obtained from porcine or bovine pancreas with an optimum pH range from 7.8 to 8.0 [206]. It cleaves at the C-terminus of hydrophobic amino acids Phe, Trp, Tyr and barely Met and Leu residues. Since the transmembrane region of membrane proteins commonly lacks tryptic cleavage sites, this enzyme works well with membrane proteins having more hydrophobic residues [188,207,208]. The chymotryptic peptides generated after proteolysis will cover the proteome space orthogonal to that of tryptic peptides both in a quantitative and qualitative manner [208,209,210]

Clostripain, commonly known as Arg-C, is a cysteine protease obtained from *Clostridium histolyticum* [211]. It hydrolyses mostly the C-terminal Arg residues and sometimes Lys residues, but with less efficiency. The peptides generated are generally longer than that of tryptic peptides. Arg-C is often used with other proteases for improving qualitative proteome data and also for investigating PTMs [184].

LysargiNase, also known as Ulilysin, is a recently discovered protease belonging to the metalloprotease family. It is a thermophilic protease derived from *Methanosarcina acetivorans* that specifically cleaves at the N-terminus of Lys and Arg residues [212]. Hence, it enabled discovery of C-terminal peptides that were not observed using trypsin. In addition, it can also cleave modified amino acids such as methylated or dimethylated Arg and Lys [212].

Peptidyl-Lys metalloendopeptidase, or Lys-N, is an metalloprotease obtained from *Grifola frondosa* [213]. It cleaves N-terminally of Lys and has an optimal activity at pH 9.0. Unlike trypsin, Lys-N is more resistant to denaturing agents and can be heated up to 70°C [184]. Peptides generated from Lys-N digestion produce more c-type ions using ETD fragmentation [214]. Hence this can be used for analysing PTMs, identification of C-terminal peptides and also for *de novo* sequencing strategies [214,215].

Pepsin A, commonly known as pepsin, is an aspartic protease obtained from bovine or porcine pancreas [216]. Pepsin was one of several proteins crystalized by John Northrop, who shared the 1946 Nobel prize in chemistry for this work [217,218,219,220]. Pepsin works at an optimum pH range from 1 to 4 and specifically cleaves Trp, Phe, Tyr and Leu [184]. Since it possess high enzyme activity and broad specificity at lower pH, it is preferred over other proteases for MS-based disulphide mapping [221,222]. Pepsin is also used extensively for structural mass spectrometry studies with hydrogen-deuterium exchange (HDX) because the rate of back exchange of the amide deuteron is minimized at low pH [223,224].

Proteinase K was first isolated from the mold *Tritirachium album* Limber [225]. The epithet 'K' is derived from its ability to efficiently hydrolyze keratin [225]. It is a member of the subtilisin family of proteases and is relatively unspecific with a preference for proteolysis at hydrophobic and aromatic amino acid residues [226]. The optimal enzyme activity is between pH 7.5 and 12. Proteinase K is used at low concentrations for limited proteolysis (LiP) and the detection of protein structural changes in the eponymous technique LiP-MS [227].

Although different specificity is useful in theory to enable improved proteome sequence coverage, there are practical challenges because most standard workflows are optimized for tryptic peptides. For example, peptides that lack a c-terminal positive charge due to arginine or lysine side chains can have a less pronounced y-ion series. This can lead to lower scoring peptide-spectra matches because some peptide identification algorithms preferentially score y ions higher.

## Peptide quantitation assays

After peptide production from proteomes, it may be desirable to quantify the peptide yeild. Quantitation of peptide assays is not as easy as protein lysate assays. BCA protein assays perform poorly with peptide solutions and report erroneous values. A simplistic measurement is to use a nanodrop device, but absorbance measurements from a drop of solution does not report accurate values either. A more standardized and reliable approach is to Fluorescamine based assay for peptide solutions for higher accuracy [228,229]. This assay is based on the reaction between a labeling reagent and the N-terminal primary amine in the peptide(s); therefore, samples must be free of amine-containing buffers (e.g., Tris-based buffer and/or amino acids). This procedure has performance similar to the Pierce Quantitative Fluorometric Peptide Assay (Cat 23290).

**Fluorescamine peptide digest assay (based on Udenfriend and Bantan-Polak papers cited above).**

This assay is scaled to small volume reactions. Use a black flat-bottom 384 well plate such as Corning 384-Well Solid Black (Cat No. 3577). Warm solubilized fluorescamine reagent to room temperature out of light.

• Dilute peptide digest standard from 1mg/ml down to 7.8ug/mL in same buffer as sample as 8 serial dilutions.
• Include buffer blank of the sample buffer.
• Dispense 35 uL of Fluorometric Peptide Assay Buffer (0.1M Sodium borate pH 8.0) to each well.
• Dispense 5 uL of sample or 5 uL of peptide standards into designated wells.
• Dispense 10 uL of Fluorescamine reagent, mix 3x by pipetting up/down.
• Incubate 5 minutes at room temperature.
• Measure fluorescence using filters at Excitation 390nm, Emission 475 nm on a BioTek Synergy plate reader.
• Excitation 360/40nm, Emission 460/40nm (fixed values)
• Gain: scale to high wells (select highest-concentration standards)
• Mirror: Top 400nm, Read Height: 7mm
• Calculate peptide concentrations using template with quadratic curve fitting and determine yield.

**Reagents:**

Fluorometric Peptide Assay Buffer: 0.1M sodium borate in Milli-Q $H_2O$, adjust to pH 8.0 with HCl

Making the Fluorescamine Reagent from Fluorescamine (Thermo Scientific Chemicals 191675000, 100mg/100mL fluorescamine in 100% HPLC grade Acetonitrile).

- For example, weigh out 25mg fluorescamine and dissolve in 25mL Acetonitrile.
- Cover in foil, store out of light at 4ºC until use.
- Keep powdered fluorescamine covered in foil out of light at room temperature.

# 5. Mass Spectrometry Methods for Peptide Quantification

## Label-free quantification (LFQ) of peptides

LFQ of peptide precursors requires no additional steps in the protein extraction, digestion, and peptide purification workflow (**Figure 3**). Samples can be taken straight to the mass spectrometer and are injected one at a time, each sample necessitating their own LC-MS/MS experiment and raw file. Quantification of peptides by LFQ is routinely performed by many commercial and freely available proteomics software (see Data Analysis section below). In LFQ, peptide abundances across LC-MS/MS experiments are usually calculated by computing the area under the extracted ion chromatograms for signals that are specific to each peptide; this involves aligning windows of accurate peptide mass and retention time. LFQ can be performed using precursor MS1 signals from DDA, or using multiple fragment ion signals from DIA (see Data Acquisition section).

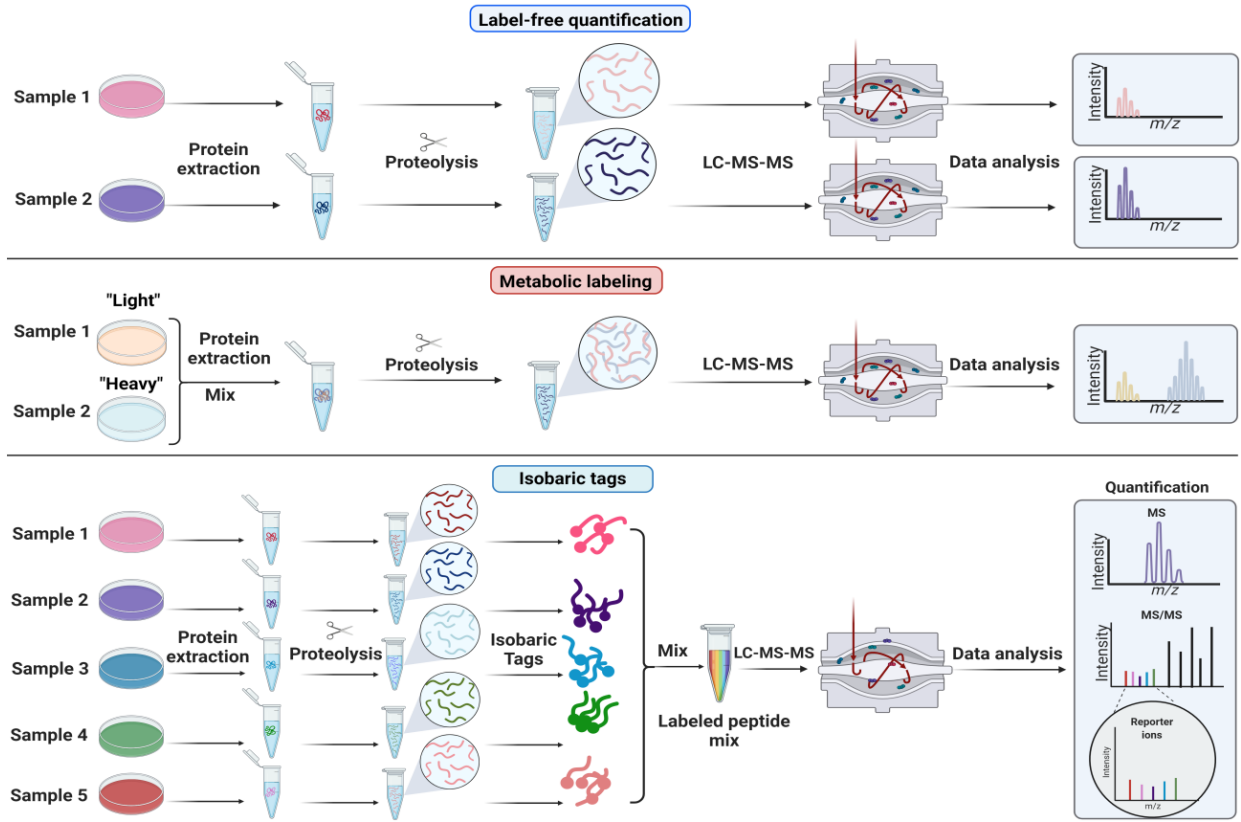

*Figure 3:* **Quantitative strategies commonly used in proteomics.** *A) Label-free quantitation. Proteins are extracted from samples, enzymatically hydrolyzed into peptides and analyzed by mass spectrometry. Chromatographic peak areas from peptides are compared across samples that are analyzed sequentially. B) Metabolic labelling. Stable isotope labeling with amino acids in cell culture (SILAC) is based on feeding cells stable isotope labeled amino acids ("light" or "heavy"). Samples grown with heavy or light amino acids are mixed before cell lysis. The relative intensities of the heavy and light peptide are used to compute protein changes between samples. C) Isobaric or chemical labelling. Proteins are isolated separately from samples, enzymatically hydrolyzed into peptides, and then chemically tagged with isobaric stable isotope labels. These isobaric tags produce unique reporter mass-to-charge (m/z) signals that are produced upon fragmentation with MS/MS. Peptide fragment ions are used to identify peptides, and the relative reporter ion signals are used for quantification.*

## Stable isotope labeling of peptides

One approach to improve the throughput and quantitative completeness within a group of samples is sample multiplexing via stable isotope labeling. Multiplexing enables pooling of samples and parallel LC-MS/MS analysis within one run. Quantification can be achieved at the MS1- or MSn-level, dictated by the upstream labeling strategy.

Stable isotope labeling methods produce peptides that are chemically identical from each sample that differ only in their mass due. Methods include stable isotope labeling in amino acid cell culture (SILAC) [230] and chemical labeling such as amine-modifying tags for relative and absolute quantification (mTRAQ) [231] or dimethyl labeling [232]. The labeling of each sample imparts mass shifts (e.g. 4 Da, 8 Da) which can be detected within the MS1 full scan. The ability to label samples in cell culture has enabled impactful quantitative biology experiments [233,234]. These approaches have nearly exclusively been performed using data-dependent acquisition (DDA) strategies. However, recent work employing faster instrumentation has shown the benefits of chemical labeling with 3-plex mTRAQ or dimethyl labels for data-independent acquisition (DIA) [235,236], an idea originally developed nearly a decade earlier using chemical labels to quantify lysine acetylation and succinylation stoichiometry [237]. As new tags with higher plexing become available, strategies like plexDIA and mDIA are sure to benefit [235,236].

## Peptide labeling with isobaric tags

Another approach to improve throughput and quantitative completeness within a group of samples is multiplexing via isobaric labels, a strategy which enables parallel data acquisition after pooling of samples. Commercial isobaric tags include tandem mass tags (TMT) [238] and isobaric tags for relative and absolute quantification (iTRAQ) [239] amongst others, and several non-commercial options have also been developed [240]. 10- or 11-plex TMT kits were recently supplanted by proline-based TMT tags (TMTpro), originally introduced as 16-plex kits in 2019 [241] and upgraded to an 18-plex platform in 2021 [242].

The isobaric tag labeling-based peptide quantitation strategy uses derivatization of every peptide sample with a different isotopic incorporation from a set of isobaric mass tags. All isobaric tags have a common structural theme consisting of 1) an amine-reactive groups (usually triazine ester or N-hydroxysuccinimide [NHS] esters) which react with peptide N-termini and ε-amino group of the lysine side chain of peptides, 2) a balancer group, and 3) a reporter ion group (**Figure 4**).

Peptide labeling is followed by pooling the labelled samples, which undergo MS and MS/MS analysis simultaneously. As the isobaric tags are used, peptides labeled with these tags give a single MS peak with the same precursor *m/z* value in an MS1 scan and identical retention time of liquid chromatography analysis. The modified parent ions undergo fragmentation during MS/MS analysis generating two kinds of fragment ions: (a) reporter ions and (b) peptide fragment ions. Each reporter ions' relative intensity is directly proportional to the peptide abundance in each of the starting samples that were pooled. As usual, b- and y-type fragment ion peaks are still used to identify amino acid sequences of peptides, from which proteins can be inferred. Since it is possible to label most tryptic peptides with an isobaric mass tag at least at the N-termini, numerous peptides from the same protein can be detected and quantified, thus leading to an increase in the confidence in both protein identification and quantification [244].

Because the size of the reporter ions is small and sometimes the mass difference between reporter ions is small (i.e., a ~6 mDa difference when using 13C versus 15N), these methods neary exclusively employ high-resolution mass analyzers, not classical ion traps [245]. There are examples, however, of using isobaric tags with pulsed q dissociation on linear ion traps

(LTQs) [246]. Suitable instruments are the Thermo Q-Exactive, Exploris, Tribrid, and Astral lines, or Q-TOFs such as the TripleTOF or timsTOF platforms [247,248].

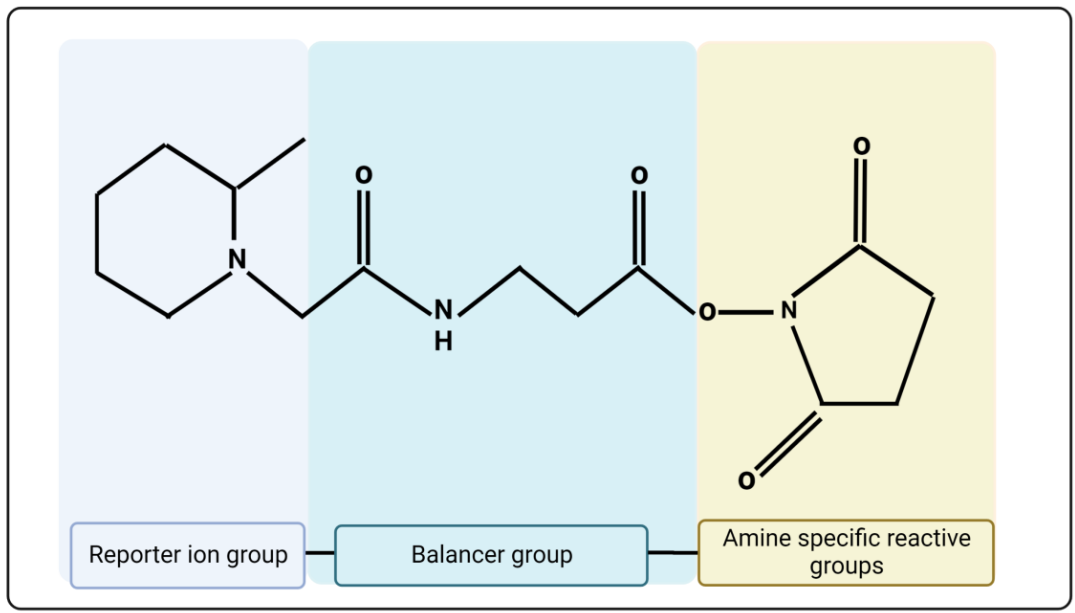

*Figure 4:* **Example chemical structure of isobaric tags "Tandem Mass Tags (TMT)".**

The following are some of the isobaric labeling techniques:

## isobaric Tags for Relative and Absolute Quantitation (iTRAQ)

The iTRAQ tagging method covalently labels the peptide N-terminus and side-chain primary amines with tags of different masses through the NHS-ester bond. This is followed by mass spectrometry analysis [249]. Reporter ions for an 8-plex iTRAQ are measured at roughly 113, 114, 115, 116, 117, 118, 119, and 121 m/z. Currently, two kinds of iTRAQ reagents are available: 4-plex and 8-plex [250]. Using 4-plex reagents, a maximum of four different biological conditions can be analyzed simultaneously (i.e., multiplexed), whereas using 8-plex reagents enables the simultaneous analysis of eight different biological conditions [251,252].

## iTRAQ hydrazide (iTRAQH)

iTRAQH is an isobaric tagging reagent for the selective labeling and relative quantification of carbonyl (CO) groups in proteins [253]. The reactive CO and oxygen groups which are generated as the byproducts of oxidation of lipids at the time of oxidative stress causes protein carbonylation [254]. iTRAQH is produced from iTRAQ and surplus of hydrazine. This reagent reacts with peptides which are carbonylated, thus forming a hydrazone group. iTRAQH is a novel method for analyzing carbonylation sites in proteins utilizing an isobaric tag for absolute and relative quantitation iTRAQ derivative, iTRAQH, and the analytical power of linear ion trap instruments (QqLIT). This new strategy seems to be well suited for quantifying carbonylation at large scales because it avoids time-consuming enrichment procedures [253]. Thus, there is no need for enriching modified peptides before LC-MS/MS analysis.

## Tandem Mass Tag (TMT)

TMT labeling is based on a similar principle as that of iTRAQ. The TMT label is based on a glycine backbone and this limits the amount of sites for heavy atom incorporation In the case of 6-plex-TMT, the masses of reporter groups are roughly 126, 127, 128, 129, 130, and 131 Da [245]. TMT works best with MS platforms which allow quantitation at the MS3 level (e.g., Thermo Fisher Orbitrap Tribrid instruments) [243,255]. In experiments performed on Q-Orbitrap or Q-TOF platforms, MS2-based sequence identification (via b- and y-type ions) and quantitation (via low m/z reporter ion intensities) is performed. In experiments performed on Q-Orbitrap-LIT platforms, MS3-based quantitation can be performed wherein the top ~10 most abundant b- and y-type ions are synchronously co-isolated in the linear ion trap and fragmented once more before product ions are scanned out in the Orbitrap mass analzer. Adding an additional layer of gas-phase purification limits the ratio distortion of co-isolated precursors within isobaric multiplexed quantitative proteomics [256,257]. Infrared photoactivation of co-isolated TMT fragment ions generates more quantitative reporter ion generation and sensitivity relative to standard beam-type collisional activation [258]. High-field asymmetric waveform spectrometry (FAIMS) also aids the accuracy of TMT-based quantitation on Tribrid systems [259]. TMT is widely used for quantitative protein biomarker discovery. In addition, TMT labeling technique helps multiplex sample analysis enabling efficient use of instrument time. TMT labelling also controls for technical variation because after samples are mixed the ratios are locked in, and any sample loss would be equal across channels. A wide range of TMT reagents with different multiplexing capabilities are available, such as TMT zero, TMT duplex, TMT 6-plex, TMT 10-plex, and TMT 11-plex. The recent addition of TMTpro tags, slight adaptations of the TMT structure based on proline, allow for higher plexing with TMTpro 16-plex [241] and now TMTpro 18-plex [242]. These TMT reagents have a similar chemical structure, which allows the efficient transition from method development to multiplexed peptide quantification [248].

## iodoTMT

IodoTMT reagents are isobaric reagents used for tagging cysteine residues of peptides. The commercially available IodoTMT reagents are iodoTMTzero and iodoTMT 6-plex [260,261].

## aminoxyTMT Isobaric Mass Tags

Also referred to as glyco-TMTs, these reagents have chemistry similar to iTRAQH. The stable isotope-labeled glyco-TMTs are utilized for quantitating N-linked glycans. They are derived from the original TMT reagents with an addition of carbonyl-reactive groups, which involve either hydrazide or aminoxy chemistry as functional groups. These aminoxy TMTs show a better performance as compared to its iTRAQH counterparts in terms of efficiency of labeling and quantification. The glyco-TMT compounds consist of stable isotopes thus enabling (i) isobaric quantification using MS/MS spectra and (ii) quantification in MS1 spectra using heavy/light pairs. Aminoxy TMT6-128 and TMT6-131 along with the hydrazide TMT2-126 and TMT2-127 reagents can be used for isobaric quantification. In the quantification at MS1 level, the light TMT0 and the heavy TMT6 reagents have a difference in mass of 5.0105 Da which is sufficient to separate the isotopic patterns of all common N-glycans. Glycan quantification based on glyco-TMTs generates more accurate quantification in MS1 spectra over a broad dynamic range. Intact proteins or their digests obtained from biological samples are treated with PNGase F/A glycosidases to release the N-linked glycans during the process of labeling using aminoxyTMT reagents. The free glycans are then purified and labeled with the aminoxyTMT reagent at the reducing end. The labeled glycans from individual samples are subsequently pooled and then undergo analysis in MS for identification of glycoforms in the sample and quantification of relative abundance of reporter ions at MS/MS level [262].

## N,N-Dimethyl leucine (DiLeu)

The N,N-Dimethyl leucine, also referred to as DiLeu, is an tandem mass tag reagent which is isobaric and has reporter ions of isotope-encoded dimethylated leucine [263]. Each incorporated label produces a 145.1 Da mass shift. A maximum of four samples can be simultaneously analyzed using DiLeu at a highly reduced cost. MS/MS analysis shows intense reporter ions i.e., dimethylated leucine a1 ions at 115, 116, 117, and 118 *m/z*. The labeling efficiency of DiLeu tags are similar to that of the iTRAQ tags. Although, DiLeu-labeled peptides offer increased confidence of identification of peptides and more reliable quantification as they undergo better fragmentation, generating higher reporter ion intensities [263].

## Deuterium isobaric Amine Reactive Tag (DiART)

DiART is an isobaric tagging method used in quantitative proteomics [264,265]. The reporter group in DiART tags is a N,N′-dimethyl leucine reporter group with a mass to charge range of 114–119. DiART reagents can a label a maximum of six samples and further analyzed by MS. The isotope purity of DiART reagents is very high hence correction of isotopic impurities is not needed at the time of data analysis [266]. The performances of DiART including the mechanism of fragmentation, the number of proteins identified and the quantification accuracy are similar to iTRAQ. Irrespective of the sequence of the peptide, reporter ions of high-intensity are produced by DiART tags in comparison to those with iTRAQ and thus, DiART labeling can be used to quantify more peptides as well as those with lower abundance, and with reliable results [264]. DiART serves as a cheaper alternative to TMT and iTRAQ while also having a comparable labeling efficiency. It has been observed that these tags are useful in labeling huge protein quantities from cell lysates before TiO2 enrichment in quantitative phosphoproteomics studies [267].

## Hyperplexing or higher-order multiplexing

Some studies have combined metabolic labels (i.e., SILAC) with chemical tags (i.e., iTRAQ or TMT) to expand the multiplexing capacity of proteomics experiments referred to as hyperplexing [268,269] or higher order multiplexing [270,271,272]. This technique combines MS1- and MS2-based quantitative methods to achieve enhanced multiplexing by multiplying the channels used in each dimension. This allows for the quantitation of proteomes across multiple samples in a single MS run. The technique uses two types of mass encoding to label different biological samples. The labeled samples are then mixed together, which increases the MS1 peptide signal. Protein turnover rates were studied using SILAC-iTRAQ multitagging [273], while various cPILOT studies employed MS1 dimethyl labeling with iTRAQ [274,275,276,277]. SILAC-TMT hyperplexing was used to study the temporal response to rapamycin in yeast [278]. SILAC-iTRAQ-TAILS method was developed to study matrix metalloproteinases in the secretomes of keratinocytes and fibroblasts [279]. TMT-SILAC hyperplexing was used to study synthesis and degradation rates in human fibroblasts [280]. Variants of SILAC-iTRAQ and BONCAT, namely BONPlex [281] and MITNCAT [282] were also developed to study temporal proteome dynamics.

# 6. Peptide/Protein Enrichment and Depletion

In order to study low abundance protein modifications, or to study rare proteins in complex mixtures, various methods have been developed to enrich or deplete specific proteins or peptides.

## Peptide enrichment

### Glycosylation

Mass spectrometry-based analysis of protein glycosylation has emerged as the premier technology to characterize such a universal and diverse class of biomolecules. Glycosylation is a heterogenous post-translational modification that decorates many proteins within the proteome, conferring broad changes in protein activity. [58,283] This PTM can take many forms. The covalent linkage of mono- or oligosaccharides to polypeptide backbones through a nitrogen atom of asparagine (N) or an oxygen atom of serine (S) or threonine (T) side-chains creates N- and O-glycans, respectively. The heterogenity of proteoglycans is not directly tied to the genome, and thus cannot be inferred. Rather, the abundance and activity of protein glycosylation is governed by glycosyltransferases and glycosidases which add and remove glycans, respectively. The fields of glycobiology and bioanalytical chemistry are intricately intertwined with mass spectrometry at the center thanks in part to its power of detecting any modification that imparts a mass shift.

Due to the myriad glycan structures and proteins which harbor them, the enrichment of glycoproteins or glycopeptides is not as streamlined as that of other PTMs [284]. The enrichment of glycoproteome from the greater proteome inherently introduces bias prior to the LC-MS/MS analysis. One must take into account which class or classes of glycopeptides they are interested in analyzing before enrichment for optimal LC-MS/MS results. Glycopeptides can be enriched via glycan affinity, for example to glycan-binding proteins, chemical properties like charge or hydrophilicity, chemical coupling of glycans to stationary phases, and by bioorthogonal, chemical biology approaches. Glycan affinity-based enrichment strategies include the use of lectins, antibodies, inactivated enzymes, immobilized metal affinity chromatography (IMAC), and metal oxide affinity chromatography (MOAC). The enrichment of glycopeptides by their chemical properties, for example by biopolymer charge and hydrophobicity, include hydrophilic interaction chromatography (HILIC), electrostatic repulsion-hydrophilic interaction chromatography (ERLIC), and porous graphitic carbon (PGC). One variation of ERLIC combines strong anion exchange, electrostatic repulsion, and hydrophilic interaction chromatography (SAX-ERLIC) has risen in popularity thanks to robustness and commercially available enrichment kits [285,286].

Chemical coupling methods most often used to enrich the glycoproteome employ hydrazide chemistry for sialylated glycopeptides. Glycan are cleaved from the stationary phase by PNGase F. The dependence of chemical coupling methods on PNGase F biases their output toward N-glycopeptides. Alkoxyamine compounds and boronic acid-based methods have also shown utility. We direct readers to several reviews on glycopeptide enrichment strategies [284,287,288,289,290]

# Phosphoproteomics

Protein phosphorylation, a hallmark of protein regulation, dictates protein interactions, signaling, and cellular viability. This post-translational modification (PTM) involves the installation of a negatively charged phosphate moiety (PO 4-) onto the hydroxyl side-chain of serine (Ser, S), threonine (Thr, T), and tyrosine (Tyr, Y), residues on target proteins. Protein kinases catalyze the transfer of PO 4- group from ATP to the nucleophile (OH) group of serine, threonine, and tyrosine residues, while protein phosphatases catalyze the removal of PO4-. Phosphorylation changes the charge of a protein, often altering protein conformation and therefore function [291]. Protein phosphorylation is one of the major PTMs that alters the stability, subcellular location, enzymatic activity complex formation, degradation of protein, and cell signaling of protein with a diverse role in cells [293]. Phosphorylation can regulate almost all cellular processes, including metabolism, growth, division, differentiation, apoptosis, and signal transduction pathways [32]. Rapid changes in protein phosphorylation are associated with several diseases [294].

Several methods are used to characterize phosphorylation using modification-specific enrichment techniques combined with advanced MS/MS methods and computational data analysis [295]. MS-based phosphoproteomics tools are pivotal for the comprehensive study for the structural and dynamics of cellular signaling networks [296], but there are many challenges [297]. For example, phosphopeptides are low stoichiometry compared to non-phosphorylated peptides, which makes them difficult to identify. Phosphopeptides also exhibit low ionization efficiency [298]. To overcome these challenges, it is important to reduce sample complexity to detect large numbers of phosphorylation sites. This is accomplished using enrichment the modified proteins and/or peptides [299,300,301]. Prefractionation techniques such as strong anionic ion-exchange chromatography (SAX), strong cationic ion-exchange chromatography (SCX), and hydrophilic interaction liquid chromatography (HILIC) are also often useful for reducing sample complexity before enrichment to observe more phosphorylation sites [302].

As with any proteomics experiment, phosphoproteomics studies require protein extraction, proteolytic enzyme digestion, phosphopeptide enrichment, peptide fractionation, LC-MS/MS, bioinformatics data analysis, and biological function inference. Special consideration is required during protein extraction where the cell lysis buffer should include phosphatase inhibitors such as sodium orthovanadate, sodium pyrophosphatase and beta-glycerophosphate [303].

Enrichment can be done at the protein level before proteolysis. Phosphoprotein enrichment typically involves the use of immobilized metal-affinity chromatography (IMAC) to selectively capture phosphorylated proteins based on their high-affinity binding to metal ions such as Ga(III), Fe(III), Zn(II) and Al(III) [304,305,306,307,308].

Enrichment is more commonly done at the peptide level because there are several advantages over phosphoprotein enrichment. First, peptides have simpler three-dimensional structures than proteins, which makes them easier to separate and analyze. Second, phosphopeptide enrichment is not hindered by small, lipophilic, and very acidic or alkaline proteins [301]. Third, prefractionation techniques such as strong anionic ion-exchange chromatography (SAX), strong cationic ion-exchange chromatography (SCX) and hydrophilic interaction liquid chromatography (HILIC) are easier to use for peptide separation than they are for protein separation, and they are more sensitive than 2D-gel electrophoresis, allowing for the identification of less abundant phosphopeptides [309]. As a result, phosphopeptide enrichment has yielded more experimental data than phosphoprotein enrichment [307]. Phosphopeptide enrichment is typically done after any isobaric labeling strategy, although several have investigated the importance of order at these stages.

Phosphopeptide enrichment often uses titanium dioxide (TiO2) [310] and/or IMAC such as Fe3+ coupled to solid-phase materials [303,306,311]. The most common cost-effective beads for phosphopeptide extraction with Ti are ReSyn and GL Sciences, and CubeBio for Fe-based beads. Often organic acids such as glutamic acid, lactic acid, glycolic acid is added to compete with acidic non-phosphopeptides for binding to the metal-ions. Carr and coworkers even demonstrated phosphoproteome analysis without any enrichment [312].

The use of Fe-IMAC column chromatography allows for the improved phosphopeptide from complex peptide mixtures [313]. Compared to other formats like StageTips or batch incubations with TiO2 or Ti-IMAC beads, Fe-IMAC columns have do not suffer from problems with poor binding or elution of phosphopeptides, and the efficiency of enrichment increases linearly with the amount of starting material [314]. Also with recent improvements to Ti based beads, the MagReSyn Ti-IMAC HP with Ti4+ attached with a flexible linker (to reduce steric hindrance) activated with phosphonate groups for Ti4+ chelation, and the MagReSyn Zr-IMAC HP, also with a flexible linker activated with phosphonate groups for Zr4+ chelation, have shown superior phosphopeptide extraction as compared to FE-IMAC.

Multiple IMAC steps can be used in parallel or sequentially to improve phosphopeptide coverage. Lai et al., (2012) showed that the combined use of Fe3+-IMAC and Ti(4+)-IMAC chromatography enables complementary identification of more phosphorylation sites than either technique alone [315]. A novel phosphopeptide enrichment technique using sequential enrichment with magnetic Fe3O4 and TiO2 particles was developed to detect mono- and multi-phosphorylated peptides [316].

More recently, the use of Src Homology 2 (SH2) domains as specific affinity reagents for phosphotyrosine is an emerging technology allowing an expanded fractionation of tyrosine phosphopeptides. Here, short chain protein domains have been constructed and affinity enhanced through yeast two hybrid screening to arrive at high affinity matrices capable of outperforming traditional IMAC approaches [317,318].

**General protocol for phosphoproteomics:**

• Collect cell line or tissue samples.
• For cell samples, suspend the cell pellet in 2% SDS lysis buffer, heat at 90°C for 5 min, and sonicate.
• For tissue samples, homogenize in liquid nitrogen and add to 4% SDS lysis buffer, then heat at 90°C for 5 min and sonicate.
• Centrifuge the samples at 3,000 rcf to remove insoluble material and remove supernatant to a new tube.
• Add 5 mM Tris (2-carboxyethyl) phosphine (TCEP) and 10 mM chloroacetamide to the samples for alkylation and reduction in the dark at room temperature for 30 minutes.
• Precipitate the protein with acetone to remove SDS.
• Dissolve the protein pellet in 8M Urea with 50 mM TEAB.
• Quantify the protein using a BCA assay.
• Perform proteolysis with trypsin using 1:100 trypsin:substrate (wt:wt). For example, if you have 1 mg of protein, add 10 ug of trypsin.
• Desalt the resulting peptides using Sep-Pak C18 columns or stage tips.
• Subject the fractions to phosphopeptide enrichment using TiO2/ Fe-IMAC beads.
• Desalt the peptides again using C18 stage tips.
• Perform LC-MS/MS.

**Tips for studying phosphorylation:**

• Cell lysates should always be prepared using phosphatase inhibitors and samples should be placed on the ice during sonication for protein extraction.
• Increase the amount of starting material of your sample for phosphoenrichment to at least 1 mg of protein or more for optimal results.
• If using anti-phosphorylation antibodies, ensure their specificity is confirmed with other methods.
• Make sure to select a suitable method for the phosphoenrichment that fits the experiment goals.
• TiO2-based phosphopeptide enrichment methods have different enrichment specificities; selecting non-phosphopeptide excluders such as glutamic acid, lactic acid, glycolic acid, and dihydroxybenzoic acid are the key part of the study [319].
• Do not use milk as a blocking agent when western blotting for phosphorylation because milk contains the phosphoprotein casein and can lead to a higher background due to non-specific binding.

# Antibody enrichments of modifications

Western blot analysis is used to detect the PTMs in a protein through the use of antibodies [320]. As an extension, pan-PTM antibodies have been used to isolate peptides bearing the PTM of interest [321]. One benefit of this approach is that peptides are less likely to experience non-specific binding than proteins [295]. Initially peptide immunoaffinity precipitation was developed to enrich for phosphotyrosine-containing peptides. This protocol was initially designed to enrich for phosphotyrosine-containing peptides [322]. Peptide immunoprecipitation yielded significantly greater coverage of the phosphotyrosine proteome than global phosphorylation enrichment strategies by enriching for a subset of the phosphoproteome. Since then, peptide immunoaffinity precipitation has been used successfully to enrich for peptides with other phosphorylation motifs [323,324] as well as peptides with other modifications such as the diglycyl-lysine residue of ubiquitin modification after trypsin proteolysis [325,326,327], acetyl-lysine [328,329,330,331,332], arginine methylation [333], tyrosine nitration [334], and tyrosine phosphorylation [335,336].

The O-linked β-D-N-acetylglucosamine (O-GlcNac) is found on serine and threonine residues of nucleocytoplasmic proteins a one of PTMs is involved in involved in the occurrence and progression of cancers in multiple systems throughout the body [337]. Anti-O-GlcNac monoclonal antibody enables enrichment from O-GlcNAcylated peptides of cells and tissues. These antibodies have high sensitivity and specificity toward O-GlcNac-modified peptides and do not identify O-GalNAc or GlcNAc in extended glycans [338].

# Abundant protein depletion (Blood samples)

Many plasma proteomics studies involve the analysis of untreated, unenriched plasma (i.e., neat plasma) [339,340]. However, the abundance range of proteins in the blood/plasma proteome exceeds 10 orders of magnitude. Due to this wide dynamic range, detection of proteins with medium and low abundance by proteomic analyses is difficult [341], and identifying protein biomarkers from biological samples such as blood is often obstructed by proteins present at higher concentrations. In fact, the top 14 most abundant proteins in human plasma constitute over 99% of the total protein mass. The removal of these high-abundant proteins enables the detection of less abundant and unique proteins. The ability to deplete abundant proteins with specificity, reproducibility, and selectivity is extremely important in proteomic studies [342].

The following are some of the methods used for abundant protein depletion:

## Dye-ligand depletion:

This method is used for the depletion of serum albumin based on the interaction between albumin and dyes like Cibacron Blue (CB) through electrostatic force, hydrogen bonding and hydrophobic interactions. The method is relatively low cost, widely available, robust and has high binding capacity. However, it lacks specificity and has varying efficiency [343,344].

## Protein-ligand depletion:

This method is used for depletion of immunoglobulins (Ig) based on the interaction between the Fragment crystallizable (Fc) region of these Igs [345] and cell wall protein A, G or A/G of Staphylococcus aureus and Streptococcus spp [346,347]. It is highly selective and has high yield and purity. However, non-specific binding may occur due to co-absorption of other proteins [348].

## Immunodepletion:

This method is used for depletion of proteins having high abundance in plasma or serum on the basis of the specific interaction of these proteins with their respective antibodies (antigen-antibody interaction) [349]. Immunodepletion has high specificity and commercial kits based on columns (Agilent MARS column) or loose beads (Thermo-Fisher High-Select depletions beads) both deplete the Top 14 most abundant protein in blood and are also readily available but expensive. For some protocols, non-specific binding to these immunodepletion columns or beads by proteins of interest is highly dependent on washing conditions [348].

## Combinatorial peptide ligand library:

This method is used for partial depletion of major proteins i.e., those with high abundance and for relative enrichment of lower and medium abundant proteins [350]. It is based on the interaction with an array of ligands which are essentially peptides of 6 amino acids in length. It is also used for normalization of the global protein abundance [351]. However, the drawbacks include non-specific binding as well as loss of proteins due to incomplete elution or inefficient binding [348].

## Precipitation:

This method of abundant protein depletion works by altering the solubility of proteins using a chemical reagent including inorganic salt solution [352], organic solvents [353], non-ionic polymer [354] and reducing agents [355]. It is extremely simple and cost-effective. However, it is less specific with a risk of protein loss, difficulty in protein resolubilization as well as time consuming [348].

## New technologies:

Newer methods of highly abundant protein depletion are based on the interaction between polymers such as bacterial cellulose nanofibers [356], cryogels [357], and nanomaterials [358]. These techniques are highly specific, relatively cheap, and very stable. They can also be reused since they have larger binding capacity and less cross-reactivity [348].

Protein enrichment/depletion strategies which make use of protein coronas [359,360] or extracellular vesicle enrichment [361] are enabling researchers to probe deeper into the plasma, serum, lymph, and cerebrospinal fluid proteomes. Automated nanoparticle (NP) protein corona-based proteomics workflows are a novel approach to perform deep blood-based proteomics analysis at unprecedented protein IDs above 6,000 proteins [362]. NPs can efficiently compress the dynamic range of protein abundances into a mass spectrometry accessible detection range and allow full automation of the protein preparation process providing a platform that can rival affinity based approaches with equivalent reproducibility and sensitivity [363].

# 7. Peptide Purification and Fractionation

## Peptide purification methods

### Solid phase extraction (SPE)

Solid phase extraction (SPE) is a common MS-based proteomics technique employed in the sample preparation. In this method, compound isolation is based on chemical and physical properties, which determines the distribution of compounds between a mobile phase (liquid) and a stationary phase (solid). After the molecules bind, washing of the bound compounds is performed and then molecules are made to elute from the stationary phase after replacing the mobile phase with the elution buffer. The material used for SPE is usually discarded after every sample and no gradient is applied for elution (single-step procedure of elution) [364]. Thus, using SPE only a specific analyte group gets separated, which depends on the stationary phase. Hence, SPE is primarily used for sample clean-up and for reducing complexity of the sample. For MS-based proteomic analysis, it is largely used to get rid of salts and other contaminants that might lead to ion suppression. The major drawback of this technique is that with SPE only a small fraction of the sample is examined because not all compounds are captured, but only those with binding capabilities same as that of the sorbent. The material for SPE is available in various types, including (micro-) columns, cartridges, plates, micropipette tips, and functionalized magnetic beads (MBs) [365,366]. Reversed-phase is the most widely used material for SPE in proteomic studies for the proteins and peptide fractionation and rarely, ion-exchange material. For the separation of glycosylated proteins and peptides, the preferred material is normal phase such as HILIC [367,368]. SPE materials which are less commonly used are silica- or polystyrene-based ones [369,370]. The other types of SPE methods are IEX, metal chelation, and affinity-based [371].

The basic idea behind the choice of binding and wash versus elution solutions for SPE is that that the binding and wash solutions should favor the interaction between the analytes of interest and the solid phase, whereas the elution solution should favor the interaction of the analyte with the liquid phase (**Figure 5**). For example, with reversed phase SPE, the solid phase is C18 or some other hydrophobic chemistry. Binding of peptides to this solid phase is based on the hydrophobicity of peptides, mostly due to their peptide backbone, but also due to the presence of amino acid side chains like leucine and phenylalanine. To encourage peptides to 'like' the stationary phase more than the liquid phase, the peptides are loaded in aqueous solution. This will enable washing of the hydrophilic contaminants like salts, small polar buffer molecules, and polar denaturants like urea. After washing the bound peptides, they can be eluted by switching the liquid phase to something hydrophobic, which allows the peptides to partition more into the liquid phase and elute from the solid phase.

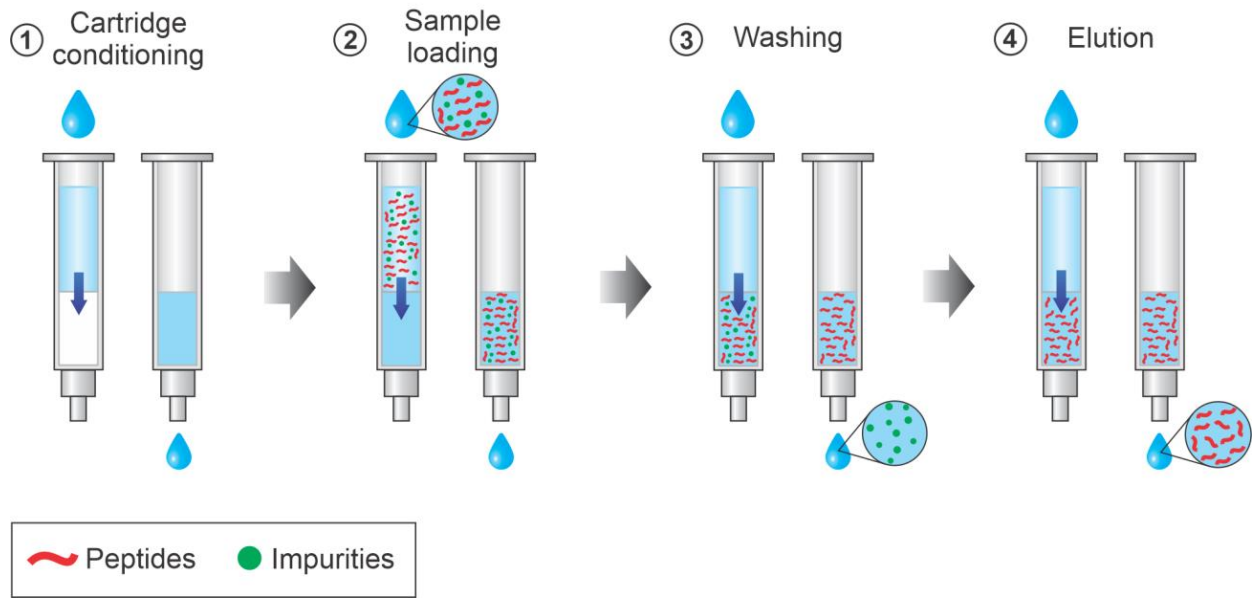

*Figure 5: **Solid phase extraction (SPE)**. SPE is a sample preparation technique that uses a solid adsorbent contained most commonly in a cartridge device to selectively adsorb certain molecules from solution. The first step is the conditioning of the cartridge which involves wetting the adsorbent to solvate its functional groups and filling the void spaces with solvent thereby removing any air in the column. This is necessary to produce a suitable environment for adsorption and thus ensure reproducible interaction with the analytes. After conditioning, the sample is loaded in the cartridge. This can be performed with the aid of positive or negative pressure to ensure a constant flow rate. In this step molecules bind the adsorbent and interferences pass through. Next, the column is washed with the mobile phase to eliminate the contaminants while ensuring the analyte remains bound. Finally, peptides are eluted in an appropriate buffer solution with polarity or charge that competes with interaction with the solid phase.*

## Specific Types of peptide purification

There are many additional peptide purification methods that are commonly used in proteomics currently. These methods include the following:

1.  StageTips, in-stagetip (iST) [372,373]
2.  SP2 or SP3 [374]
3.  Suspension trapping (S-trap) [61]

# Peptide fractionation methods

The number of peptides produced from proteolysis of the whole proteome is immense. Thus, after peptides are cleaned from interferences, they are often fractionated into subsets to enable increased proteome coverage. The characterization of the whole proteome is expected from higher order organisms, and with rising interest in post-translational modifications, an elaborate coverage of protein sequence is required. There are different methods for peptide fractionation as follows:

## Ion-exchange chromatography (IEC)

This method involves the separation based on contrasting electric charge [375]. In this approach, the mechanism of analyte retention is based on the principle of electrostatic attraction between the sample and the stationary phase functional groups (FGs), having opposite charges. IEC is classified into two types: cation-exchange and anion-exchange chromatography. In cation-exchange chromatography, at an acidic pH, the negatively charged functional groups such as sulfates are attracted to positively charged peptides, whereas, in anion-exchange chromatography, positively charged FGs such as quaternary ammoniums are attracted to peptides with negative charge at an alkaline pH. These techniques are further classified into: strong (cation [SCX] and anion [SAX] exchange), and weak exchangers (cation [WCX] and anion [WAX] exchange), based on the type of FG attached [376]. These functional groups are most commonly supported in resins made up of silica and synthetic polymers, however, some inorganic materials are sometimes used [375]. In the IEC method, peptide elution is performed using a mobile phase with higher ionic strength, to ensure peptide partition into the liquid phase. SCX along with a salt gradient/plug is a routinely used proteomics technique. In the SCX method, peptides are resolved according to their net charge, in which the peptide with the lowest positive charge is eluted first. Increasing the salt concentration decreases the peptide retention time due to competition with the electrostatic interactions between the peptides and the solid phase. However, SCX resolution is limited compared to reversed phase chromatography and will thus limit the suitability of this technique for complex mixtures [377].

## Reversed-phase chromatography (RPLC)

Reversed-phase chromatography is the most commonly used chromatographic technique which separates molecules in solution having neutral pH based on their hydrophobicity. The separation occurs on the basis of the partition coefficient of analytes between the mobile phase and the hydrophobic stationary phase. Highly polar peptides elute before the ones having less polarity because of the strong interaction with the hydrophobic functional groups forming a layer similar to a liquid around the silica resin [378]. RPLC has been widely used in separation of peptides because of its compatibility with gradient elution and aqueous samples and its retention mechanism, which modulates separation owing to changes in the properties like pH, additives and organic modifier [379]. Numerous factors influence the capacity of chromatographic peaks, such as temperature, column length, stationary phase, particle size, mobile-phase ion-pairing reagent, mobile-phase modifier, and gradient slope [380]. Usually online RPLC is done at acidic pH to ensure peptide ionization, but it can be paired with offline high pH RPLC and multiple fraction concatenation to produce orthogonal separation due to altered ionization of amino acids changing peptide hydrophobicity [381].

## Hydrophilic interaction liquid chromatography (HILIC)

Inverse-gradient chromatography was the forerunner to HILIC [382] HILIC is similar in its principle to normal-phase chromatography where the stationary phase is polar and the intitial solvent conditions are nonpolar. Gradient elution in HILIC is accomplished by increasing the polarity of the mobile phase, by decreasing the concentration of organic solvent, i.e., in the "opposite" direction compared to RPLC separations. With charged HILIC stationary phases there is also a possibility of increasing the salt or buffer concentration during a gradient to disrupt electrostatic interactions with the solute [383,384]. Thus, the peptides with less polarity elute before the more polar peptides. It is used for the separation of hydrophilic peptides and polar analytes [385]. This separation is achieved by a stationary phase that is hydrophilic in nature, for example: cyano-, diol-, amino- bonded phases [386], and an organic and hydrophobic mobile phase [383]. HILIC can also be used for enrichment and targeted proteomic analysis of PTMs, such as glycosylation, N-acetylation and phosphorylation, which increase the polarity of peptides and therefore also their retention on HILIC [378].

## Isoelectric focusing (IEF)

IEF is a type of high-resolution (HR) technique of electrophoresis used for the separation as well as concentration of peptides that are amphoteric in nature on the basis of their isoelectric point (pI) using a solution without buffer consisting of either carrier ampholytes or a gel with immobilized pH gradient (IPG). After IEF separation, the separated amphoteric peptides in the liquid phase are recovered for further analysis by RPLC-MS/MS [387]. Along with being a technique with improved resolution and capacity, for separation of peptides, IEF provides with additional information on physicochemical properties of the peptides, for example: peptide iso electric point (pI) which acts as a tool for validation and filtration for identifying MS/MS peptide sequence during the step of database search [388]. The IEF system is not only used for increasing the coverage of proteome but also in quantitative label-free [389] and stable isobaric labeling experiments [388]. IEF and gel-based separations have fallen out of favor in the last decade due to improvements in liquid chromatography.

## Electrostatic repulsion-hydrophilic interaction chromatography (ERLIC)

ERLIC is a method based on use of a weak anion exchange column operated at low pH with high organic solvent enabling isocratic elution [390]. Acidic peptides are retained by electrostatic interaction, basic and neutral peptides are retained through hydrophilic interaction made favorable by high organic solvent. This improves retention of acidic peptides and reduces retention of basic peptides compared to normal HILIC [391].

# 8. Liquid Chromatography (LC)

Chromatography is the physical sorting of a mixture of molecular species that are dissolved in a mobile phase through the strength of binding, or affinity, to the chromatographic column's stationary phase [392]. The mobile phase is pressure driven through the column and molecular species, or analytes, that have a strong affinity to the stationary phase are retained, or slowed, while those with a weak affinity pass through quickly. Thusly the analytes are separated by order of elution from the column. Chromatography can exploit most physical properties of the analytes, including ionic charge (anion/cation exchange chromatography), hydrogen binding (hydrophilic interaction), and size (size exclusion chromatography, capillary electrophoresis). In some chromatographic separations the mobile phase composition is adjusted by mixing two or more buffers at different ratios to influence the strength of affinity of individual analytes to the stationary phase and exquisitely regulate retention.

Mass spectrometers suffer from ion suppression, a phenomenon where the over-abundance of one or a few species within the ion population entering the mass spectrometer masks the presence of less abundant species [393]. Complex biological samples, such as tissue, cell lysate, or physiological fluids contain a wide dynamic range of molecule concentrations that span many orders of magnitude. The physical separation of analytes from biological samples by LC reduces the complexity of the ion population presented to the mass spectrometer at a given time, thus allowing the instrument to carry out the necessary fragmentation scans to identify and quantify the detectable species. Therefore, one major benefit of LC is that it allows detection of low abundant analytes in other elution windows.

The field of proteomics predominantly separates peptides using reversed phase liquid chromatography [394,395,396]. Reversed stationary phase is most commonly composed of microscopic (1-3 μm) silica beads coated with covalently bound long (e.g. C18) hydrophobic alkyl chains. The hydrophobic side chains of certain residues and the peptide backbone bind to this stationary phase through non-polar interactions. These interactions are strong in an aqueous solvent but are disrupted when the organic composition of the solvent is increased. Thus, in a reversed phase separation the proportion of non-polar, or organic, solvent in the mobile phase is gradually increased to release analytes from the stationary phase based on the strength of hydrophobic binding: weakly bound hydrophilic analytes elute with a low organic level in the mobile phase and strongly bound hydrophobic analytes only elute when the organic composition reaches a higher percentage. By far the most popular combination of solvents for peptide analysis is water and acetonitrile with dilute acid modifier (such as 0.1% formic acid or 0.5% acetic acid). The programmed rate at which the proportion of organic solvent is increased in the mobile phase is called the "gradient", which you will often find described in the methods sections for reversed phase separations.

## LC considerations regarding Electrospray Ionization (ESI)

LC is paired to MS through ESI, and LC parameters greatly influence ESI. The analytes are eluted in a liquid mobile phase and must be released into the gas phase as charged ions for detection by mass spectrometry. This is achieved by spraying the eluent from the chromatographic separation through a narrow nozzle under a high voltage potential (1,000-4,000 volts) between the nozzle, or emitter, and the mass spectrometer inlet. The eluent is sprayed as a mist of small charged droplets that explode into smaller droplets as the solvent evaporates and the repelling columbic force of the charged analytes increases [397]. The droplets become progressively smaller until individual analyte molecules are ejected. The ejected analytes are ionized by the retained charge and can thus be manipulated by the electric fields in the mass spectrometer to measure their mass and perform the necessary fragmentations to elucidate structure. The chromatographic flowrate (the volume of mobile phase driven through the chromatographic column per unit time e.g. uL/min) dictates the efficiently of electrospray ionization (proportion of analytes eluting from the column that are ionized and into the gas phase) and is thus a key consideration for sensitivity of analysis [398]. Reduced flowrates generate smaller droplets which degrade into ejected charged analytes rapidly, thus resulting in more detectable analytes and higher ionization efficiency. Electrospray ionization efficiency is also aided by an inert sheath gas, high temperature, and reduced pressure between the nozzle and ion lensing elements, thus decent sensitivity can still be achieved at high flowrates. For more detailed discussion of ionization, see the "Ionization" section.

# Quality Attributes of Chromatographic Separation

The quality of chromatographic separation defines the number of analytes that are identified and quantified by LC-MS analysis. The theory around chromatographic separation was developed when LCs were paired with spectrophotometer detectors that only measure the combined signal intensity from all co-eluting analytes. The ability of MS to simultaneously detect the masses of individual components re-defines the significance of certain LC attributes. For those looking for mathematical descriptions of chromatographic quality, refer to the "Van Deemter equation", which we do not cover here to maintain simplicity [399]. The following attributes are the most important to consider in LC-MS.

## Chromatographic Resolution

Chromatographic resolution is defined as the ability to fully resolve adjacent chromatographic peaks containing analytes with nearly equal affinities to the solid phase. In mass spectrometry analytes are distinguished by mass even if they are not resolved by LC. Thus in LC-MS, the more relevant, but closely related concept is the peak width at the half maximum (FWHM) of the peak. A low FWHM indicates a sharp elution peak. In a sharp peak the entirety of the analyte population is electrosprayed into the mass spectrometer in a short time thus increasing the signal. Low FWHM of high abundance species also confines their ionization suppression to narrow time windows, which means a lower number of co-eluting analytes are hidden. Conversely, high FWHM means that the analyte signal is spread out over time, thus reducing sensitivity. Furthermore, at a high FWHM, high abundance species mask analytes through ion suppression over a larger portion of the separation.

## Peak Capacity

Peak capacity is defined as the maximal number of peaks that ideally can be completely resolved in a pre-established time window. A long separation in which FWHM remains low would have a large peak capacity and thus allow identification of many species. Unfortunately increasing the length of a reversed phase gradient also increases the FWHM due to an increase in diffusion, which results in a diminishing return for longer analytical methods. A longer separation provides more time and opportunities for the mass spectrometer to sample each analyte to acquire fragmentation spectra required for identification and the selection of gradient length should consider both the desired throughput and the speed of the MS data acquisition strategy.

## Reproducibility and Robustness

Reproducibility is defined as the ability to repeatedly obtain the same measurement for the same analytes each time that the analysis is repeated. In liquid chromatography this means that each analyte should elute at nearly the same retention time (the time elapsed since the start of the analysis until the analyte's elution from the chromatographic column) with the same peak width. Robustness is the ability of the system to maintain reproducible performance despite nonoptimal conditions. The most typical obstacles to robustness are mechanical wear of the system components and the analytical column, fouling of the system by contaminants introduced in the samples, and clogging due to accumulation of contaminants. High flow methods tend to be more robust due to reduced impact of pump and plumbing configurations and changes in dwell volumes, and the wider bore of the components used is more resilient to clogging. However, higher flowrate comes at the cost of reduced sensitivity due to reduced ionization efficiency at higher flow rates and increases in the overall peak volume at constant sample loading, thus nanoflow (100-300 nL/min flowrate) chromatography remains a widely utilized strategy in proteomics. For applications where sample is not limited, slightly higher amounts of applied samples can take advantage of robustness of higher flow rates in the microflow range using newer optimized electrospray sources [400].

### Throughput and Instrument Utilization

Throughput is the number of samples that are analyzed in a given timeframe, for example samples per day. High throughput is required to analyze thousands of samples that truly represent biological diversity in a timely manner. Increasing throughput means less data are collected for individual samples. Furthermore, many steps in the LC process are required for sample analysis in which no useful data is collected including sample injection, and system cleaning and equilibration, which reduce the ratio of data collected to instrument operation time, or instrument utilization. The ability to perform these steps while a different sample is analyzed, or parallelization, increases instrument utilization and the amount of data collected by several minutes which is a significant increase when several samples are analyzed per hour.

# Trapping and Pre-Columns

Trapping and pre-columns are short chromatographic columns that are used to increase robustness of an LC-MS system. A pre-column is connected directly to the front of the analytical column and is intended to be disposable and to absorb contaminants and protect the analytical column. The trapping column is a connected indirectly to the analytical column through a valve. The valve can be switched to redirect the flow through the trapping column away from the analytical column. This allows analytes to be loaded on the trapping column while analytes that are hydrophilic and poorly retained are washed away and do not contaminate the analytical column or the mass spectrometer. This process is referred to as desalting, and once it is complete, the valve configuration is changed to connect the trapping column to the analytical column, and analytes captured on the trapping column can be eluted off the trap and through the analytical column for analysis by MS. Certain trapping columns can be operated in both directions, which allows aggregates to be flushed away when the trapping column is cleaned in the reverse direction. Additionally trapping columns are shorter and have less backpressure so they can be loaded with sample quickly at a fast flowrate. Whereas loading the sample directly on the analytical column requires a slower flowrate. Two trapping columns can be used in tandem to provide parallelization, while one trapping column is cleaned and loaded with samples the second trapping column is in line with the analytical column analyzing the sample that was loaded on it in the previous run [401,402].

# Multi Dimensional LC

Depth of profiling has previously been increased by combining two or more orthogonal LC separations. Orthogonal in this context means that each separation sorts the analytes into different populations [403]. For example, a separation based on positive charge (strong cation exchange, SCX) separates analytes based on positive charge, and when paired with reversed phase chromatography results a higher peak capacity and more analytes identified. The first highly popular method was multidimensional protein identification technology (MudPIT), which used online separation by SCX followed by C18 reversed phase [404]. However, the resolution of peptide separation by SCX is low, leading to the presence of peptides in many fractions. The currently accepted most popular method for two-dimensional separation combines iterative reversed phase at different high and then low pH to sort analytes by changes in hydrophobicity due to changes in amino acid side chain ionization. Although the separations are not entirely orthogonal, multiple fraction concatenation across the high pH elution can produce entirely orthogonal peptide sets [405]. In recent years the focus of proteomics has shifted from deep profiling of fewer samples to rapid profiling of large cohorts. Thus, lengthy multidimensional methods have been replaced with single shot experiments only using one dimension of high resolution reversed phase separation [406]. However peak capacity is regained by using ion mobility spectrometry (separation of ionized peptides in the gas phase).

# 9. Peptide Ionization

Until the early 1990s, peptides analysis by mass spectrometry was challenging. Hard ionization techniques in use at the time, like fast atom bombardment, were not directly applicable to peptides without destroying or breaking them. The soft ionization techniques however, revolutionized the proteomics field and it became possible to routinely ionize and analyze peptides using MALDI and ESI techniques at high-throughput scale. These two techniques were so impactful that the 2002 Nobel Prize in Chemistry was co-awarded to John Fenn (ESI) and Koichi Tanaka (MALDI) "for their development of soft desorption ionization methods for mass spectrometric analyses of biological macromolecules" [407].

## MALDI

The term "Matrix-assisted LASER desorption" was coined by Hillenkamp and Karas in 1985, although this orignal paper only applied the technique to dipeptides [408]. It was Koichi Tanaka who first applied this idea to proteins above 10,000 Daltons in size and published a paper in the Proceedings of the 2nd Japan-China Joint Symposium on Mass spectrometry in 1987 (Tanaka, K., Ido, Y., Akita, S., Yoshida, Y. and Yoshida, T. (1987) Detection of high mass molecules by laser desorption time-of-flight mass spectrometry. Proceedings of the 2nd Japan-China Joint Symposium on Mass spectrometry, 185-187), and then in a follow-up paper published in 1988

[7]. A few months later, Karas and Hillenkamp also demonstrated MALDI applied to proteins above 10kDa with MALDI [409]. This resulted in some controversy about who should have won the Nobel prize [410] as it was felt by the community that Hillenkamp and Karas had provided the technology several years before but it was Koichi Tanaka that was the first to apply the MALDI technology to proteins a year before Hillenkamp and Karas.

MALDI first requires the peptide sample to be co-crystallized with a matrix molecule, which is usually a volatile, low molecular-weight, organic aromatic compound (**Figure 6**). Some examples of such compounds are cyno-hydroxycinnamic acid, dihyrobenzic acid, sinapinic acid, alpha-hydroxycinnamic acid, ferulic acid etc [411]. Subsequently, the analyte is placed in a vacuum chamber in which it is irradiated with a LASER, usually at 337nm [412]. This laser energy is absorbed by the matrix, which then transfers that energy along with its free protons to the co-crystalized peptides without significantly breaking them. The matrix and co-crystallized sample generate plumes, and the volatile matrix imparts its protons to the peptides as it gets ionized first. The weak acidic conditions used as well as the acidic nature of the matrix allows easy exchange of protons for the peptides to get ionized and fly under the electrical field in the mass spectrometer. These ionized peptides generally form the metastable ions, most of them will fragment quickly [413]. However, it can take several milliseconds and the mass spectrometry analysis can be performed before this time. Peptides ionized by MALDI almost always take up a single charge and thus observed and detected as [M+H]+ species.

## MALDI Mechanism

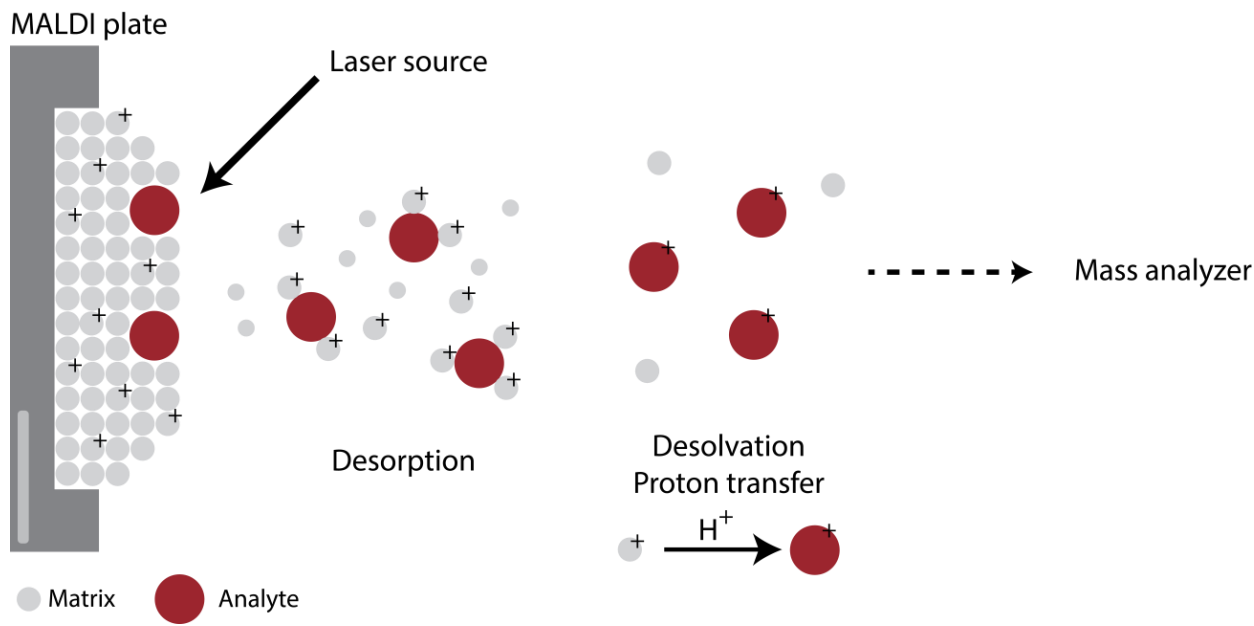

Figure 6: **MALDI** The analyte-matrix mixture is irradiated by a laser source, leading to ablation. Desorption and proton transfer ionize the analyte molecules that can then be accelerated into a mass spectrometer.

# Electrospray Ionization

ESI was first applied to peptides by John Fenn and coworkers in 1989 [6]. Concepts related to electrospray Ionization (ESI) were published at least as early as 1882, when Lord Rayleigh described the number of charges that could assemble on the surface of a droplet [397]. ESI is usually coupled with reverse-phase liquid-chromatography of peptides directly interfaced to a mass spectrometer. A high voltage (~ 2 kV) is applied between the spray needle and the mass spectrometer (**Figure 7**). As solvent exits the needle, it forms droplets that take on charge at the surface, and through a debated mechanism, those charges are imparted to peptide ions. The liquid phase is generally kept acidic to help impart protons easily to the analytes.

Tryptic peptides ionized by ESI usually carry one charge one the side chain of their C-terminal residue (Arg or Lys) and one charge at their n-terminal amine. Peptides can have more than one charge if they have a longer peptide backbone, have histidine residues, or have missed cleavages leaving extra Arg and Lys. In most cases, peptides ionized by ESI are observed at more than one charge state. Evidence suggests that the distribution of peptide charge states can be manipulated through chemical additives [414].

## Electrospray Mechanism

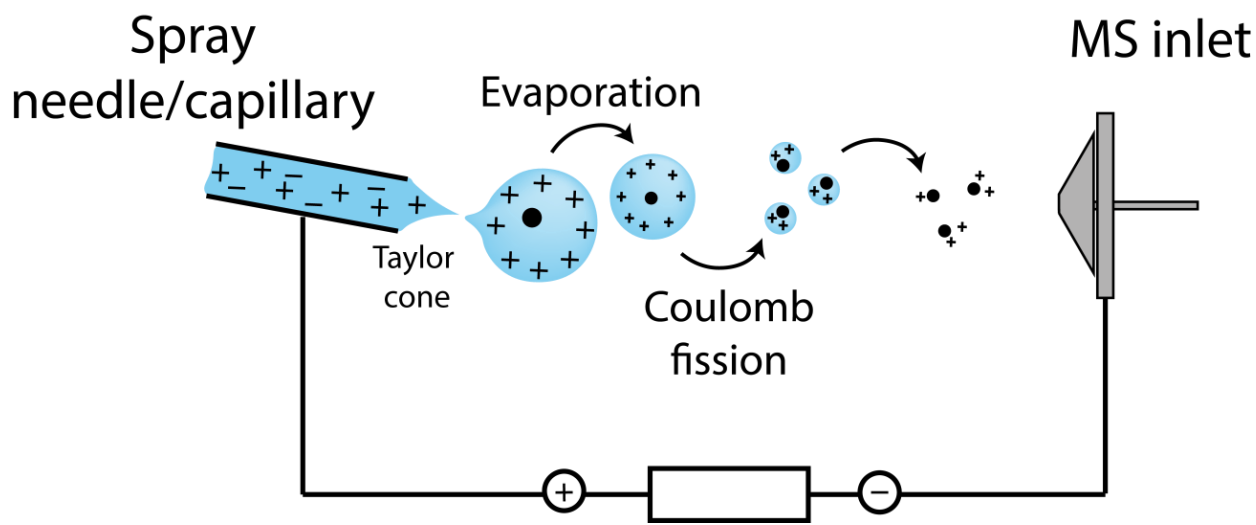

*Figure 7:* ***Electrospray Ionization*** *Charged droplets are formed, their size is reduced due to evaporation until charge repulsion leads to Coulomb fission and results in charged analyte molecules.*

The main goal of ESI is the production of gas-phase ions from electrolyte ions in solution. During the process of ionization, the solution emerging from the electrospray needle or capillary is distorted into a Taylor cone and charged droplets are formed. The charged droplets subsequently decrease in size due to solvent evaporation. As the droplets shrink, the charge density and Coulombic repulsion increase. This process destabilizes the droplets until the repulsion between the charges is higher than the surface tension and they fission (Coulomb explosion) [415] [416]. Typical bottom-up proteomics experiments make use of acidic analyte solutions which leads to the formation of positively charged analyte molecules due to an excess presence of protons.

# 10. Types of Mass Spectrometers used for Proteomics

## Mass spectrometry

Mass spectrometry is a science of ions; mass spectrometers serve as sophisticated instruments for determining the masses of compounds and elements. Mass spectrometry can therefore be likened to an ultra-precise weigh scale that can differentiate mass variations down to a single electron, or even lighter. Since J.J. Thomson's initial exploration in 1912, the field of mass spectrometry has undergone numerous improvements, spanning from isotope assessment to the interpretation of biomacromolecules [417], all thanks to the combined efforts of diverse fields like chemistry, physics, electronic engineering, and computer science. Nowadays, with the rapid improvement of sensitivity, mass resolution, tandem mass spectrometry methods and ion dissociation methods, mass spectrometers have evolved as a core tool for proteomic (and metabolomic) analysis. It is precisely the widespread application of mass spectrometry in proteomics analysis that has given rise to more instrument manufacturers and a greater diversity of mass spectrometer types. This also brings a happy annoyance to many beginners or researchers in other fields who have no background in mass spectrometry: which manufacturer and which type of mass spectrometry should I choose to analyze my samples? Here, to help new learners build a basic understanding faster, we will briefly introduce some basic concepts, common types of mass spectrometers, and their suitable application scenarios.

## Mass Spectrometer Structure and Basic Principles

The fundamental principle of mass spectrometry revolves around specific physical processes that can be described by various mathematical formulas. Since this article serves as a guide for those new to the field, particularly those from a biology background, we've chosen to steer clear of delving too deeply into intricate mathematical and physical explanations. However, for those keen on a deeper understanding, we've included references pertaining to these foundational principles. Our focus lies on introducing fundamental concepts and outlining the typical workflow in mass spectrometry.

The process of mass spectrometry (MS) is to generate gas phase ions from compounds in samples by any suitable method, to separate these ions by their $m/z$ ratio, and then detect them by their respective $m/z$ and abundance. The successful implementation and demonstration of this process requires participation of five fundamental systems (**Figure 8**):

### 1) The ion source.

The ion source is where gas phase ions are generated. Common methods of ionization in mass spectrometry include electrospray ionization (ESI), matrix assisted laser desorption ionization (MALDI), atmospheric pressure chemical ionization (APCI), electron ionization (EI) and chemical ionization (CI) [418,419]. For proteomic analysis, soft ionization methods such as ESI and MALDI are the most widely applied techniques [6,7], which will be discussed in next chapter in more detail.

## 2) The mass analyzer.

The mass analyzer is where gas phase ions are separated according to their *m/z* ratio based on physical principles. There are several types of mass analyzers applied in mass spectrometry, including the quadrupole, linear ion trap and three-dimensional ion trap, Orbitrap, Fourier transform-ion cyclotron resonance (FT-ICR), time-of-flight (TOF), and the magnetic sector analyzers [420,421], each with unique advantages and applications (Table 12.1). For proteomic analysis, tandem mass spectrometry, which involves combining two or more mass analyzers, is typically used to achieve precursor selection, structural analysis, improved sensitivity, and better mass resolution [422]. The mass analyzer is the core component of a mass spectrometer, it is also the most important factor that we need to take into consideration when choosing a mass spectrometer for a specific project.

## 3) The detector.

The detector is where ions are detected and their respective *m/z* values and abundances are recorded, generating a mass spectrum. Common types of ion detectors, including the Electron Multiplier (EM), Photomultiplier Tube (PMT), Microchannel Plate (MP) and Faraday Cup (FC), along with a summary of their strengths and limitations, are illustrated in Table 12.2. It is worth noting that Orbitrap and FT-ICR mass analyzers don't use conventional detectors as listed above. Instead, these analyzers detect an image current produced by oscillating ions [423,424,425]. In both mass analyzers, the detector is essentially measuring an electrical current (or more accurately, a voltage that's proportional to the current) that's induced by the motion of the ions. This signal is then processed to extract the frequencies of oscillation and Fourier-transformed into a mass spectrum, which is quite different from other types of detectors that count individual ions or particles striking a surface. Longer transients generate higher resolution spectra.

## 4) The vacuum system.

This is designed to maintain a high-vacuum environment for ions' transmission inside the instrument. The vacuum system consists of different type of pumps including roughing vacuum pumps (rotary vane pumps, scroll pumps) and high-vacuum pumps (turbo molecular pumps, diffusion pumps). Maintaining a high vacuum is essential to reduce collisions between analyte ions and inert gas molecules during their transmission from one region of the mass spectrometer to another, or during oscillations within a mass analyzer. Collisions within the vacuum chamber may lead to unstable ion trajectories and poorer transmission efficiency, in turn leading to lower resolving powers and poorer sensitivities. Even so, some inert gas is intentionally plumbed into the mass spectrometer either for collisional activated dissociation (CAD), typically with nitrogen, helium, or argon, or to dampen ions' energy so that they don't fly out of the back of the instrument. FT-ICR and Orbitrap mass analyzers require higher vacuum in the 10-9 to 10-11 Torr range, while TOFs require medium vacuum in the 10-7 to 10-8 Torr range, and quadrupole and ion trap insturments require a relatively low vacuum in the 10-5 to 10-6 Torr range.

## 5) The control system.

This is needed to regulate and coordinate the various parts of the mass spectrometer to ensure seamless functioning. This typically includes ion source control, mass analyzer control, detector control, data acquisition control, interfacing with auxiliary systems (such as a liquid chromatograph and gas chromatograph), and modules for instrument diagnostics and calibration.

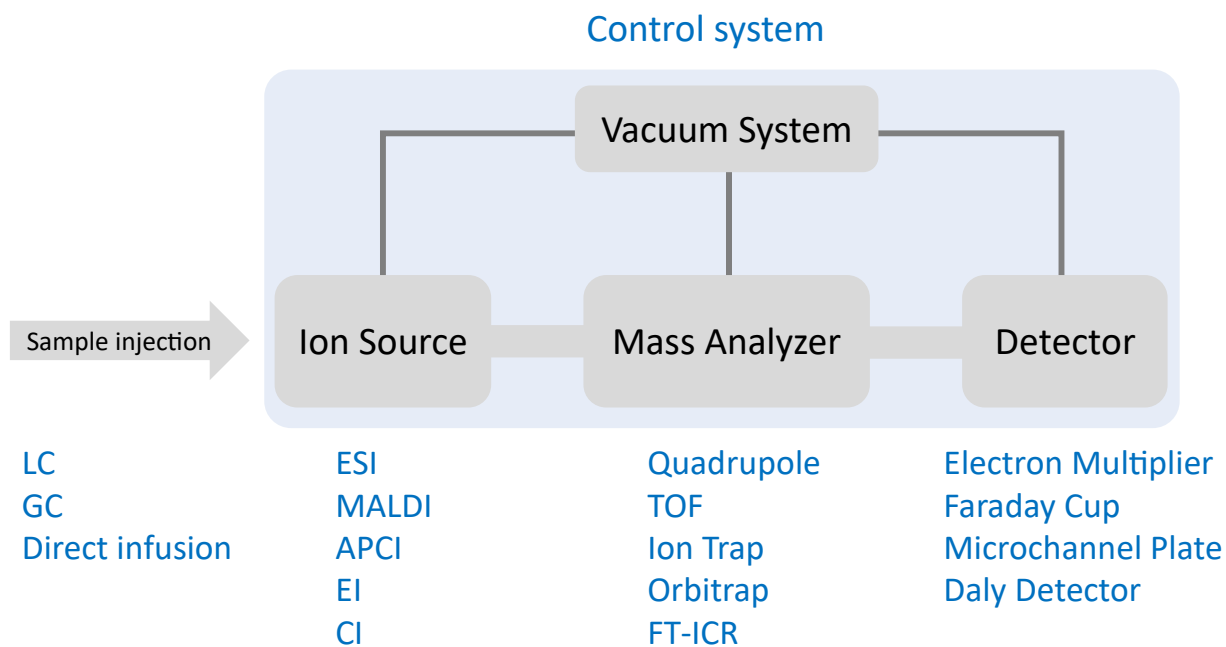

*Figure 8:* **Diagram of typical mass spectrometer modules.** *Systems must have an ion source, mass analyzer, detector, vacuum system, and control system.*

Table 10-1 Common mass analyzers.

| Type | Acronym | Principle | Characteristics |
|---|---|---|---|
| Time-of-flight | TOF | Time dispersion of a pulsed ion beam; separation by the time it takes for ions to travel a fixed distance | High-speed analysis, large mass range and good sensitivity. Suited for fast data acquisition and high-throughput applications. Modern TOF systems usually can achieve mass resolution well over 10,000 (m/$\Delta$m) or even higher. |
| Linear quadrupole | Q | Continuous ion beam in linear radio frequency quadrupole field; separation due to instability of ion trajectories | High transmission efficiency, simple design, good sensitivity, and tunable mass range; relatively low mass resolution ranges from several hundreds to a thousand; often used in tandem mass spectrometry (MS/MS) experiments |
| Quadrupole ion trap | QIT | Traps ions by electromagnetic fields; separation in three-dimensional radio frequency quadrupole field by resonant excitation | Efficient for fragmenting ions and structural elucidation, higher sensitivity, and relatively compact which good for benchtop instruments. Relatively a low mass resolution around 1,000 - 3,000. |
| Fourier transform-ion cyclotron resonance | FT-ICR | Traps ions in a strong magnetic field by Lorentz force; separation by cyclotron frequency, image current detection and Fourier transformation of transient signal | Ultimate high mass resolution (up to 1,000,000), making it ideal for elemental and isotopic analysis. Large size, low speed, and expensive in terms of both initial purchase cost and ongoing operation and maintenance costs. |
| Orbitrap | Orbitrap | Axial oscillation in inhomogeneous electric field; detection of frequency after Fourier transformation of transient signal | Extremely high resolution and accuracy (up to 500,000), capable of resolving complex mixtures with high sensitivity. Relatively low speed, expensive in terms of both initial purchase cost and ongoing operation and maintenance costs. Need high vacuum. |

Table 10.2 Common detectors.

| Type | Principle | Characteristics |
|------|-----------|-----------------|
| Electron Multiplier | Amplifies signals by utilizing a sequence of dynodes that emit secondary electrons when struck by an incident electron, creating a cascading effect. This results in an amplified output current at the final anode, proportional to the intensity of the initial signal. | Very good signal amplification to even one electron (may cause more noise dependent on gain), high sensitivity, need high vacuum and high voltage, expensive. Limited dynamic range, finite lifespan and need to be replaced periodically |
| Faraday Cup | Charged particles, such as ions or electrons, enter the cup and transfer their charge to it, causing a change in electric potential that can be measured over time to infer the number of particles. | Suitable for particles and charge state detection. Simplicity and robustness, Wide dynamic range, no need for high voltage and high vacuum. Lower sensitivity. Sensitive to Secondary Emission directional sensitivity (direction of incoming particles) |
| Microchannel Plate | Similar to electron multiplier, a two-dimensional matrix or "plate" of many tiny, parallel, hollow channels made from a type of glass that can generate secondary electron emissions upon incident particles striking the channel walls. These secondary emissions create an electron avalanche down the channels and amplifies the original signal. | Signal Amplification (Not as good as electron multiplier, but lower noise), Spatial Resolution ability, shorter life expectancy due to channel aging and depletion of the secondary emission material, smaller and cheaper than electron multiplier |
| Daly Detector | Directing ions onto a surface (Doorknob) to trigger the emission of electrons, which are then accelerated towards a phosphor screen to produce photons, that are subsequently detected and amplified by a photomultiplier tube, thereby converting the ion signal into a measurable electrical signal. | High gain, ruggedness, wide dynamic range, suitable for high mass and high energy ions. Limited mass resolution, larger size and need high voltage, finite lifespan. |

# Types of mass spectrometers used for proteomics.

Typically, mass spectrometers are named based on the abbreviations of their principal or tandem mass analyzers. This naming convention stems from the fact that the mass analyzer forms the core component of a mass spectrometer, and it also dictates key performance attributes such as mass resolution, scanning speed, sensitivity, and cycle time. These performance metrics, in turn, determine what type of analysis we can conduct, its speed and its accuracy. Next, we will focus on introducing several classic tandem mass spectrometry types commonly used in proteomics.

## 1. Triple quadrupole (QqQ).

Triple quadrupole mass spectrometer (often abbreviated as QqQ, QQQ, TQ, or TQMS) is a type of tandem mass spectrometer where three quadrupole mass analyzers are combined in series

(**Figure 9**). Each quadrupole is essentially a set of four parallel metal rods to which radio frequency (RF) and direct current (DC) voltages are applied to each opposing pair of rods. The QqQ operates in a synchronized manner to isolate ions of interest (according to the Mathieu function) in the first quadrupole, induce fragmentation with inert gas in the second, and then detect the resulting product ions in the third quadrupole. Specifically, the first quadrupole (Q1) is a mass filter, where ions of a specific *m/z* are selected from the incoming ion beam. This is achieved by adjusting the voltage applied to the pair rods within the quadrupole, allowing ions with a particular *m/z* value to pass through while deflecting others. The second quadrupole (Q2), also known as the collision cell, is where selected ions from Q1 are fragmented into product ions. This fragmentation happens due to the collisions between inert gas molecules (nitrogen, argon, or helium) and ions, which causes the ions to break up (fragment) into smaller pieces (fragment ions). For more detail about peptide fragmentation, see the Tandem Mass Spectrometry section. This process is known as collision-induced dissociation (CID) [426,427]. The Q2 is usually only subjected to RF potential and does not filter ions; instead, it transmits the product ions to the third quadrupole. In some tandem mass spectrometry, hexapoles or octupoles are also used to replace a quadrupole as the collision cell. Lastly, the third quadrupole (Q3) acts as a secondary mass filter, similar to Q1, but with the purpose of selecting specific fragment ions produced in the collision cell while excluding other ions. The chosen ions are then directed to the detector, where their abundance is measured (**Figure 9**). This process, involving precursor ion selection, precursor ion fragmentation, and product ion detection, is a general operating principle in tandem mass spectrometry and determines what kind of scan mode you can utilize. While discovery-based proteomics approaches can be performed on triple-quadrupole systems, selected ion monitoring (SRM), also refered to as multiple reaction monitoring (MRM) has contirbuted to their popularity. A key characteristic and advantage of QqQ is the flexibility of choosing various scan modes [427,428,429], such as the following.

## 1. Product Ion Scan:

Q1 is set to filter a specific precursor ion, which is then fragmented in Q2. Q3 scans the full range of product ion masses. This mode is usually used to identify the structure of a particular compound.

## 2. Precursor Ion Scan:

Q3 is set to filter a specific product ion. Q1 scans the full range of precursor ions, that when fragmented in Q2, yield the selected product ion. This mode is used to find compounds that yield a specific fragment ion, which can be particularly useful when looking for compounds with a common structural motif.

## 3. Neutral Loss Scan:

Both Q1 and Q3 scan the full range of ions, but with a mass difference equal to a specific "neutral loss". This mode is used to identify compounds that, when fragmented, lose a specific neutral molecule.

## 4. Multiple/Selected Reaction Monitoring (M/SRM):

Both Q1 and Q3 are set to filter specific ions (precursor and product, respectively). This highly selective mode is used for quantitative analysis of specific compounds, offering excellent sensitivity and specificity [430,431].

The triple quadrupole mass spectrometer is a highly versatile instrument, capable of both qualitative and quantitative analysis. Enke and Yost at Michigan State University developed the first commercial triple-quadrupole mass spectrometer in the late 1970s [432]. QqQ is particularly well-suited for targeted quantitative analysis due to its high sensitivity, selectivity, and dynamic range, which has made it a go-to instrument in areas such as drug metabolism studies, environmental monitoring, food safety analysis, pharmaceuticals, and clinical diagnostics [433,434,435,436].

However, Quadrupoles suffer from inherent limitations in mass resolution due to the constraints of principles and precision in mechanical manufacturing. Consequently, they face difficulties in accurately identifying unknown molecules within complex mixtures and thus not appropriate for applications like structure analysis and biomarker discovery.

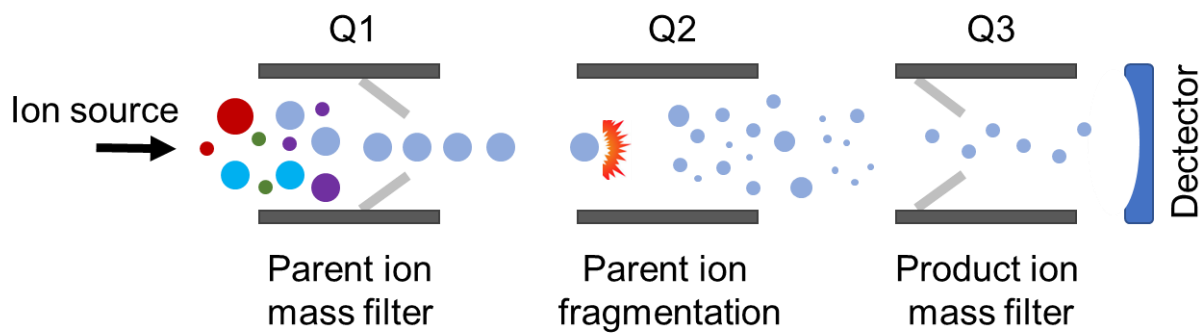

*Figure 9:* **Schematic diagram of typical QqQ system.** *Three quadrupoles enable precursor selection, fragmentation, and the fragment ion selection.*

# 2. Q-TOF

Even though quadrupoles face difficulties in accurately identifying unknown peptides within complex mixtures due to its mass resolution, they serve effectively as mass filters, making them an excellent choice for combining with other high-resolution mass analyzers to form tandem mass spectrometry systems. One commonly used approach is Quadrupole-Time-of-Flight Mass Spectrometer (Q-TOF-MS), a 'hybrid' device, integrating quadrupole techniques with a time-of-flight mass analyzer. W.E. Stephens constructed and published the design of the first time-of-flight (TOF) analyzer in 1946 [437,438]. The principle of TOF is quite straightforward: ions of different *m/z* are imparted with the same initial kinetic energy (E = Uq = ½ mv2) and then separated over time as they travel along a field-free drift path of known length. If all ions begin their flight simultaneously, or at least within a short enough time span, the lighter ions will reach the detector before the heavier ones due to their faster velocity (V)[439]. Based on this principle, the *m/z* of different ions can be calculated according to the order in which they reach the detector. Similarly, we can easily conclude that the longer the drift path, the higher of the mass resolution can reach if keep the response time of detector the same. In fact, in pursuit of higher mass resolution, researchers have indeed built time-of-flight (TOF) drift tubes that are tens of meters long. However, apparently, this is not practical for widely application in a regular lab place. An alternative way to expand drift length and achieve higher resolution is to apply reflector (often called a reflectron). The principles and advantages of using a reflector can be summarized as follows.

Under ideal circumstances within a TOF mass spectrometer, ions sharing the same $m/z$ would reach the detector concurrently post-acceleration, thus generating a sharp peak on the mass spectrum. However, the inherent oscillation path variability of ions within the mass spectrometer makes it challenging to maintain uniform initial kinetic energy amongst all ions, leading to peak broadening and a substantial reduction in mass resolution. The reflector is designed to rectify this issue. Comprising a series of electrodes that set to different voltages, the reflector generates a retarding electric field that reverses ion trajectories back through the flight tube. Notably, the reflector is engineered such that ions carrying lower kinetic energy delve less into the reflector and have a reduced flight path, while those with higher kinetic energy permeate more deeply and follow a longer flight path. This equalizes the variances in initial kinetic energy, enabling ions of the same $m/z$ to hit the detector almost simultaneously, thereby enhancing the resolution of TOF.

Furthermore, the usage of reflector effectively expands the flight path length within the same physical confines, resulting in superior ion separation and consequently, higher resolution. This reflection comes at the cost of some ion loss, and therefore some sensitivity loss. As such, reflecting TOFs are the basis of most commercial instruments currently in use.

The construction of a Q-TOF bears significant resemblance to a triple-quadrupole mass spectrometer, with the critical distinction that the third quadrupole has been replaced by a time-of-flight tube. **Figure 10** delineates the schematic of a typical Quadrupole-Time-of-Flight (Q-TOF) mass spectrometer, which comprises three fundamental components:

## 1. Quadrupole mass analyzer (Q).

This part of the instrument is basically the same to the Q1 in QqQ, which selects specific $m/z$ values to pass through by applying a combination of DC and RF voltages across the rods.

## 2. Collision cell.

Here, selected ions undergo collision-induced dissociation (CID) by interacting with a neutral gas, leading to their fragmentation into smaller constituents. This process yields structural information about the original molecules. Usually, quadrupole, hexapole, or even octopoles are used as the collision cell for better focusing and transporting.

## 3. Time-of-Flight (TOF) mass analyzer.

Upon exiting the collision cell, the fragmented ions are reaccelerated into the ion modulator region of the time-of-flight analyzer. There, they undergo pulsing by a strong electric field

(typically 20 kV or higher) and get accelerated to a field free drift tube, and then reflected to the detector.

TOFs generally offer mass resolutions surpassing 50,000, rendering it a reliable instrument for identifying unknown compounds. Moreover, the rapid travel time of ions in the vacuum tube (at the nanosecond level) confers the Q-TOF with distinctive benefits in short gradient and high-throughput analyses [440,441,442]. Another advantage of TOF is its broad mass range, which allows for the detection of large proteins, nanoclusters, and even large particles [443,444,445]. However, it should be noted that due to ion numbers and detector limitations, mass resolution is typically difficult to maintain over a wide mass range.

Presently, Q-TOF related instruments are available from all leading instrument manufacturers, and the main models are listed below: Sciex: "TripleTOF® 6600+", "TripleTOF® 5600+" System and "X500R QTOF" System. Bruker Corporation: "Impact II", "timsTOF" series, "microTOF-Q III", "ultrafleXtreme-MALDI-TOF/TOF" and "maXis II". Agilent Technologies: "Agilent 6530 Accurate-Mass Q-TOF", "Agilent 6545 Accurate-Mass Q-TOF", and "Agilent 6550 iFunnel Q-TOF". Waters Corporation: "SYNAPT G2-Si HDMS", "Xevo G2-XS QToF" and "SYNAPT XS".

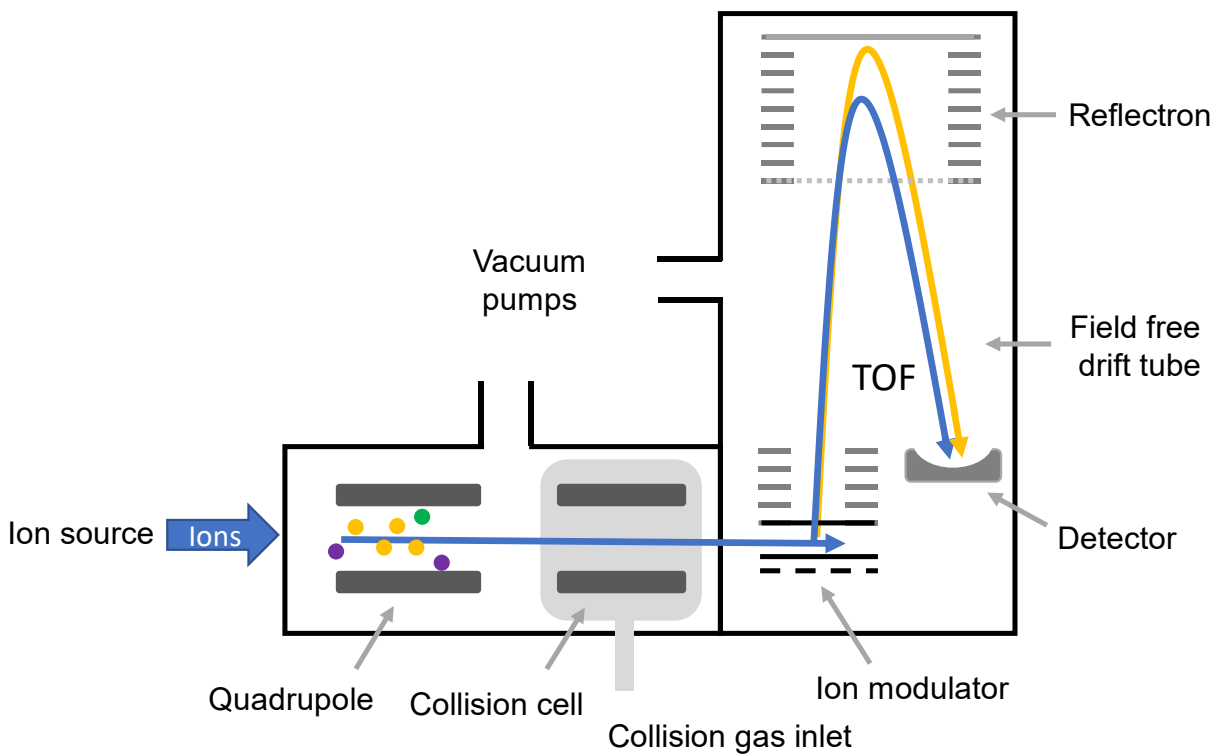

*Figure 10:* **Schematic diagram of a typical quadrupole time-of-flight mass spectrometer.** *Like a QQQ, a Q-TOF will have two quadrupoles for selection and fragmentation followed by the TOF for the final higher resolution separation and detection.*

# 3. Q-Orbitrap

Mass spectrometry that uses Orbitrap as the core mass analyzer is another critical pillar in the field of proteomics. In the late 20th century, Russian scientist Alexander Makarov invented the Orbitrap [446], which is a novel mass analyzer that operates based on the principle of electrodynamic ion trapping and Fourier Transform. The orbitrap consists of two main components: an inner spindle-like electrode and a coaxial outer barrel-like electrode (**Figure 11A**). The ions are trapped in an orbit around the spindle electrode due to the electrostatic attraction. Once inside, the ions begin oscillating along the central axis of the device, or "orbiting", due to the electric field formed by the inner and outer electrodes. The oscillation frequency of an ion is inversely proportional to the square root of its mass-to-charge ratio. The frequency at which each ion oscillates induces an image current on the detector, which can be measured and transformed into a mass spectrum using Fourier transform.

The biggest difference between Orbitrap and other mass spectrometers (TOF, Q) is that it does not use ions to hit an induction device like an electron multiplier. One of the main advantages of the Orbitrap is its ultra-high mass resolution, often exceeding 240,000 or even higher. This gives the Orbitrap a significant superiority in the identification of unknown molecules such as peptides and metabolites [421,447]. Moreover, Orbitrap spectrometers are also appreciated for their compact structure, small size, robustness, and reliability. Just like the Q-TOF, the Orbitrap is also usually used for tandem mass spectrometry. **Figure 11B** demonstrates a typical 2D schematic diagram of Q-Orbitrap. Ions first pass through an ion optics module, which consists of a high-capacity ion transfer tube (HCTT), an electrodynamic ion funnel (EDIF), and an advanced active beam guide (AABG). These are designed to capture ions, reduce ion losses, prevent neutrals and high-velocity clusters from entering the quadrupole, and increase sensitivity. The ions are then segmented by the quadrupole for precursor ion selection, and the selected ions are trapped by the ion-routing multipole for higher energy collisional dissociation. Finally, the fragmented ions are captured once again by the C-trap and injected into the Orbitrap batch-by-batch for accurate mass-to-charge analysis. Overall, this process still follows the logical sequence of precursor ions selection, precursor ions fragmentation, and fragment ions detection.

Compared to a TOF, one disadvantage of the Orbitrap is its longer cycle time (AGC pre-scan, ion injection, ion isolation, ion activation and mass analysis, usually >100ms), which is a negative factor for the currently favored short gradient, high-throughput analysis. Another minor flaw of Orbitrap is the challenge encountered when trying to pair it with MALDI. This primarily stems from the fact that MALDI uses a pulsed ionization technique, whereas Orbitrap operates continuously. This mismatch can lead to inefficiencies and challenges in coupling the two techniques. At present, the Orbitrap still dominates important applications in almost all aspects of proteomics including biomarker discovery [448], post-translational modification (PTM) analysis [27,449], quantitative proteomics (LFQ, TMT, iTRAQ)[20,21,450], protein-protein interaction studies [451] and structural proteomics [452,453]. It can perform both top-down and bottom-up analyses owing to its broad mass range, and is suitable for both Data-Dependent Acquisition (DDA) and Data-Independent Acquisition (DIA) methods. Right now, the Orbitrap is still under patent protection and only one company, ThermoFisher, is allowed to manufacture related products. Classic models from ThermoFisher include Orbitrap Ascend Tribrid, Orbitrap

Eclipse™ Tribrid™, Orbitrap Fusion™ Lumos™, Orbitrap Exploris series (120, 240, 480) and Q Exactive™ series.

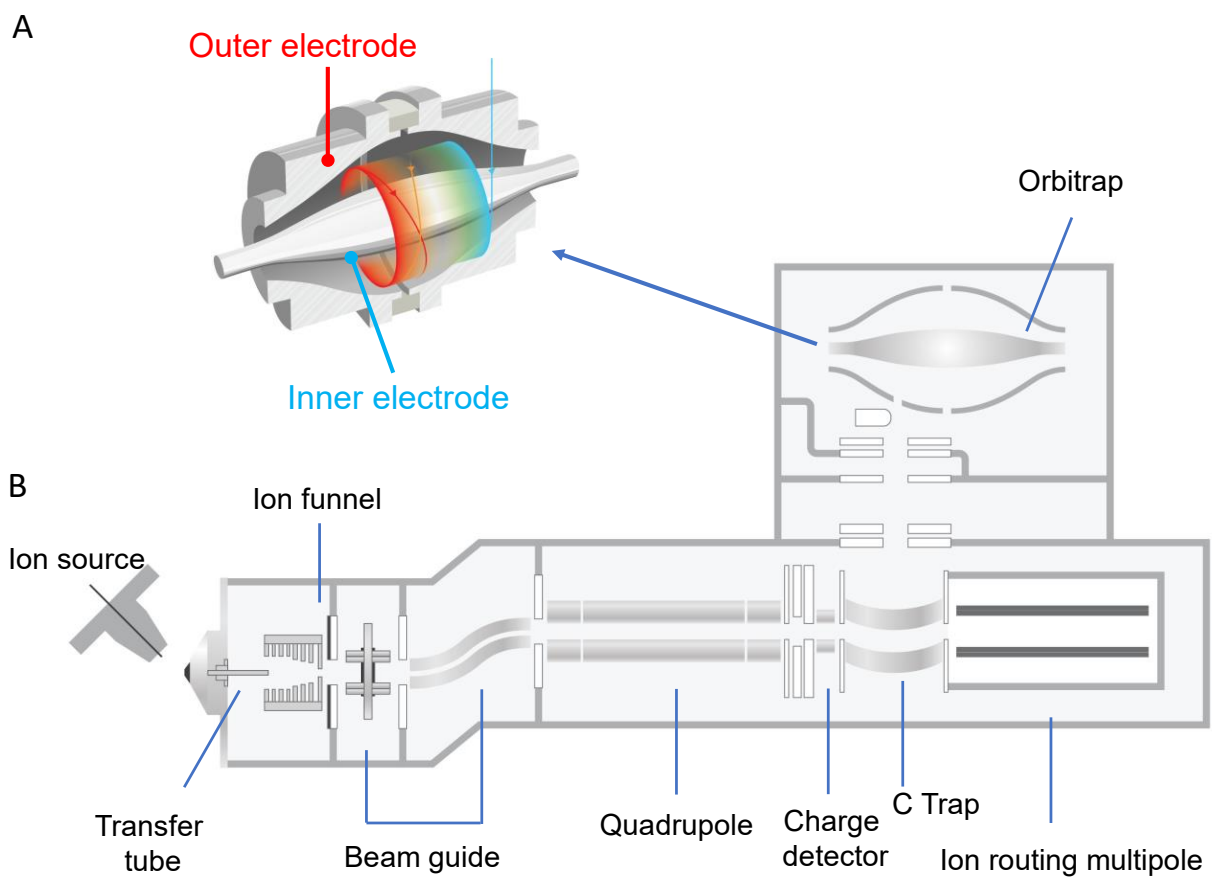

*Figure 11:* **Schematic diagram of orbitrap.** *(A) Close up of an Orbitrap. (B) General schematic of complete Q-Orbitrap system.*

# 4. Quadrupole Fourier Transform Ion Cyclotron Resonance (Q-FT-ICR)

The Fourier Transform Ion Cyclotron Resonance (FT-ICR) mass spectrometer is a type of mass spectrometry that uses magnetic fields to separate ions based on their mass-to-charge ratio. FT-ICR was first invented in 1974 by Alan G. Marshall and Melvin B. Comisarow from the University of British Columbia [454] and is widely recognized for its high mass resolution and precision, making it a highly valuable tool in many scientific fields including proteomics, metabolomics, petroleum analysis, and environmental science. The central feature of an FT-ICR mass spectrometer is a superconducting magnet coupled with an ICR cell (**Figure 12A**). This magnet creates a strong and homogeneous magnetic field in which ions are injected. Once the ions are inside ICR cell, under the influence of the strong magnetic field, they follow a circular path with a very small orbital radius at a specific frequency directly proportional to their mass-to-charge ratio. At this point, no detectable image current signal is generated by detector plates located inside the ICR cell. To improve the signal, a voltage is applied by excitation plates and resonance occurs when the frequency of the strong magnetic field matches the cyclotron frequency of the ions. The ions absorb radio frequency energy, which increases the radius of their circular path, and consequently, the excited ions move closer to the detector plates and generate a current. The resulting signal is an oscillating pattern or a time-domain signal.

Similar to Orbitraps, this time-domain signal is then transformed into a frequency-domain signal using Fourier transform, hence the name Fourier Transform ion cyclotron resonance (ICR). The Fourier transformed data forms a mass spectrum where each peak corresponds to a specific ion present in the sample. One of the most important advantages of FT-ICR mass spectrometry is its exceptionally high mass resolution and mass accuracy, even for large and complex molecules. This enables precise identification and characterization of a wide range of compounds in complex mixtures [455,456]. Moreover, FT-ICR mass spectrometry can be used for multiple stages of mass analysis (MSn), including tandem mass spectrometry (MS/MS), providing detailed information about the structure of ions. Another significant benefit of FT-ICR is its broad mass range, making it possible to identify macromolecules like proteins for top-down proteomics [19,457].

Despite its advantages, FT-ICR mass spectrometry is not without challenges. The technique requires high-performance superconducting magnets, which are expensive for both initial purchase and further maintenance. This is because FT-ICR requires liquid nitrogen and liquid helium cooling systems to keep the magnet at a sufficiently low temperature to maintain its superconducting state. Moreover, the device demands high vacuum conditions and careful temperature control to maintain the stability of the magnetic field and the ion trajectories. A schematic representation of a Q-FT-ICR system is shown in **Figure 12B**. In congruence with the tandem mass spectrometers elucidated earlier, ions pass through an array of ion optics modules which designed for ion focusing and purification. Following this, the ions are selectively filtered by the first quadrupole. After this filtration, precursor ions undergo fragmentation in the collision cell, which can be a quadrupole, hexapole, or octopole. The fragmented ions are subsequently re-concentrated by the ensuing focusing lens. Ultimately, these fragmented ions are trapped, excited, and detected within the ICR cell. At present, commercial FT-ICR mass spectrometers are available in both Thermo Fisher Scientific ("LTQ FT Ultra" and "LTQ FT Ultra Hybrid" systems) and Bruker Daltonics ("solariX" and "apex" series).

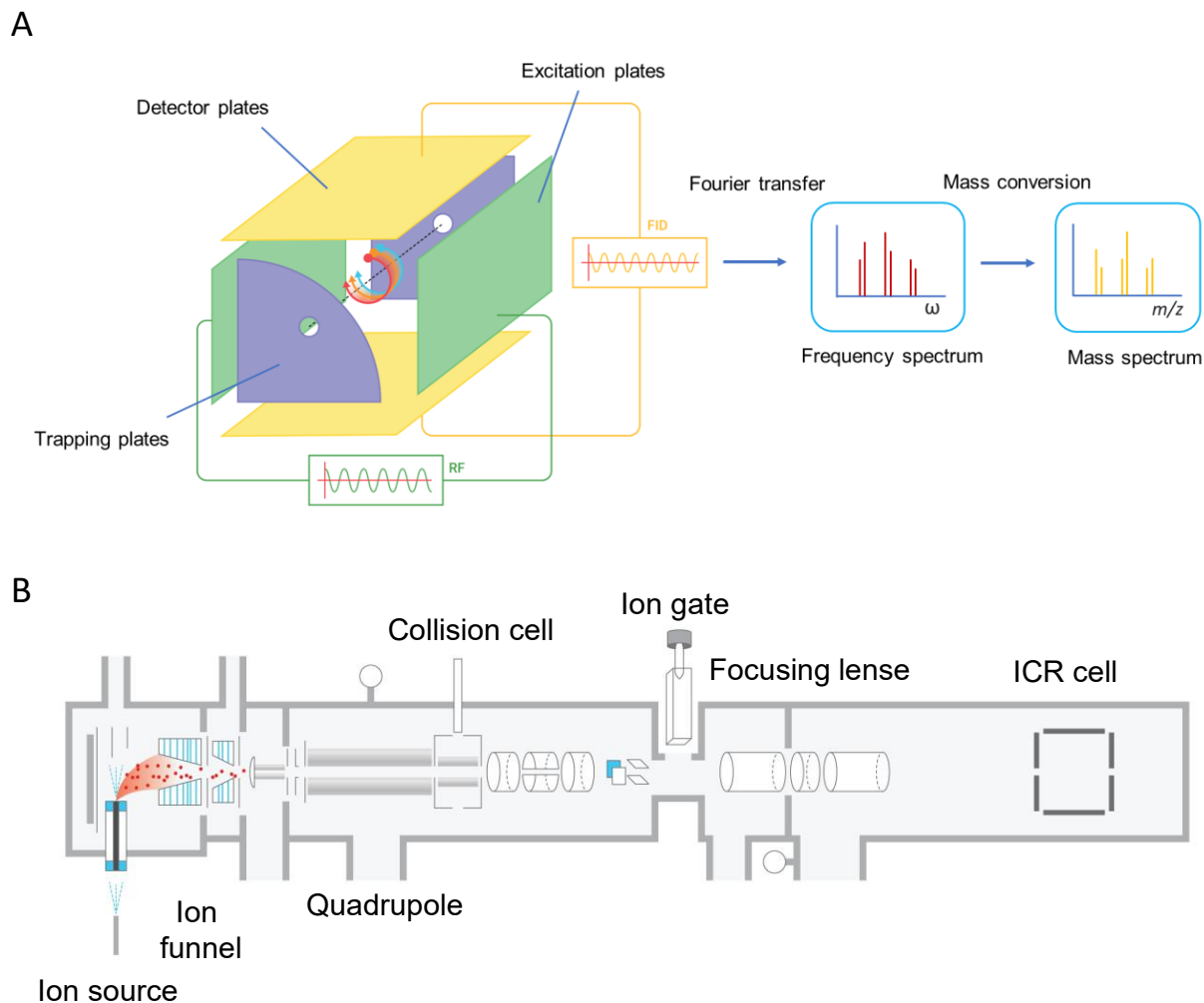

*Figure 12:* **Schematic of FT-ICR.** *(A) Typical FT-ICR cell. (B) Example of complete FT-ICR system.*

# 5. Ion mobility

In the context of omics research, a fundamental task is the separation, identification, and quantification of molecules in complex mixtures. Mass spectrometry alone can only provide two-dimensional data including mass-to-charge ratio and their intensity. Liquid chromatography contributes to the separation of compounds and further provides the third dimension of information, retention time (RT), which make LC-MS evolves as the "golden standard" for proteomic analysis [458,459]. Despite the substantial improvements in mass spectrometry resolution and liquid chromatography consistency, accurately identifying extremely similar molecules such as isomers with LC-MS remains a challenge. Ion mobility mass spectrometry (IM-MS), a technique that utilizes electric fields to transport analytes through a buffer gas, is beneficial for separating and identifying ions based on their size, shape, and charge state. This technique provides the fourth dimension of information, collision cross section (CCS), which allows for more comprehensive characterization of molecules [460]. Apparently, multi-dimensional data is always beneficial for us to understand things comprehensively and accurately, thus getting closer to the truth.

In terms of mass spectrometry based proteomic analysis, adding CCS data can help us better separate, identify, and quantify peptides.

The core principle of ion mobility spectrometry is to separate ions in an inert gas under the influence of an electric field (E), and then measure the amount of time it takes for each ion to pass through drift tube, which is defined to be the steady-state drift velocity (Vd) correlated to the specific analyte's mobility (K), as shown in Eq. 1. Vd = KE (Eq.1) While the primary measurement in IMS analyses is the mobility (K), for many analytical applications, it has become routine to convert K into the calculated collision cross-section value (CCS or Ω) using Mason-Schamp equation (Eq. 3) [461].

$$\Omega = (3/16 \; [\![(2\pi/\mu KT)]\!]^{(1/2)} \; ze)/N0K0 \; (Eq.2)$$

The components of the equation are defined as follows: e, charge of an electron; z, ion charge; $N_0$, buffer gas density; μ, reduced mass of the collision partners; $k_b$, Boltzmann's constant; and T, the drift region temperature. Although the Mason-Schamp equation isn't universally embraced, it is currently the primary formula the community uses to compute CCS. In basic terms, the CCS serves as a standard metric for the size in the gas phase, generally expressed in units of square Angströms ($Å^2$). However, according to the Eq.2, parameters including gas composition, working pressure, temperature within the mobility region, path of analyte movement, and the strength of the applied field can influence the final CCS value and may differ for each specific IMS platform. Hence, direct comparison of CCS value between different platforms often requires calibration.

Generally, ion mobility techniques can be categorized into three separation concepts: (1) temporally dispersive, (2) spatially dispersive, and (3) ion confinement (trapping) and selective release (**Figure 13A**) [458]. Temporally dispersive methods produce an arrival time spectrum based on differences in the time it takes for ions to traverse a similar gas-filled drift region under the influence of an electric field. Time-dispersive technique inherently provides an extensive examination of all signals detected during a given observation window. However, a fundamental limitation of this wide-ranging analysis is the diminished sensitivity linked to a single time dispersion occurrence, which usually requires many (10−100) events to be aggregated to achieve statistically significant ion mobility measurements. In contrast, spatially dispersive methods separate ions based on mobility differences (charge, shape and size), leading them on distinct drift paths or trajectories, but without significant time differences. A characteristic of spatially dispersive techniques is the scanning of voltage to obtain a broad-band ion mobility spectrum. Types of spatially dispersive ion mobility include High Field Asymmetric Waveform Ion Mobility Spectrometry (FAIMS), uniform-field differential mobility analyzers (DMA), and the newly introduced scanned frequency ion mobility filter called transverse modulation ion mobility spectrometry (TMIMS). Ion confinement and release strategies are recently developed techniques which trap ions in a pressurized drift cell by electric field, and then release them based on mobility distinctions. This technique relies on the ability to control the position of ions under elevated pressure conditions using precisely adjustable electrodynamic fields. It requires a precise fabrication craft and more complicated control system. While it has only been perfected recently, typical products like trapped ion mobility spectrometry (TIMS) [462,463] and cyclic traveling wave IMS have become commercially available [464]. Table 10.3 summarized typical ion mobility separation techniques, their separation concept, electric field direction, gas flow direction, strengths, and drawbacks. Also, for three categories of ion mobility techniques, we have selected a typical technique from each for brief introduction.

Table 10.3 Typical ion mobility separation techniques.

| Separation concept | Ion mobility techniques | Ion movement direction | Electric field direction | Drift Gas direction | Characteristics |
|---|---|---|---|---|---|
| Temporally Dispersive | drift tube IMS (DTIMS) | → | → | # | High mobility resolution (need long drift tube), direct measurement of CCS. Low speed, large size, low sensitivity, |
| Temporally Dispersive | traveling wave IMS (TWIMS) | → | →→→ | # | High mobility resolution, faster than DTIMS. Low sensitivity, large size, low sensitivity, traveling electric field waves. |
| Spatially Dispersive | high-field asymmetric IMS (FAIMS) | → | ↑↓ | → | Good as mass filter, Fast. No CCS measurement (Compensation voltage instead), low mobility resolution. |
| Spatially Dispersive | transverse modulation IMS (TMIMS) | → | → and ↑↓ | # | Transverse Modulation, compact instrumentation, orthogonal Separation, fast and high resolution |
| Confinement and Selective Release | trapped ion mobility spectrometry (TIMS) | → | ← | → | High mobility resolution, compact instrumentation, high sensitivity, high speed, high ion utilization rate |
| Confinement and Selective Release | multi-pass cyclic traveling wave IMS | → | →→→ | # | High mobility resolution, improved Signal-to-Noise ratio and sensitivity, versatility (from small molecules to large biomolecules) and adjustability (number of passes can often be adjusted) |

'#' means stationary drift gas; →, ←, ↑↓ indicates drift gas direction or electric force direction; →→→ represents a wave and gradient electric field.

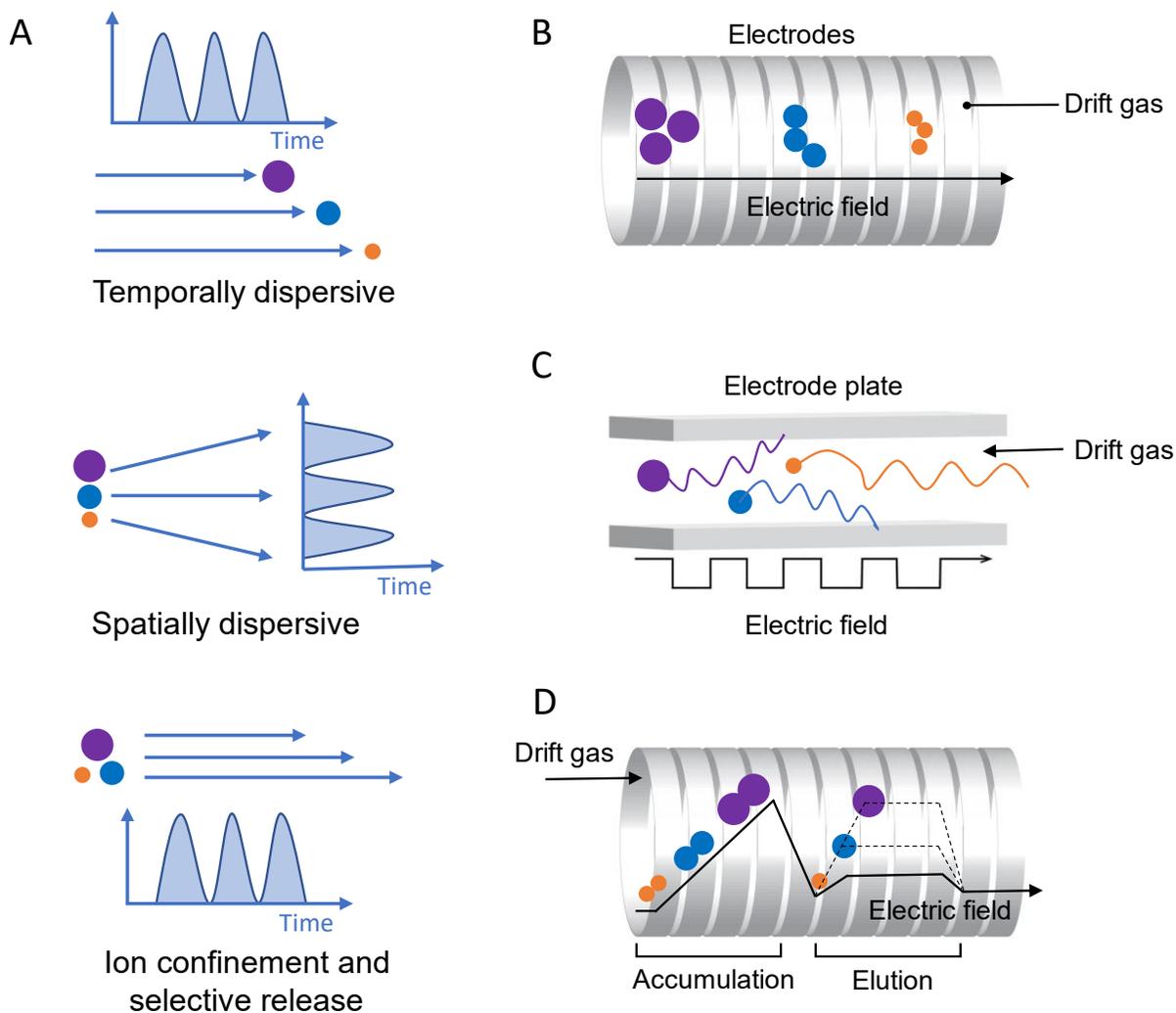

*Figure 13: **Ion Mobility.** (A) Conceptional diagram of three types of ion mobility strategies. (B) Schematic of drift tube ion mobility spectrometry. (C) Schematic of high field asymmetric waveform ion mobility spectrometry (FAIMS). (D) Schematic of trapped ion mobility spectrometry (TIMS).*

## 10.5.1. Drift Tube Ion Mobility Spectrometry (DTIMS)

The principle of Drift Tube Ion Mobility Spectrometry (DTIMS) is based on the differential migration (time) of ions through a neutral buffer gas (commonly helium or nitrogen) under the influence of a weak uniform electric field (typically tens of V/cm). The mobility (K) of an ion is proportional to its drift velocity (V) and inversely proportional to the strength of the applied electric field (E). For ions with same charge states, the drift velocities are primarily determined by their collisional interactions with a buffer gas, namely, mainly affected by their shape and size. To illustrate this process, imagine two objects with identical mass: a solid metal ball and a feather. Due to its lower density, the feather should have a larger volume than the ball. When both are dropped from the same height, the solid ball reaches the ground before the feather because of air resistance. This observation doesn't contradict Newton's law of universal gravitation, as we have accounted for air resistance. In the context of DTIMS, the buffer gas in the drift tube acts as the "air resistance", while the uniform electric field represents the "gravity". Hence, ions with the same mass-to-charge ratio are separated based on their shape and size. This capability allows DTIMS to distinguish between isomeric compounds with identical masses but different structural configurations, given that these isomers might have distinct interactions with the drift gas. Also, follow the intuition of the free fall example, in DTIMS, smaller ions will move faster and hit the detector earlier than larger ions in DTIMS (**Figure 13B**).

DTIMS possess the strengths including high resolving power and allows for straightforward measurement of an ion's CCS from first principles[465,466]. However, DTIMS also suffers from disadvantages including: 1) separation time is too long for all ions passing through the drift tube, relative to the accumulation time, which decreases the duty cycle. 2) A longer drift tube or higher pressure is needed for greater resolving power. However, this inevitably increases ion diffusion and ion losses unless ion focusing techniques are employed. 3) Segmentation and collision between ions and gas molecules during the traveling process in drift tube reduces the sensitivity. Continual advancements in DTIMS design and application of ion focusing techniques further pushed the mobility resolution of same DTIMS platforms to 100 to 250 (t/Δt) range or even greater.

## 10.5.2 High Field Asymmetric Waveform Ion Mobility Spectrometry (FAIMS)

High Field Asymmetric Waveform Ion Mobility Spectrometry (FAIMS) represents a distinct version of spatially dispersive ion mobility spectrometry. This technique differentiates ions utilizing a pronounced asymmetric oscillating electric field combined with a moving gas. The principle of FAIMS is based on the different trajectories of ions as they move through a high asymmetric electric field, which are determined by their physical structure and charge states [467,468,469]. In FAIMS, gas-phase ions are carried by a flow of carrier gas between two electrodes in a direction orthogonal to the direction of asymmetric electric field (E). The asymmetric waveform electric field is typically characterized by a short, high-voltage pulse of one polarity followed by a longer, lower-voltage pulse of the opposite polarity. An ion's mobility within such an electric field is determined by its charge state, its physical structure, and the properties of the surrounding gas it moves through. Once the ions are subjected to an asymmetric electric field, the ions will alternate between travelling toward one electrode or the other as the field oscillates in polarity, resulting in a curved trajectories between the electrodes. Some ions move more in the high field relative to the low field, and vice versa (**Figure 13C**). To differentiate between ions, a so-called "compensation voltage" (CV), which is a DC offset voltage that compensates for the differential ion movement in the high and low fields, is applied [470]. In this case, only ions with a specific response to the changing electric field and those that match the applied compensation voltage (CV) will have a zero net movement and are able to traverse the drift region to the detector, while others hit the electrode plate and be neutralized. By scanning or modulating the CV, different ion species can be selectively transmitted through the FAIMS device. In contrast to drift tube IMS in which the ion stream is sampled in discrete packets and all ions reach the detector, FAIMS is a continuous filtration technique that allows uninterrupted sampling of the ion stream, but only for a selected subset of the ion population. One of the primary advantages of this continuous collection technique is greatly increase the signal-to-noise ratio for the ion(s) of interest by removing unwanted chemical noise, which make

FAIMS more similar to a m/z filter than other ion mobility spectrometry tools. FAIMS also has the advantage of operating at atmospheric pressure. Drawbacks of FAIMS, however, are that it does not produce any CCS values and it has relatively low resolution separations. Commercial FAIMS products from vendors including Thermo Fisher and Waters are available now.

### 10.5.3 Trapped ion mobility spectrometry (TIMS)

Trapped ion mobility spectrometry (TIMS) is a common type of ion mobility which uses ion refinement and release strategy [471]. The basic idea behind TIMS is a combination of traditional ion mobility spectrometry and ion trapping techniques. Instead of driving ions through a drift tube filled with stationary gas, TIMS holds the ions stationary in a drift cell under a moving buffer gas and then releases them by adjusting electric fields (voltages on electrodes). This process was realized by applying two different electric fields: 1) Radially confining pseudopotential. An RF (radio frequency) voltage is applied to the electrodes of the TIMS analyzer to generate a radially confining pseudopotential, which essentially no axial component and only use for "focusing" ions in the central region of TIMS tube, preventing them from diffusion or hitting electrodes. 2) Axially electric field. An axially electric field gradient, produced by superimposing DC potentials on tunnel electrodes, is applied for "trapping" ions based on the equilibrium between the force of drift gas and the opposing force from the electric field gradient, which is stronger at the entrance and becomes progressively weaker moving deeper into the tunnel.

As a result, once ions entered the device, lower mobilities ones are trapped at positions where the magnitude of axially electric field is larger, while higher mobilities ones are confined to deeper positions of tunnel where axially electric field is lower. Then, after enough ions have been accumulated in the TIMS tunnel, additional ions are prevented from entering the tunnel region and residing ions are trapped for a short time (usually few milliseconds) which can be defined by users. Finally, the magnitude of axially electric field is decreased at a user defined rate so that ions are eluted as an order of mobilities value (K) from high to low (**Figure 13D**). The axially electric field gradient is set by a resistor divider. Importantly, like other ion mobility strategies, the resolving power of TIMS is highly dependent on the length of the gas column through which the ions traverse. In TIMS, ions are trapped in a specific location while buffer gas continuously flows past them. Thus, the resolving power achieved by TIMS depends on the "quantity" of gas, specifically the length of the gas column, that passes by the ions during the separation time. This offers the direct benefit of allowing the analyzer to maintain a compact physical size (around 5 cm) and achieve a high resolving power (R ~ 300), while the analytical gas column – the portion that flows during an analysis – can be extensive (up to 10 m) and tailored to the user's needs. Moreover, by leveraging the "trapping" capability (trapping time) of TIMS and the high scanning speeds of TOF, platforms such as TIMS-Q-TOF can implement a full duty cycle acquisition protocol known as Parallel Accumulation-Serial Fragmentation (PASEF) [462,472]. This is particularly meaningful for identifying more peptides within a given time frame, such as capture more precursors from co-eluted peptides in the same liquid chromatography peak. Currently, Bruker is the primary provider of commercial mass spectrometers that utilize TIMS-tof technology. (TIMS-tof pro, TIMS-tof pro2, SCP. etc.).

### 10.5.4 Structures for Lossless Ion Manipulation (SLIM)

A final type of ion mobility spectrometry discussed here is Structures for Lossless Ion Manipulation (SLIM), invented by Richard Smith and colleagues at Pacific Northwest National Labs [473]. SLIM uses printed circuit boards to confine ions in long path lengths for high resolution ion mobility. Ions can be passed through the board multiple times to achieve path lengths of several meters to over 1 km for high-resolution IMS separation [474,474]. This technology is currently under commercial development by Mobilion, in a platform named "Mobie" [475].

# 11. Tandem Mass Spectrometry and Peptide Fragmentation

## Tandem Mass Spectrometry

Tandem MS, where precursor ions are selected and fragmented to generate an MS/MS spectrum containing peptide-derived product ions, is a fundamental process in modern proteomics [476,477]. This is largely because intact peptide mass alone cannot unambiguously provide a peptide's sequence [478]; however, MS/MS spectra provide more information due to predictable fragmentation behavior of peptide ions to generate sequence-informative fragments [476,479]. Some more advanced proteomic acquisition methods use MS1-only feature detection in combination with retention time to maximize information used for downstream quantitation [480]. In most of these, identifications are fundamentally based on MS/MS spectra, either acquired as part of a specific LC-MS/MS analysis that contains the MS/MS spectra themselves or on a spectral library of MS/MS spectra acquired previously [481,482]. True MS1-only methods that use only accurate mass and retention time for identification have been discussed, but these have yet to be widely adopted [480].

The value of MS/MS spectra for peptide identification comes from predictable fragmentation behavior of peptide ions to generate sequence-informative fragments [476,479]. Multiple dissociation methods exist to generate product ions in MS/MS spectra through various mechanisms (**Figure 14**). In non-modified peptides, the most labile bonds are typically peptide bonds (i.e., amide bonds) between amino acids. Depending on where peptides dissociate along the peptide backbone, the fragments are assigned different ion types (**Figure 14A**). Fragment ion nomenclature was first developed by Roepstorff and Fohlman in 1984 [483] and then refined by Biemann in 1990 [484]. The main ion types are the fragments that contain the original peptide N-terminus (i.e., a-, b-, and c-type ions), or the original peptide C-terminus (i.e., x-, y-, and z-type ions). The number associated with each fragment ion indicates how many amino acids from each terminus are included.

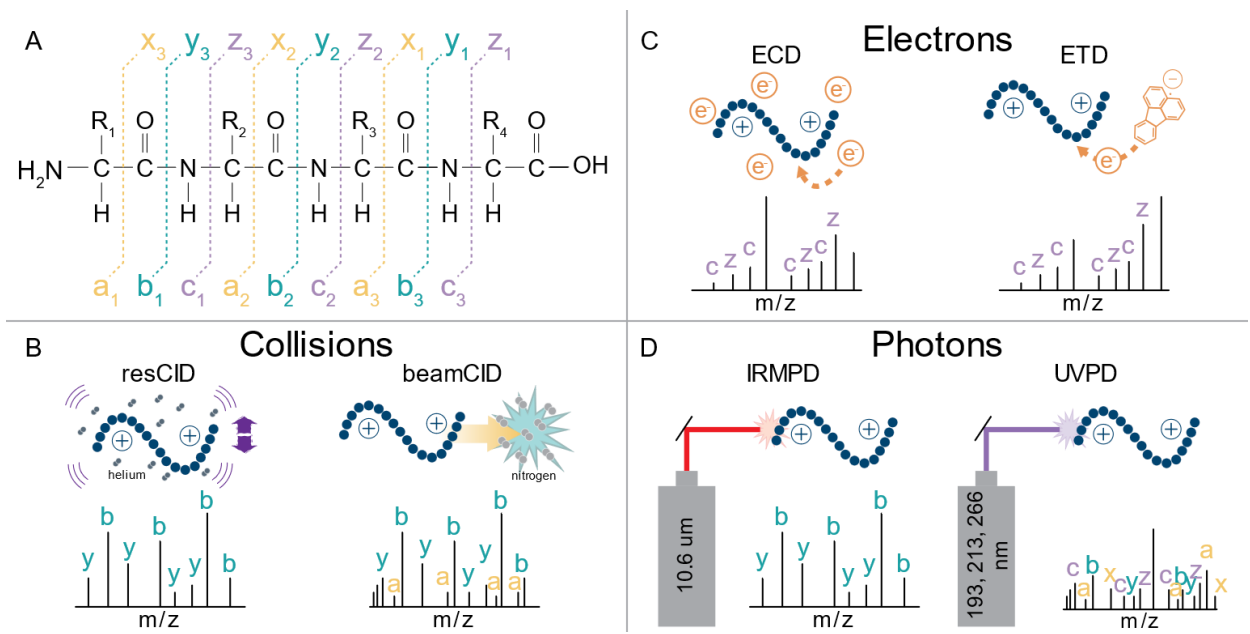

*Figure 14: **Peptide Fragmentation Methods.** (A) Sequence-informative fragment ions are termed a/x-, b/y-, and c/z-type fragments depending on which bond along the peptide backbone breaks. Fragments that explain the intact N-terminus of the peptide are a-, b-, and c-type ions, while x-, y-, and z-type ions explain the intact C-terminus of the peptide. Other panels show common dissociation methods, including collision, electron, and photon-based fragmentation. (B) Resonant collision-induced dissociation (resCID) and beam-type CID (beamCID) both produce mainly b/y-type sequencing ions through collisions with background gases like helium and nitrogen that increase the internal energy of peptide cations. (C) Electron capture and electron transfer dissociation (ECD and ETD) generate mainly c/z-type fragments through electron-mediated radical driven cleavage of the peptide backbone. (D) Infrared multi-photon dissociation (IRMPD) is a slow heating method similar in dissociation mechanism to resCID, but very different in implementation due to the IR lasers required (often with lower energy 10.6 micron photons). Ultraviolet photodissociation (UVPD) can use a range of wavelengths (popular options shown) to introduce higher energy photons to peptide cations, causing vibrational and electronic excitation that can generate all major fragment ion types depending on wavelength used.*

One of the earliest and most ubiquitous peptide fragmentation methods is collision-induced dissociation (CID, also called collisionally-activated dissociation, CAD) [426] (**Figure 14B**). Here, collisions with inert gas molecules are used to increase the internal energy of peptide ions to reach bond dissociation energies that fragment them into products. Various inert gases can be used; helium, nitrogen, and argon are the most common. Preferences for which gas is used is often a function of how much energy per collision is desired. Two main versions of CID are used in proteomics, with the most common being beam-type CID (beamCID, sometimes called higher-energy collisional dissociation, HCD) [485,486]. BeamCID typically uses nitrogen or argon as a collision gas, and peptide ions are accelerated into a collision cell filled with several mTorr of bath gas. The kinetic energy used to accelerate precursor ions (often generated using direct current voltage differentials between the source of the ions and the collision cell) determines the energy imparted through collisions with the bath gas, which in turn governs their fragmentation behavior.

Since in non-modified peptides the most labile bonds are typically peptide bonds (i.e., amide bonds) between amino acids, the increase in internal energy from beamCID generates b- and y-type ions that represent this peptide bond cleavage, as shown in Biemann fragment ion nomenclature (**Figure 14A**). b-type ions provide sequence information for fragments that have an intact N-terminus, while y-type ions denote fragment ions with an intact C-terminus. Collisions in beamCID cause near instantaneous generation of primary fragment ions. Because the increase in internal energy happens rapidly before energy can be redistributed, beamCID can generate fragments that are not necessarily derived from cleavage of the most labile bonds (e.g., PTM-modified peptides, discussed below), but spectra are often dominated by b/y-type ions from amide bond cleavage (**Figure 14B**). BeamCID can also generate secondary fragments, such as immonium ions from side chain losses [487] or a-type fragment ions that come from water loss from b-type ions due to multiple collision events (note: a-type ions can form as primary fragmentation products in other dissociation methods). The simplicity of beamCID, which simply requires an rf-only collision cell, has made it widely implemented on most instrument platforms used in modern proteomics.

A second form of CID is called resonant CID (resCID), where the internal energy of peptide ions is slowly increased through multiple low-energy collisions. Here, helium gas is most often used, as it imparts less energy per collision, and activation typically happens in ion trap devices where supplemental frequencies can be used to excite ions. In other words, ions are trapped using axial rf-frequencies, and an additional rf-frequency is applied to the electrodes of the ion trap [488]. This supplemental rf is selected to have a frequency resonant with the fundamental frequency of the ions to be fragmented, as determined by the Mathieu equations, which excites the ions of interest so that they have increased kinetic energy as they move in the ion trap [489,490]. The increased kinetic energy creates more collisions with the background helium gas to slowly build up the internal energy of the precursor ions until the dissociation energy of the most labile bond is reached, causing fragmentation. Once ions dissociate, the fragments have different $m/z$ values than the precursor ions, meaning they fall out of resonance with the supplemental rf and are no longer activated. Thus, resCID typically fragments only the most labile bonds in precursor ions and does not have secondary fragmentation behavior. As above, for non-modified peptide ions, this typically generates sequence-informative b- and y-type product ions. For modified peptides where the bonds connecting the modification to an amino acid are more labile than peptide bonds (e.g., phosphopeptides and glycopeptides), resCID MS/MS spectra can be dominated by product ions only of the PTM-loss rather than sequence-informative fragment ions, although many factors govern this behavior [491,492]. Because of this, and because this method requires an ion trap device with the ability to apply supplemental rfs, resCID is less prevalent than beamCID. For both beamCID and resCID, the mobile proton model has been widely accepted to explain fragmentation behavior [493], and this largely predictable behavior has greatly helped in manual and algorithm-assisted spectral interpretation.

Despite the utility and broad adoption of CID, alternative dissociation methods have been explored for a variety of uses, including applications where CID is inadequate for the experimental question [494,495,496]. The most popular of these alternative dissociation methods are electron-based dissociation (ExD) approaches, which include electron capture dissociation (ECD) and electron transfer dissociation (ETD). In both of these, peptide cations capture thermal electrons (ECD [497]) or abstract an electron from a reagent anion (ETD [498]) to generate radical-driven dissociation of the N-Ca bond that predominantly generates sequence-informative c- and z-type product ions (**Figure 14C**). The mechanisms of ExD methods have been widely explored [499,500], and the preferential cleavage of N-Ca bonds along the peptide backbone have been particularly useful for PTM-modified species because the modifications remain largely intact even during peptide backbone bond fragmentation. ExD methods have shown promise for analysis of numerous PTMs, including phosphorylation, glycosylation, ADP-ribosylation, and more [501,502]. Electron-based dissociation is also more suitable than collision-based dissociation for MS analyses of intact proteins [503,504] and larger oligonucleotides [505,506,507,508,509,510]

Two fundamental challenges exist with ExD methods. First, ExD implementation requires instruments that can manipulate cations and anions (or free electrons) within the same scan sequence and can trap both simultaneously for electron capture/transfer events to occur. This has been successfully accomplished on a number of instruments, including FT-ICR systems, ion traps, ToFs with quadrupole ion traps, and hybrid Orbitrap instruments, but it is not a ubiquitous feature of all platforms. That said, several exciting advances in recent years have made ExD methods more accessible on numerous instrument configurations [501,502,511,512,513]. A second challenge is the dependence of ExD dissociation efficiency on precursor ion charge density [514]. ExD methods generally produce robust fragmentation for charge dense precursor ions (i.e., those with relatively low *m/z* values and higher z). Alternatively, precursors with low charge density (i.e., higher *m/z* values) have relatively condensed secondary gas-phase structure that leads to non-covalent interactions. Even in the cases when ExD methods drive peptide backbone cleavage, product ions (i.e., c- and z-type fragments) are held together by the non-covalent interactions so that few (or no) sequence-informative product ions are produced. This process is called non-dissociative electron-capture/transfer (ECnoD/ETnoD) [515]. Several strategies to mitigate ECnoD/ETnoD have been successfully explored, including supplemental activation of product ions with resCID (ETcaD [516]) or beamCID (EThcD [517,518]), supplemental activation with infrared photons (AI-ECD [519,520] and AI-ETD [521,522,523,524]) or ultraviolet photons (ETuvPD [525]), and use of higher energy electrons [513,526,527]. Despite their successes, these methods still require instrumentation capable of ExD in addition to extra hardware needed for a given strategy (e.g., a CO2 laser in AI-ETD [528]). As with ExD in general, recent advances in supplemental activation strategies for ExD are making these tools more accessible [501,502].

Photoactivation is another family of alternative dissociation strategies that has been steadily gaining popularity [529,530]. Infrared multi-photon dissociation (IRMPD) is canonically the photodissociation method used in early proteomic applications [530], but ultraviolet photodissociation (UVPD) has been the more widely used approach in the recent decade [531]. IRMPD functions similarly to resCID; it is a slow heating approach that causes vibrational excitation due to absorption of low energy photons, generally 10.6 µm photons from a CO2 laser [532,533]. Predominant fragments are b- and y-type fragments, although secondary fragmentation occurs because fragment ions remain in the photon path after the initial dissociation event (**Figure 14D**). Despite limited use in the past decade, recent work shows that

IRMPD, or more generally activation with IR photons, may still have value in the proteomics toolkit [258,523]. UVPD has been explored with a number of wavelengths, including 157 nm, 193 nm, 213 nm, 266 nm, and 355 nm [534,535,536,537,538,539]. Higher-energy UVPD approaches, like 193 and 213 photons, are typically used for underivatized peptide and protein ions [531], while others, like 266 and 355 nm, can be used for directed fragmentation at specific residues with natural chromophores (e.g., tyrosine) or exogenously added chromophore tags [540,541]. UVPD with 193 and 213 generate multiple fragment types, including sequence-informative a-, b-, c-, x-, y-, and z-ions in addition to other fragmentation pathways, which occur through vibrational and electronic excitation [542]. UVPD has been explored for bottom-up proteomic applications, but its more impactful utility, arguably, has been realized in intact protein characterization [543]. The laser needed for UVPD (i.e., the photon wavelength desired) determines much about its implementation. 193 nm photons are typically generated using an Excimer laser with ArF gas [544], while 213 nm photons can be generated with a solid-state laser that is easier to integrate into an instrument platform and maintain [536,545]. That said, 213 nm photons tend to provide more directed, preferential cleavage pathways compared to 193 nm photons that cleave more broadly in non-directed fashion [546]. Outside of ExD and photoactivation approaches, other alternative dissociation methods have been explored for various proteomic applications, although they are not as widely adopted at ExD and UVPD methods [529].

# 12. Data Acquisition

Hybrid mass spectrometers used for modern proteome analysis offer the flexibility to collect data in many different ways. Data acquisition strategies differ in the sequence of precursor scans and fragment ion scans, and in how analytes are chosen for MS/MS. Constant innovation to develop better data collection methods improves our view of the proteome, but many method options may confuse newcomers. This section provides an overview of the general classes of data collection methods.

Data acquisition strategies for proteomics fall into one of two groups.

1. Data dependent acquisition (DDA), in which the exact scan sequence in each analysis depends on the data that the mass spectrometer observes.
2. Data independent acquisition (DIA), in which the exact scan sequence in each analysis DOES NOT depend on the data; the collected scans are the same whether you inject yeast peptides, human peptides, or a solvent blank.

DDA and DIA can both be further subdivided in to targeted and untargeted methods.

# DDA

In most cases, the peptide masses that will be observed are not known before doing the experiment. Data collection methods must account for this. DDA was invented in the early 1990s, which enabled collecting MS/MS spectra for observed peptides as they eluting from the LC column [547,548,549].

## Untargeted DDA

A common method currently used in modern proteomics is untargeted DDA. The MS collects precursor (MS1) scans iteratively until precursor mass envelopes meeting certain criteria are detected. Criteria for selection are usually specific charge states and a minimum signal intensity. When those ions meet these criteria, the MS selects those masses for fragmentation.

Because ions are selected as they are observed, repeated DDA of the same sample will produce a different set of identifications. This stochasticity is the main drawback of DDA. To ameliorate this issue, often strategies are used to transfer identifications between multiple sample analyses. This transfer of IDs across runs is known as "match between runs", which was originally made famous by the processing software MaxQaunt [550,551]. There are several other similar tools and strategies, including the accurate mass and time approach [552], Q-MEND [553], IDEAL-Q [554] and superHIRN [555]. More recent work has introduced statistical assessment of MBR methods using a two-proteome model [556]. Statistically controlled MBR is currently available in the IonQuant tool [557].

Because DDA is required for quantification of proteins using isobaric tags like TMT, this stochasticity of DDA limits the ability to compare quantities across batches. For example, if you have 30 samples, you can use two sets of the 16-plex kit to label 15 samples in each set with one channel labeled by a pooled sample to enable comparison across the groups. When you collect DDA data from each of those sets, each set will have MS/MS data from an overlapping but different set of peptides. If one set has MS/MS from a peptide but the other set does not, then that peptide cannot be quantified in the whole sample group. This limits the number of quantified proteins in large TMT experiments with multiple batches.

## Targeted DDA

Targeted DDA is not common in modern proteomics. In targeted DDA, in addition to general criteria like a minimum intensity and a certain charge state, the mass spectrometer looks for specific masses. These masses might be previously observed signals that were previously missed by MS/MS [558,559]. In these studies, the sample is first analyzed by LC-MS to detect precursor ion features with some software, and then subsequent analyses target those masses for fragmentation with inclusion lists until they are all fragmented. This was shown to increase proteome coverage.

## DDA methods for modifications

Resonant CID [560] and beam-type HCD [561] are the most popular methods for unmodified and modified peptides due to their speed, accessibility, and efficiency. Due to the weak phosphoester bond relative to the peptide backbone, resonant CID usually produces spectra that are dominated by only the neutral loss of the phosphate. For this reason, the optimal dissociation methods for phosphopeptide identification and phosphosite localization include HCD or ExD-based methods, discussed later in more depth [562,563]. ExD methods generate phosphopeptide MS/MS spectra with many c- and z•-type fragment ions for peptide sequencing and localization of labile phosphate modifications, typically disrupted with CID [498]. Gas-phase phosphate rearrangement induced by collisional activation represents a glaring challenge for the field and several have explored site localization in the face of rearrangement [564,565,566].

Advanced data acquisition schemes trigger predetermined MS/MS events when a specific fragment ion or neutral loss is detected in a spectrum. Certain decision-tree strategies have arisen to increase data acquisition efficiency, including pseudo-MS3 scans which are triggered on detection of phosphate losses [492] and the use of site-specific x-type ions [567]. For example, when linear ion traps were the main proteomics workhorses, resonant CID analysis of phosphopeptides would result in predominantly neutral loss of the phosphate with limited sequence ion information. To gain sequence ions in these experiments, instruments could be set to isolate a loss of 98 Daltons for MS3 activation [568,569]. The newer collisional dissociation technique HCD, or beam-type collisional activation, significantly improves the detection of peptide fragments with the phosphorylation intact on fragment ions, and thus, this neutral loss scanning technique is no longer common.

Recently developed approaches to phosphopeptide identification include DIA-based phosphoproteomics with Spectronaut [570,571], "plug-and-play" high-resolution MS [572], SureQuant for phosphotyrosine [573], PIQED for direct identification and quantification of phosphorylation from DIA without a prior spectral library [449], and FAIMS front-end separations which yield 15-20% more phosphosite identifications than non-FAIMS experiments [469]. For quantification of phosphoproteins, Hogrebe et al. investigated several of the most common strategies and concluded that TMT-based MS2 strategies may be the current best approach [574].

Additionally, while less commonly modified than serine and threonine, histidine [575,576,577], arginine [578], and tyrosine [573,579,580] phosphorylation also represent intriguing cell signaling biology. Going forward, we expect that faster instruments will enable investigations of high phosphoproteomic depth and reproducibility in rapid timeframes, such that many proteomes can be analyzed for temporal and spatial insight.

A similar product-dependent MS/MS triggering strategy was introduced for N-linked glycopeptides [581]. Collisional dissociation of glycosylated peptides produces oxonium ions, for example at *m/z* 204.09 (HexNAc) or *m/z* 366.14 (HexHexNAc). If oxonium ions from the fragmented glycan are detected among the most abundant fragment ions of the HCD spectra, then an ETD scan is triggered. This ETD scan provides information about the peptide sequence, while the original HCD scan provides glycan structure information.

# DIA

The simplest method to operate a mass spectrometer is to have predefined scans that are collected for each sample analysis. This is data-independent acquisition (DIA); the scans that are collected do not depend on the data that the instrument observes. Thus, the scan sequence is repetitive, looping through binned windows of predetermined width, and/or a predetermined *m/z* range. Although simple in terms of data collection, when the scan sequence includes MS/MS, sophisticated software is required to analyze the data. Like DDA, DIA can also be either targeted or untargeted [582]: The two targeted DIA methods are selected reaction monitoring (SRM) or multiple reaction monitoring (MRM), and untargeted DIA (uDIA) is often referred to simply as "DIA" or "SWATH" (Sequential Window Acquisition of All Theoretical Mass Spectra) (**Figure 15**).

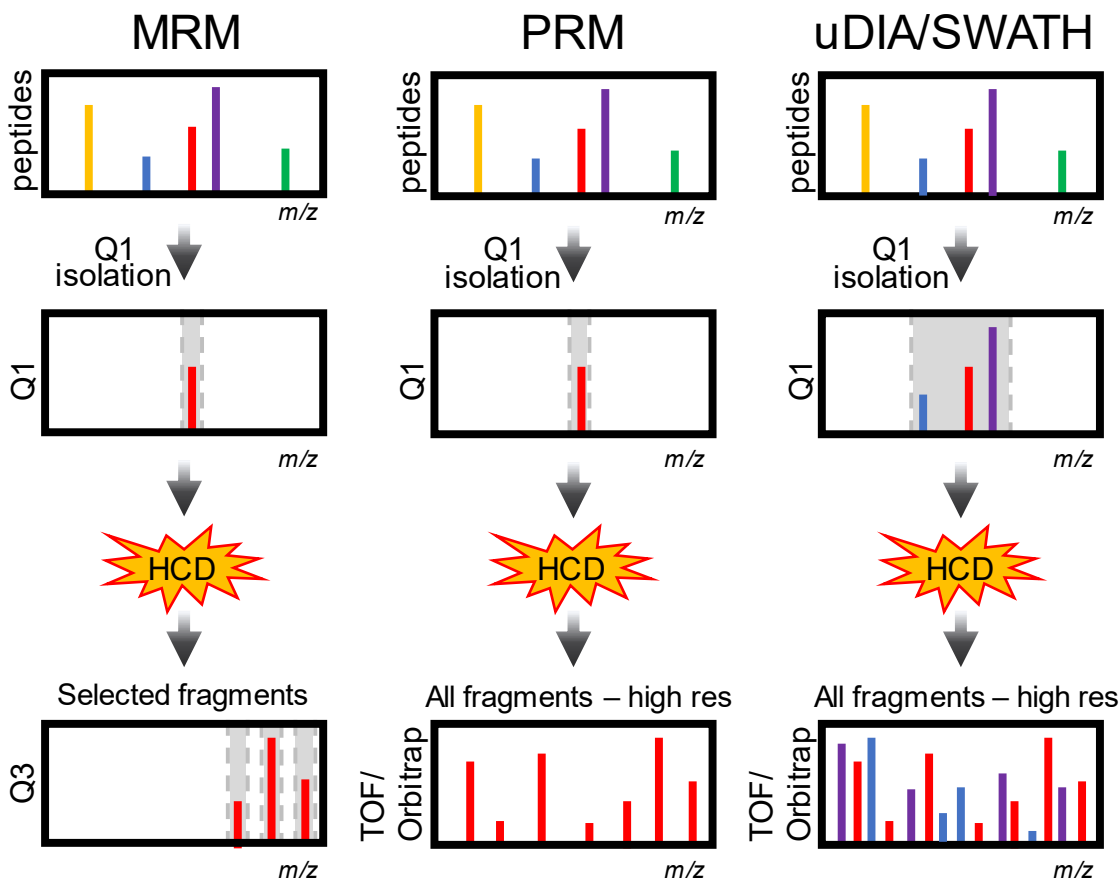

*Figure 15: **Types of DIA.** A) SRM/MRM. Peptides are ionized by ESI and although there are many peptides entering the mass spectrometer at any time, the first quadrupole (Q1) isolates one mass, which is then fragmented by HCD. Fragment masses from the peptide are then selected in the third quadrupole (Q3). This leads to very low noise and high sensitivity. B) PRM. Like MRM, peptides are selected in the first quadrupole, but this analysis is done on a high-resolution instrument like an Orbitrap or TOF. Selectivity is gained by exploiting the high mass accuracy and resolution to monitor multiple fragment ions. C) uDIA/SWATH. Like MRM and PRM, peptides are isolated with Q1, but in this case a much wider isolation window is used. This usually results in co-isolation of many peptides simultaneously. Fragments from many peptides are measured with high resolution and high mass accuracy. Special software is used to get peptide identities and quantities from the fragment ions.*

## Targeted DIA

The first type of targeted DIA is called SRM or MRM [583]. The popularity of this method in the literature peaked in 2014, with just under 1,500 documents on PubMed that year resulting from a search for "MRM". In this strategy, the QQQ MS is set so that the first quadrupole selects the precursor mass of the peptide(s) of interest, the second quadrupole fragments the peptide, and the third quadrupole monitors the product of specific fragments from that peptide. This strategy is very sensitive and has the benefit of very low noise. The fragments monitored in Q3 are chosen such that it is unlikely these fragments could arise from another peptide. Usually at least a few transitions are monitored for each peptide in order to get multiple measures for that peptide.

An early example of MRM applied to quantify c-reactive protein was in 2004 [584]. Around the same time, SRM was combined with antibody enrichment of peptides from target proteins [585]. This approach was popular for analysis of plasma proteins [430]. These early examples led to many more studies that used QQQ MS instruments to get accurate quantitation of many proteins in one injection [586,587]. Scheduling MRM measurement when chromatography is stable additionally enabled better utilization of instrument duty cycle and therefore monitoring of more peptides per injection [588]. Efforts even developed libraries of transitions that allow quantification of any protein in model organisms [589].

Another similar targeted DIA method is called parallel reaction monitoring (PRM) [590]. Instead of using a QQQ instrument, PRM uses a hybrid MS with a quadrupole and a high-resolution mass analyzer, such as an Q-TOF or Q-Exactive. The idea is that instead of monitoring specific fragments in Q3, the high mass accuracy can be used to filter peptide fragments for high selectivity and accurate quantification. Studies have found that PRM and MRM/SRM have comparable dynamic range and linearity [591].

## Untargeted DIA

There were many implementations of uDIA over the years, starting in 2003 by Purvine et al from the Goodlett lab [592]. In this first work they demonstrated uDIA using a Q-TOF with in source fragmentation and showed that extracted ion chromatograms of precursor and fragment ions matched in shape suggesting that this could be used to identify and quantify peptides. The following year, Venable et al from the Yates lab introduced uDIA with an ion trap [593]. Subsequent methods include MSE [594], PAcIFIC [595], all ions fragmentation (AIF) [596]. Computational methods were also developed to automate interpretation of this data, such as DeMux [596], XDIA [597], and ETISEQ [598].

The paper that is often cited for uDIA that led to widespread adoption was by Gillet et al. from the Aebersold group in 2012 [599]. In this paper they branded the idea as SWATH. Widespread adoption may have been facilitated by the co-marketing of this idea by ABSciex as a proteomics solution on their new 5600 Q-TOF (called "tripleTOF" despite containing only one TOF, likely a portmanteau of "triple quadrupole" and "Q-TOF"). Importantly, in the Gillet et al. paper the authors described a computational method to extract information from SWATH where peptides of interest were queried against the data. They also demonstrated the application of SWATH to measure proteomic changes that happen in diauxic shift, and showed that SWATH can reveal modified peptides, in this case a methionine oxidation.

There are also many papers describing uDIA with orbitraps. One early example described combining random isolation windows together and then demultiplexing the chimeric spectra [600]. In another landmark paper, over 6,000 proteins were identified from mouse tissue by at least 2 peptides [601]. In 2018, the new model orbitrap at that time (HF-X) enabled identification of nearly 6,000 human proteins in only 30 minutes. Currently orbitraps have all but replaced the Sciex Q-TOFs for DIA data collection.

A new direction in uDIA is the addition of ion separation by ion mobility. This has appeared in two forms. On the timsTOF, diaPASEF makes use of the trapped ion mobility to increase speed and sensitivity of analysis [602]. On the orbitrap, the combination of FAIMS and DIA has enabled the identification of over 10,000 proteins from one sample, which is a major milestone [406].

# 13. Analysis of Raw Data

The goal of basic data analysis is to convert raw spectral data into identities and quantities of peptides and proteins that can be used for biologically focused analysis (**Figure 16**). This step may often include measures of quality control, cross-run data normalization, quantification on different levels (precursor, peptide, protein), protein inference, PTM (post translational modification) localization and also first steps of data analysis, such as statistical hypothesis tests.

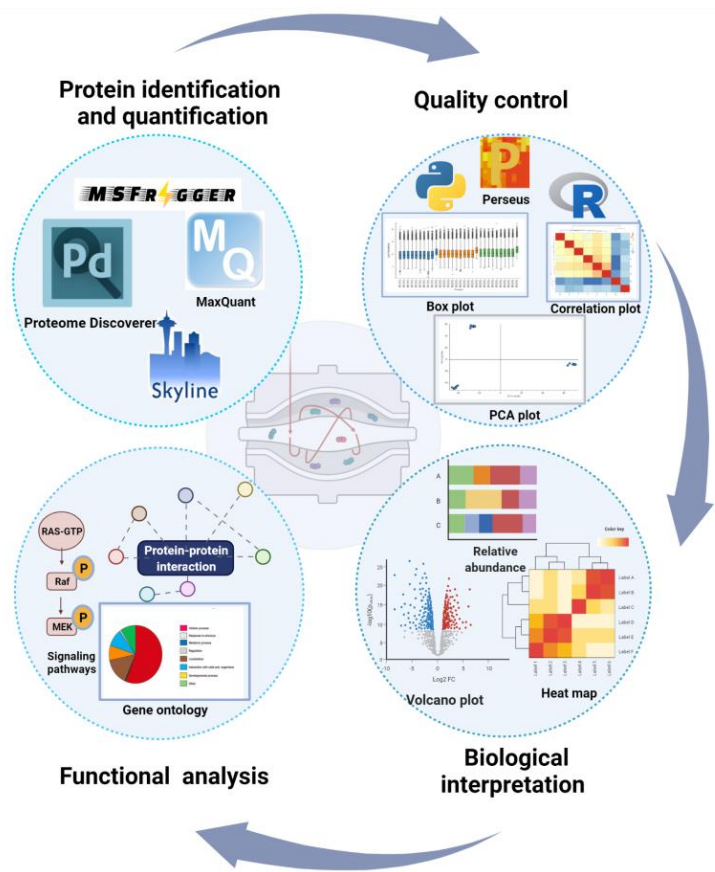

Figure 16: **Proteomics Data Analysis and Biological Interpretation.** The process begins with protein identification and quantification using tools such as Proteome Discoverer, Spectronaut, Spectromine, MS Fragger, MaxQuant, and Skyline. Quality control measures ensure data integrity, leading to a biological interpretation of the results. Differential expression analyses

*may include relative abundance charts, heat maps, and volcano plots. Functional analysis encompasses gene ontology, protein-protein interactions, and signaling pathways.*

In typical bottom-up proteomics experiments, proteins are digested into peptides and further analyzed with LC-MS/MS systems. Peptides can have different PTMs and ionize differently depending on their length and amino acid distributions. Therefore, mass spectrometers often record different charge and modification states of one single peptide. The entity that is recorded on a mass spectrometer is usually referred to as a precursor ion (peptide with its modification and charge state). This precursor ion is fragmented, and the precursor or peptide sequences are obtained though spectral matching. The quantity of a precursor is estimated with various methods. The measured precursor quantities are combined to generate a peptide quantity. Peptides are also often combined into a protein group through protein inference, which combines multiple peptide identifications into a single protein identification [603,604]. Protein inference is still a challenge in bottom-up proteomics.

Due to the inherent differences in the data structures of DDA and DIA measurements, there exist different types of software that can facilitate the steps mentioned above. The existing software for DDA and DIA analysis can be further divided into freeware and non-freeware:

## 15-1 DDA freeware:

| Name | Publication | Website |
|---|---|---|
| MaxQuant | Cox and Mann, 2008[605] | MaxQuant |
| MSFragger | Kong et al., 2017[606] | MSFragger |
| Mascot | Perkins et al., 1999[607] | Mascot |
| MS-GF+ | Kim et al., [608] | MS-GF+ |
| X!Tandem | Craig et al., [609,610] | GPMDB |
| Comet | Eng et al., 2012[611] | Comet |

## 15-2 DIA freeware:

| Name | Publication | Website |
|---|---|---|
| MaxDIA | Cox and Mann, 2008[605] | MaxQuant |
| Skyline | MacLean et al., 2010[612] | Skyline |
| DIA-NN | Demichev et al., 2019[613] | DIA-NN |
| EncyclopeDIA | Searle et al., 2018[614] | EncyclopeDIA |

## 15-3 Targeted proteomics freeware:

| Name | Publication | Website |
|---|---|---|
| Skyline | MacLean et al., 2010[612] | Skyline |

## 15-4 DDA non-freeware:

| Name | Publication | Website |
|---|---|---|
| ProteomeDiscoverer | | ProteomeDiscoverer |
| Mascot | Perkins et al., 1999[607] | Mascot |
| Spectromine | | Spectromine |
| PEAKS | Tran et al., 2018[615] | PEAKS |

## 15-5 DIA non-freeware:

| Name | Publication | Website |
|---|---|---|
| Spectronaut | Bruderer et al., 2015[616] | Spectronaut |
| PEAKS | Tran et al., 2018[615] | PEAKS |

## 15-6 Data Summary and Interpretation

| Name | Publication | Website |
|---|---|---|
| SpectroDive | | Biognosys |

## 15-7 Data Summary and Interpretation

| Name | Publication | Website |
|---|---|---|
| Peptide Shaker | Vaudel *et al.*, 2015[617,618] | PeptideShaker, Peptide Shaker Online |

# Analysis of DDA data

DDA data analysis either directly uses the vendor proprietary data format directly with a proprietary search engine like Mascot, SEQUEST (through Proteome Discoverer), Paragon (through Protein Pilot), or it can be processed through one of the many freely available search engines or pipelines, for example, Comet, MaxQuant, MSGF+, X!Tandem, Morpheus, MS-Fragger, and OMSSA. Tables 15-1 and 15-4 give weblinks and citations for these software tools.  For analysis with freeware, raw data is converted to either text-based MGF (mascot generic format) or into a standard open XML format like mzML [619,620,621]. The appropriate FASTA file containing proteins predicted from that organism's genome is chosen as a reference database to search the experimental spectra. All search parameters like peptide and fragment mass errors (i.e. MS1 and MS2 tolerances), enzyme specificity, number of missed cleavages, chemical artefacts (fixed modifications) and potential biological modifications (variable/dynamic modifications) are specified before executing the search. The search algorithm scores each query spectrum against its possible peptide matches [622]. A spectrum and its best scoring candidate peptide are called a peptide spectrum match (PSM). The scores reflect a *goodness-of-fit* between an experimental spectrum and a theoretical one and do not necessarily depict the correctness of the peptide assignment.

For evaluating the matches, a decoy database is preferred as a null model for peptide matching. A randomized or reversed version of target database is used as a nonparametric null model. The decoy database can be searched separate from the target database (Kall's method)[623] or it can be combined with the target database before search (Elias and Gygi method)[624]. Using either separate method or concatenated database search method, an estimate of false hits can be calculated which is used to estimate the false discovery rate (FDR) [625]. The FDR denotes the proportion of false hits in the population accepted as true. For Kall's method: the false hits are estimated to be the number of decoys above a given threshold. It is assumed that the number of decoy hits that pass a threshold are the false hits. A similar number of target population may also be false. Therefore, the FDR is calculated as [626]:

$$FDR = \frac{DecoyPSMs + 1}{TargetPSMs}$$

For Elias and Gygi Method, the target population in which FDR is estimated changes. The target and decoy hits coming from a joint database compete against each other. For any spectrum, either a target or a decoy peptide can be the best hit. It is argued that the joint target-decoy population has decoy hits as confirmed false hits. However, due to the joint database search, the target database may also have equal number of false hits. Thus, the number of false hits is multiplied by two for FDR estimation.

$$FDR = \frac{2 * DecoyPSMs}{Target + DecoyPSMs}$$

# Integrated MS Data Analysis Platforms

Given the complexity of proteomic data analysis and the requirement for many steps to get from raw data to quantified proteins, there are some integrated software enviroments that easily allow users to complete everything in one place.

## Trans-Proteomic Pipeline (TPP)

The Trans-Proteomic Pipeline (TPP) is a free and open-source mass spectrometry data analysis suite for end-to end analysis that remains in continual development to provide ever expansive data analysis capabilities since its inception over twenty years ago [73,627,628,629,630,631,632,633,634,635,636]. The current release provides tools for mass spectrometry spectral processing, spectrum searching, search validation, abundance computation, protein inference, and statistical evaluation of the data to ensure controlled false-discovery rates. Many of the tools include machine-learning modeling to extract the most information from datasets and build robust statistical models to compute probabilities that derived information is correct.

One of the major advantages of TPP is its ability to be deployed in a wide variety of environments, from personal Windows laptops to extensive large Linux clusters for automated use within cloud computing environments. While the command-line interfaces are appreciated by many power users, others prefer a graphical user interface (GUI), which is provided by the TPP GUI called Petunia, allowing users to use the TPP from any web browser on any platform. Petunia has the advantages that the same exact GUI is available on a modest Windows laptop, a powerful expandable Linux server shared by a research group, or a remote cloud computing instance running on Amazon Web Services (AWS) [631].

The TPP incudes many statistical validation tools such as PeptideProphet [637], ProteinProphet [638], iProphet [630], and PTMProphet [635], where Bayesian machine learning techniques are applied to the various search engine scores to model the correct and incorrect assignment distributions and then use these models to assign a probability of being correct based on these learned models. With these tools it is possible to validate search engine results on large-scale datasets and in short order, enabling users to select probability thresholds based on a selected tolerable false discovery rate (FDR). The TPP is made fully interoperable via the open XML-based formats pepXML and protXML for different aspects of processing data-dependent acquisition (DDA), and Data-independent acquisition (DIA) proteomics data, resulting in a complete suite of tools for processing the increasingly larger datasets from start to finish.

DIA workflows are supported via the DISCO tool which reads mzML files containing the instrument-produced spectra and uses signal processing approaches to isolate the fragment ions in the multiplexed MS2 spectra that correlate with precursors in the MS1 and writes the results to new mzML files that may then be searched with standard DDA search engines and downstream tools, including target-decoy analysis. This provides a comprehensive analysis of DIA data without the need for building a spectral library first.

From its inception, the TPP has been and will always be free and open-source software, allowing anyone to use it without cost and to inspect its source code, alter the source code for their own needs, or even incorporate parts of it into their own products. Others have performed these tasks and include various analysis routines as addons such as TAILS N-terminomics analysis [639], quantitation analysis with PyQuant [640], SimPhospho [641], WinProphet [642], ProtyQuant [643], and inclusion of R-tools for metaproteomic analysis [644]. As a collection of individual tools, they are easily amenable to pipelining in a very flexible manner to support a huge variety of combinations and workflows, and a custom program may easily be inserted into the pipeline to support technology development.

## Search engines supported by TPP

The heart of MS proteomics DDA data continues to be the "search engine" that interprets collections of mass spectra to determine the peptide or peptides that yielded them. Spectral library search engines and de novo search engines, which are less common, are also available and are included in software suites such as the Trans Proteomic Pipeline. A sequence search engine most commonly used is the open-source version of SEQUEST called Comet, which is actively maintained and updated with new functionality as needs arise. For spectral library searching, SpectraST uses an approach where new spectra are matched against a library of previously identified spectra in the form of a spectral library [645]. This approach is much faster, more sensitive, and more specific than sequence database searching, although is only as good as the reference spectral library provided. There is renewed interest in spectral libraries because of data-independent acquisition (DIA) approaches being increasingly deployed and therefore the quality and coverage of libraries is paramount and likely to improve in the coming years, aided by the standard spectral library format being developed by the PSI [646]. For de novo sequence analysis, Novor [647] and Casanovo [648] are very fast and capable de novo sequence search engines that are available.

For chemical crosslinking proteomics analysis, open-source programs such as Kojak [72,73] are available for standard or cleavable MS2-based crosslinking techniques. Crosslinking-based MS analyses are employed to elucidate protein-protein interactions and facilitate protein structure and topology predictions. Kojak is designed to identify two independent peptides covalently bonded with a crosslinker and fragmented in a single MS2 scan event using a database search approach. Kojak algorithm also includes support for cleavable cross-linkers, and identification of cross-links between 15N-labeled homomultimers and is integrated into the Trans-Proteomic Pipeline, enabling access to dozens of additional tools, in particular, the PeptideProphet and iProphet tools for validation of cross-links improve the sensitivity and accuracy of correct cross-link identifications at user-defined thresholds. Development of Kojak has continued over the last ten years culminating in many improvements and new features. These improvements include support for additional open formats and standards, further refinement to the search algorithm for efficiency, E-values to normalize the scores of the results, support for cleavable cross-linkers, and methods to identify cross-links between homomultimer subunits.

For open modification database searching, programs such as Magnum ([649]) are also now available which is specialized in identification of non-peptide masses that are bound to peptides. The tool is capable of identifying xenobiotic mass adducts, in addition to PTMs that were uncharacterized in the search parameters.

## Strategies for analysis of DIA data

DIA data analysis is fundamentally different from DDA data analysis because, instead of a single MS/MS spectrum for each peptide, we can observe the elution of peptide fragments for any peptide over chromatography time. There are two general approaches for peptide identification from DIA data: peptide-centric and spectrum-centric.

Peptide-centric approaches looks for evidence of specific peptides that are in some assay library of MS/MS spectra. That library could be predicted spectra (e.g., using Prosit) [650], or previously measured spectra (e.g., from a organism-wide knowledge base) [651]. Examples of software that perform peptide-centric analysis include OpenSWATH [652], Spectronaut [601], csoDIAq [653], and DIA-NN [613].

Spectrum-centric approaches instead ask if there is evidence for any peptide based on analysis of the observed spectra. Examples of spectrum-centric approaches include DIA-Umpire [654] and PECAN [655]. Spectrum-centric approaches may assemble pseudo-MS/MS spectra from co-elution of fragments that can then be used with any DDA database search [654]. Spectrum-centric may be less sensitive at peptide identification than peptide-centric approaches.

# Quality control

Quality control should be a central aspect of any mass spectrometry-based study to ensure reproducibility of generated results. There are two types of quality controls that can be conducted for any kind of mass spectrometry experiment. The first one is focused on monitoring the performance of the instruments themselves (e.g. HPLC and mass spectrometer), whereas the second one is focused on your experiments. For further reading, we recommend to take a look at issue 11 on quality control published in the journal *Proteomics* in 2011 [656], especially the review by Köcher *et al.* [657], as well as the review published by Bittremieux *et al.* in 2017 [658].

## Instrument Performance

It is generally advisable to monitor instrument performance regularly. Instrument calibrations in regular intervals help ensure that performance is maintained. Often basic calibration and sensitivity can be checked by direct infusion of a standard. During the calibration you can check injection times (for ion trap instruments) and intensity of the ions in the calibration mix.

After ensuring good calibration and signal with the simple calibration mixture, it is advisable to analyze complex samples, such as tryptic digests of whole-cell lysates (e.g. HeLa cells, HEK cells, yeast, etc.) or tryptic digests of purified proteins. The additional check with a complex sample ensures all aspects of the system are working together correctly, especially the liquid chromatography and emitter. These digests should be analyzed after every instrument calibration and periodically between samples when acquiring more extensive batches. Data measured from tryptic digests should be analyzed by the software of your choice and the numbers of identified peptide precursors and proteins can be compared with previous controls for consistency.

Another strategy is to analyze digested purified proteins, which easily enable discovery of retention time shifts and mass accuracy problems. In case you are working with a Thermo mass spectrometer, you can open the acquired .raw file directly either in FreeStyle or in Qual Browser and look for specific *m/z* values of your peptides. Looking at the intensity of the extracted peaks will help identify sensitivity fluctuations.

Carry-over between different measurements can be identified from blank measurements which are subsequently analyzed with your search software of choice. Blank measurements can be injections of different buffers, water or the starting conditions of your liquid chromatography. In case of increased detection of carry-over, injections with trifluoroethanol can be performed.

Another factor to take into consideration is the stability of your electrospray. Electrospray stability tends to worsen over time as columns wear, as well as when measuring samples with residual contaminants, such as salts or detergents. You will notice spray instabilities either in the total ion chromatogram (TIC) as thin spikes with short periods of no measured signal or if you install cameras at your ESI source. Suboptimal spray conditions will usually result in droplets forming on the emitter, being released into the mass spectrometer (also referred to as

"spitting"). Real-time quality control software (listed in the table below) can help you identify instrument issues right away.

## Data Quality Control

Apart from instrument performance, any kind of data analysis should have proper quality control in place to identify problematic measurements and to exclude them if necessary. It is recommended to develop a standardized system for data quality control early on and to keep this consistent over time. Adding indexed retention time (iRT) peptides can help identify and correct gradient and retention time inconsistencies between samples at the data analysis stage. Decoy searches help monitor and control the false-discovery rate. Including common contaminants, such as keratins, in the FASTA files used for searches can help identify sample preparation issues. Other parameters to check in your analysis are the consistency of the number of peptide-spectrum matches, identified peptides and proteins over all samples of your study, as well as your coefficients of variation between your replicates. Before and after data normalization (if normalization is performed) it is good to compare the median intensities of all measurements to identify potential measurement or normalization issues. Precursor charge distributions, missed cleavage numbers, peak width, as well as the number of points per peak are additional parameters that can be checked. In case you are analyzing different conditions, you can perform hierarchical clustering or a principal component analysis to check if your samples cluster as expected.

## Quality Control Software

## Raw file and real-time analysis

| Name | Supported instrument vendors | Website/Download | publication | Note |
|------|------------------------------|------------------|-------------|------|
| QuiC | Thermo Scientific, AB SCIEX, Agilent, Bruker, Waters | QuiC | | requires Biognosys iRT peptides |
| AlphaPept | Thermo Scientific, Bruker | AlphaPept | [659] | |
| RawMeat 2.1 | Thermo Scientific | RawMeat | | |
| rawDiag | Thermo Scientific | rawDiag | [660] | |
| rawrr | Thermo Scientific | rawrr | [661] | |
| rawBeans | Thermo or mzML | rawBeans | [662] | |
| SIMPATIQCO | Thermo Scientific | SIMPATIQCO | [663] | |
| QC-ART | | QC-ART | [664] | |
| SprayQc | Thermo Scientific, AB SCIEX, extensible to other instrumentation | SprayQc | [665] | |
| Metriculator | | Metriculator | [666] | |
| MassQC | | MassQC | | |
| OpenMS | | OpenMS | [667] | |

## Search result QC

| Name | Website/Download/publication | publication | Note |
|------|------------------------------|-------------|------|
| MSStats | MSStats | [668] | can use output from MaxQuant, Proteome Discoverer, Skyline, Progenesis, Spectronaut |
| MSStatsQC | MSStatsQC | [669] | |
| PTXQC | PTXQC | [670] | requires MaxQuant search engine output |
| protti | protti | [671] | |

# Quantitative Proteomic Data Analysis Best Practices

This section aims to provide an overview of the best practices when conducting large scale proteomics quantitative data analysis. The proteome is more complex than the genome and transcriptome due to PTMs and splicing, therefore careful selection of data analysis techniques is required to make capture the true biological signals in the data [672]. A well-established workflow for proteomic data analysis does not currently exist [672]. Analyzing proteomic data requires knowledge of a multitude of pre-processing techniques where order matters, and it can be challenging knowing where to start. This review will cover tools to reduce bias due to nonbiological variability, statistical methods to identify differential expression and machine learning (ML) methods for supervised or unsupervised interpretation of proteomic data.

## Data Transformation

Peptide or protein quantities are generally assumed to be logarithm (log) transformed before any subsequent processing [673,674,675,676]. Log transformation allows data to more closely conform to a normal distribution and reduces the effect of highly abundant proteins [674]. Many normalization techniques also assume data to be symmetric, so log transformation should precede any downstream analysis in these cases [674]. If there are missing values present, a simple approach would be to use log(1+x) to avoid taking the log of zero. After the transformation, zero quantities will remain as zero and the other quantities should be large enough that adding one will have a minor effect.

## Data Normalization

Data normalization, the process for adjusting data to be comparable between samples, should be performed prior to batch correction and any subsequent data analysis [674,677]. Normalization removes systematic bias in peptide/protein abundances that could mask true biological discoveries or give rise to false conclusions [678]. Bias may be due to factors such as measurement errors and protein degradation [674], although the causes for these variations are often unknown [676]. As data scaling methods should be kept at a minimum [679], a normalization technique well suited to address the nuances specific to one's data should be selected. The assumptions for a given normalization technique should not be violated, otherwise choosing the wrong technique can lead to misleading conclusions [680]. There are a multitude of data normalization techniques and knowing the most suitable one for a dataset can be challenging.

Visualization of peptide or protein intensity distributions among samples is an important step prior to selecting a normalization technique. Normalization is suggested to be done on the peptide level [679]. If the technical variability causes the peptide/protein abundances from each sample to be different by a constant factor, and thus intensities are graphically similar across samples, then a central tendency normalization method such as mean, median or quantile normalization may be sufficient [674,679]. However, if there is a linear relationship between bias and the peptide/protein abundances, a different method may be more appropriate. To visualize linear and nonlinear trends due to bias, we can plot the data in a ratio versus intensity, or a M (minus) versus A (average), plot [674,681]. Linear regression normalization is an available technique if bias is linearly dependent on peptide/protein abundance magnitudes [674,675]. Alternatively, local regression (LOESS) normalization assumes nonlinearity between protein intensity and bias [675]. Another method, removal of unwanted variation (RUV), uses information from negative controls and a linear mixed effect model to estimate unwanted noise, which is then removed from the data [682].

If sample distributions are drastically different, for example due to different treatments or samples are obtained from various tissues, one must use a method that preserves the heterogeneity in the data, including information present in outliers, meanwhile reducing systematic bias [679]. For example, Hidden Markov Model (HMM)-assisted normalization [679], RobNorm [683] or EigenMS [684] may be suitable for this type of data. These techniques assume error is only due to the batch and order of processing. The first method that addresses correlation of errors between compounds by using the information from the variation of one variable to predict another is systematic error removal using random forest (SERRF) [685]. SERRF, among 14 normalization methods, was the most effective in significantly reducing systematic error [685].

Studies aiming to compare these methods for omics data normalization have come to different conclusions. Ranking of different normalization methods can be done by assessing the percent decrease in median log2(standard deviation) and log2 pooled estimate of variance (PEV) in comparison to the raw data [686]. One study found linear regression ranked the highest compared to central tendency, LOESS and quantile normalization for peptide abundance normalization for replicate samples with and without biological differences [674]. A paper comparing multiple normalization methods using a large proteomic dataset found that mean/median centering, quantile normalization and RUV had the highest known associations between proteins and clinical variables [673]. Rather than individually implementing normalization techniques, which can be challenging for non-domain experts, there are several R and Python packages that automate mass spectrometry data analysis and visualization. These tools assist with making an appropriate selection of a normalization technique. For example, NormalyzerDE, an R package, includes several popular methods for normalization and differential expression analysis of LC-MS data [687]. AlphaPeptStats [688], a Python package, allows for comprehensive mass spectrometry data analysis, including normalization, imputation, batch correction, visualization, statistical analysis and graphical representations including heatmaps, volcano plots, and scatter plots. AlphaPeptStats allows for analysis of label-free proteomics data from several platforms (MaxQuant, AlphaPept, DIA-NN, Spectronaut, FragPipe) in Python but also has web version that does not require installation.

## Data Imputation

Missing peptide intensities, which are common in proteomic data, may need to be addressed, although this is a controversial topic in the field. Normalization should be performed before imputation since bias may not be removed to detect group differences if imputation occurs prior to normalization [676]. Reasons for missing data include the peptide not being biologically present, being present but at too low of a quantity to be detected, or present at quantifiable abundance but misidentified or incorrectly undetected [676]. If the quantity is not at the detectable limit, the quantity is called censored and these values are missing not at random [676]. Imputing these censored values will lead to bias as the imputed values will be overestimated [676]. However, if the quantity is present at detectable limits but was missed due to a problem with the instrument, this peptide is missing completely at random (MCAR) [676]. While imputation of values that are MCAR using observed values would be a reasonable approach, censored peptides should not be imputed because their missingness is informative [676]. Peptides MCAR are a less frequent problem compared to censored peptides [676]. Understanding why the peptide is missing can be challenging [676], however there are techniques such as maximum likelihood model [689] or logistic regression [690] that may distinguish censored versus MCAR values.

Commonly used imputation methods for omics data are random forest (RF) imputation[691], k-nearest neighbors (kNN) imputation [692], and single value decomposition (SVD) [693]. Using the mean or median of the non-missing values for a variable is an easy approach to imputation but may lead to underestimating the true biological differences [676]. Choice of the appropriate imputation method is critical as how these missing values are filled in has a substantial impact on downstream analysis and conclusions [694]. In one study, RF imputation was the most accurate among nine imputation methods across several combinations of types and rates of missingness and does not require preprocessing (e.g., does not require normal distribution) for metabolomics data [695]. Another study found RF, among eight imputation methods, had the lowest normalized root mean squared error (NRMSE) between imputed values and the actual values when MCAR values were randomly replaced with missing values, followed by SVD and KNN using metabolomics data [694]. Lastly, a study found RF also had the lowest NRMSE when comparing seven imputation methods using a large-scale label-free proteomics dataset [696].

## Batch Correction

Normalization is assumed to occur prior to batch effect correction [679]. Batch effect correction is still a critical step after normalization as proteins may still be affected by batch effects and diagnosing a batch effect may be easier once data is normalized [679]. Prior to performing any statistical analysis of data, we must start with distinguishing signals in the data due to biological versus batch effects. A batch effect occurs when differences in preparation of samples and how data was acquired between batches results in altered quantities of peptides (or genes or metabolites) which results in reduced statistical power in detecting true differences [679,679]. This non-biological variability originates from the time of sample collection to peptide/protein quantification [673] and is often a problem when working with large numbers of samples, involving multiple plates run by different technicians, on different instruments and/or using different reagent batches [697]. Results from these different batches ultimately need to be aggregated and data analysis to be performed on the whole dataset, so it may be difficult to measure and then control for exact changes due to non-biological variability once the data has been aggregated [673]. Batch correction methods remove technical variability, however they should not remove any true biological effect [673,697]. Although it is agreed upon that these biases should be accounted for to prevent misleading conclusions, there is no one gold standard batch correction method.

Batch effects can manifest as continuous, such as from MS signal drift, or as discrete, such as a shift that affects the entire batch [679]. To visualize batch effects, one can plot the average intensity per sample in the order each was measured by the MS to see if intensities are shifted in a certain batch [679]. Measuring protein-protein correlations is another method to check for batch effects; if proteins within a batch are more correlated compared to those from other batches, there are likely batch effects[679]. Prior to batch correction, one should ensure the experimental design is not irreversibly flawed due to batch effects and whether a change in design should be implemented. Studies spanning multiple days and experiments involving samples from different centers are vulnerable to batch effects [698]. One example of technical variability that may irreversibly flaw an experiment would be running samples at varying time points, or 'as they came in' [699]. This problem can be circumvented by balancing biological groups in each batch [699]. Additionally, collection of samples at different institutions introduces non-biological variability due to differences in a multitude of conditions such as collection protocols, storage, and transportation [673]. A solution to this problem would be to evenly distribute samples between centers or batches [673].

There are several batch correction methods, the most popular method being Combating Batch Effects When Combining Batches of Gene Expression Microarray Data (ComBat), originally designed for genomics data [697,700]. ComBat uses Bayesian inference to estimate batch effects across features in a batch and applies a modified mean shift, but requires peptides to be present in all batches which can lead to loss of a large number of peptides [697]. Out of six batch correction methods using microarray data, ComBat was the best in reducing batch effects across several performance metrics and was effective using high dimensional data with small sample sizes [701]. ComBat may be more suitable for small datasets when the source of batch effects are known [697]. However if potential batch variables are not known or processing time or group does not adequately control for batch effects, surrogate variable analysis (SVA) may be used where the source of batch effect is estimated from the data [697,698]. A third option for batch effect correction uses negative control proteins to estimate unwanted variation, called "Remove Unwanted Variation, 2-step" (RUV-2) [702]. There are many additional batch effect correction methods for single cell data, such as mutual nearest neighbors [703], or Scanorama, which generalizes mutual nearest neighbors matching [704].

## Quality Control

Prior to conducting any statistical analysis, the raw data matrix should be compared to the data after the above-described pre-processing steps have been performed to ensure bias is removed. We can compare data using boxplots of peptide intensities from the raw data matrix versus corrected data in sample running order to look at batch associated patterns; after correction, we should see uniform intensities across batches [679]. We can also use clustering methods such as Principal Component analysis (PCA), Uniform Manifold Approximation and Projection (UMAP), or t-SNE (t-Distribute Stochastic Neighbor Embedding) and plot protein quantities colored by batches or technical versus biological samples to see how proteins cluster in space based on similarity. We can measure the variability each PC contributes; we want to see similar variability among all PCs, however if see one PC contributing to overall variability highly then means variables are dependent [705]. tSNE and UMAP allow for non-linear transformations and allow for clusters to be more visually distinct [705]. Grouping of similar samples by batch or other non-biological factors, such as time or plate, indicates bias [679]. Quantitative measures of whether batch effects have been removed are principal variance components analysis (PVCA), which provides information on factors driving variance between biological and technical differences, and checking correlation of samples between different batches, within the same batch and between replicates. When batch effects are present, samples in the same batch will have higher correlation than samples from different batches and between replicates [679]. Once batch effects are removed, proteins in the same batch should be correlated at the same level with proteins from other batches [679]. Similarity between technical replicates can be measured using pooled median absolute deviation (PMAD), pooled coefficient of variation (PCV) and pooled estimate of variance (PEV); high similarity would mean batch effects are removed and there is low non-biological effects [675].

Lastly, it is also important to show that batch correction leads to improvement in finding true biological differences between samples. We can show the positive effect that batch correction has on the data by demonstrating reproducibility after batch correction. One way to provide evidence for reproducibility is to show that prior to batch correction, there was no overlap between differentially expressed proteins between groups in one batch with those found between the same groups in another batch and, after batch correction, the differentially expressed proteins between the groups become the same between batches [679]. This applies generally datasets with large numbers (e.g., hundreds) of samples to allow for meaningful statistical comparisons [679].

# Statistical Analysis

Once the above pre-processing steps have been applied to the dataset, we can investigate which proteins discriminate between groups. There is an urgent need for biomarkers for disease prediction and there is large potential for protein based biomarker candidates [706]. However, omics datasets are often limited due to having many more features than number of samples, which is termed the 'curse of dimensionality'[672]. Attributes that are redundant or not informative can reduce the accuracy of a model [707].

Univariate statistical tests including t-tests and analysis of variance (ANOVA) provide p-values to allow ranking the importance of variables [672]. T-tests are used in pairwise comparisons, and ANOVA is used when there are multiple groups to ask whether any group is different from the rest. After ANOVA, the Tukey's posthoc test can reveal which pairwise differences are present among the multiple groups that were compared. Wilcoxon rank-sum tests can be used if the data are still not normal after the above approaches and therefore violates the assumptions required for a t-test. Kruskal-Wallis test is the non-parametric version of ANOVA useful for three or more groups when assumptions of ANOVA are violated.

Data can be reduced using a feature selection method, which includes either feature subset selection, where irrelevant features are removed, or feature extraction, where there is a transformation that generates new, aggregated variables and do not lead to loss of information [672]. An example of a commonly used multivariate feature extraction method using proteomic data is principal component analysis (PCA) [672].

Proteomics data analyses commonly involve multiple testing, which can lead to false positives (i.e., a p-value will appear significant by chance) [705] and multiple testing correction should be applied to main the overall false positive rate at less than a specified cut-off [672]. Benjamini-Hochberg correction is less stringent than the Bonferroni correction, which leads to too many false negatives, and thus is a more commonly used multiple testing correction method [672].

Volcano plots allow visualization of differentially abundant proteins by displaying the negative log of the adjusted p-value as a function of the log fold change, a measure of effect size, for each protein. Points with larger y axis values are more statistically significant and those further away from zero on the x axis have a larger fold change.
There are two methods for identifying differentially expressed proteins. The first method involves a combined adjusted p-value cut-off (y axis) and fold change cut-off (x axis) to create a 'square cut-off'[705]. The second involves a non-linear cut-off, where a systematic error is added to all the standard deviations used in the t-tests [705].

There are other statistical tests to consider for quantitative proteomics data. Another popular statistical method in proteomics when dealing with high dimensional data is lasso linear regression, which removes regression coefficients from the model by applying a penalty parameter [708]. Bayesian models are an emerging technique for protein based biomarker discovery that are more powerful than standard t-tests [708] and have outperformed linear models [708,709]. Bayesian models incorporate external information into the prior distribution; for example, knowledge of peptides that usually have more technical variability are assigned a less informative prior [708]. Prior to implementing machine learning (ML), one can start with the simpler models, such as linear regression or naïve Bayes [706].

## Machine Learning Tools

Despite ML methods being highly effective in finding signals in a high dimensional feature space to distinguish between classes [706], the application of ML to proteomic data analysis is still in its early stages [706] as only 2% of proteomics studies involve ML [710].

Supervised classification is the most common type of ML used for proteomic biomarker discovery, where an algorithm has been trained on variables to predict the class labels of unseen test data [707]. Supervised means the class labels, such as disease versus controls, is known [710]. Decision trees are common model choice due to their many advantages: variables are not assumed to be linearly related, models are able to rank more important variables on their own, and interactions between variables do not need to be pre-specified by the user [708]. There are three phases of model development and evaluation [711]. In the first step, the dataset is split into training and testing splits, commonly 70% training and 30% testing. Second, the model is constructed using only the training data, which is further subdivided into training and test sets. During this process, an internal validation strategy, or cross-validation (CV), is employed [672]. Commonly used CV methods in proteomics are k-fold and leave-one-out cross-validation[672]. The final step is to evaluate the model on the testing set that was held-out in step one. There should not be overlap between the training and testing data, and the testing data should only be evaluated once after all training has been completed. The dataset used for training and testing should be representative of the population that is to be eventually tested. If underrepresented groups are lacking from models during training, these models will not generalize to these populations [712]. Proteomic data and patient specific factors derived from the electronic health record (EHR) like age, race, and smoking status can be employed as inputs to a model [706]. However, addition of EHR data may not be informative in some instances; in studying Alzheimer's Disease, adding these patient specific variables were informative for non-Hispanic white participants, but not for African Americans [712].

A common mistake in proteomics ML studies is allowing the test data to leak into the feature selection step [710,713]. It has been reported that 80% of ML studies on the gut microbiome performed feature selection using all the data, including test data [713]. Including the testing data in the feature selection step leads to development of an artificially inflated model [713] that is overfit on the training data and performs poorly on new data [706]. Feature selection should occur only on the training set and final model performance should be reported using the unseen testing set. The number of samples should be ten times the number of features to make statistically valid comparisons, however this may not be possible in many cases [714]. If a study is limited by its number of samples, one can perform classification without feature selection [713].

Pitfalls also arise when a ML classifier is trained using an imbalanced dataset [711]. Proteomics biomarker studies commonly have imbalanced groups, where the number of samples in one group is drastically different from another group. Most ML algorithms assume balanced number of samples per class and not accounting for these differences can lead to reduced performance and construction of a biased classifier [715].

Care should be practiced when choosing an appropriate metric when dealing with imbalanced data. A high accuracy may be meaningless in the case of imbalanced classification; the number of correction predictions will be high even with a blind guess for the majority class [716]. F1 score, Matthews correlation coefficient (MCC), and area under the precision recall curve (AUPR) are preferred metrics for imbalanced data classification [717,718].
MCC, for example, is preferred since it is only high if the model predicts correctly on both the positive and negative classes [711]. Over- and under-sampling to equalize the number of samples in classes are potential methods to address class imbalance, but can be ineffective or even detrimental to the performance of the model [715]. These sampling methods may lead to a poorly calibrated model that overestimates the probability of the minority class samples and reduce the model's applicability to clinical practice [716].

# 14. Protein Sequence Databases

## What are they and where do you get them?

### Protein Database Sources and Types

Many mass spectrometry-based proteomic techniques use search algorithms that require a defined theoretical search space to identify peptide sequences based on precursor mass and fragmentation patterns, which are then used to infer the presence and abundance of a protein. The search space is calculated from the potential proteins in a sample, which includes the proteome (often a single species) and expected contaminants. This is called database searching and the flat file of protein sequences in FASTA format acts as a protein database. In this section, we will describe major resources for proteome FASTA files (protein sequence collections), how to retrieve them, and suggested best practices for preserving FASTA file provenance to improve reproducibility.

In general, FASTA sequence collections can be retrieved from three central clearing houses: UniProt, RefSeq, and Ensembl. These will be discussed separately below as they each have specific design goals, data products, and unique characteristics. It is important to learn the following three points for each resource: the source of the underlying data, canonical versus non-canonical sequences, and how versioning works. These points, along with general best practices, such as using a taxonomic identifier, are essential to understand and communicate search settings used in analyses of proteomic datasets. Finally, it is critical to understand that sequence collections from these three resources are not the same, nor do they offer the same sets of species.

Key terminology may vary between resources, so these terms are defined here. The term "taxon identifier" is used across resources and is based on the NCBI taxonomy database. Every taxonomic node has a number, e.g., *Homo sapiens* (genus species) is 9606 and Mammalia (class) is 40674. This can be useful when retrieving and describing protein sequence collections. Another term used is "annotation", which has different meanings in different contexts. Broadly, a "genome annotation" is the result of an annotation pipeline to predict coding sequences, and often a gene name/symbol if possible. Two examples are MAKER [719] and the RefSeq annotation pipeline [720]. Alternatively, "protein annotation" (or gene annotation) often refers to the annotation of proteins (gene products) using names and ontology (i.e., protein names, gene names/symbols, functional domains, gene onotology, keywords, etc.). Protein annotation is termed "biocuration" and described in detail by UniProt [721]. Lastly, there are established minimum reporting guidelines for referring to FASTA files established in MIAPE: Mass Spectrometry Informatics that are taxon identifier and number of sequences [722,723]. The FASTA file naming suggestions below are not official but are suggested as a best practice.

## UniProt

The Universal Protein Resource (UniProt) [724], has three different products: UniProt Knowledgebase (UniProtKB), the UniProt Reference Clusters (UniRef), and the UniProt Archive (UniParc). The numerous resources and capabilities associated with the UniProt are not explored in this section, but these are well described on UniProt's website. UniProtKB is the source of proteomes across the Tree of Life and is the resource we will be describing herein. There are broadly two types of proteome sequence collections: Swiss-Prot/TrEMBL and designated proteomes. The Swiss-Prot/TrEMBL type can be understood by discussing how data is integrated into UniProt. Most protein sequences in UniProt are derived from coding sequences submitted to EMBL-Bank, GenBank and DDBJ. These translated sequences are initially imported into TrEMBL database, which is why TrEMBL is also termed "unreviewed". There are other sources of protein sequences, as described by UniProt [725]. These include the Protein Data Bank (PDB), direct protein sequencing, sequences derived from the literature, gene prediction (from sources such as Ensembl) or in-house prediction by UniProt itself. Protein sequences can then be manually curated into the Swiss-Prot database using multiple outlined steps (described in detail by UniProt here [726]) and is why Swiss-Prot is also termed "reviewed". Note that more than one TrEMBL entry may be removed and replaced by a single Swiss-Prot entry during curation. A search of "taxonomy_id:9606" at UniProtKB will retrieve both the Swiss-Prot/reviewed and TrEMBL/unreviewed sequences for Homo sapiens. The entries do not overlap, so users often either use just Swiss-Prot or Swiss-Prot combined with TrEMBL, the latter being the most exhaustive option. With ever-increasing numbers of high-quality genome assemblies processed with robust automated annotation pipelines, TrEMBL entries will contain higher quality protein sequences than in the past. In other words, if a mammal species has 20 000 to 40 000 entries in UniProtKB and many of these are TrEMBL, users should be comfortable using all the protein entries to define their search space (more on this later when discussing proteomes at UniProtKB). Determining the expected size of a well-annotated proteome requires additional knowledge, but tools to answer these questions continue to improve. As more and more genome annotations are generated, the backlog of manual curation continues to increase. However, automated genome annotations are also rapidly improving, blurring the line between Swiss-Prot and TrEMBL utility.

The second type of protein sequence collections available at UniProtKB are designated proteomes, with subclasses of "proteome", "reference proteome" or "pan-proteome". As defined by UniProt, a proteome is the set of proteins derived from the annotation of a completely sequenced genome assembly (one proteome per genome assembly). This means that a proteome will include both Swiss-Prot and TrEMBL entries present in a single genome annotation, and that all entries in the proteome can be traced to a single complete genome assembly. This aids in tracking provenance as assemblies change, and metrics of these assemblies are available. These metrics include Benchmarking Universal Single-Copy Ortholog (BUSCO) score, and "Completeness" as Standard, Close Standard or Outlier based on the Complete Proteome Detector (CPD). Given the quality of genome annotation pipelines, using a proteome as a FASTA file for a species is the preferred method of defining search spaces now. Outside of humans, no higher eukaryotic Swiss-Prot sequence collections are complete enough for use in proteomics analyses, but this does not mean that the available Swiss-Prot plus TrEMBL protein sequence collection precludes accurate proteomic data analysis. Lastly, the difference between reference proteome and proteome is used to highlight model organisms or organisms of interest, but not to imply improved quality. UniProt also has support for the concept of "pan proteomes" (consensus proteomes for a closely related set of organisms) but this is mostly used for bacteria (e.g., strains of a given species will share a pan proteome).

When retrieving protein sequence collections as Swiss-Prot/TrEMBL or designated proteomes, there is an option of downloading "FASTA (canonical)" or "FASTA (canonical & isoform)". The later includes additional manually annotated isoforms for Swiss-Prot sequences. Each Swiss-Prot entry has one canonical sequence chosen by the manual curator. Any additional sequence variants (mostly from alternative slicing) are annotated as differences with respect to the canonical sequence. Specifying "canonical" will select only one protein sequence per Swiss-Prot entry while specifying "canonical & isoforms" will download additional protein sequences by including isoforms for Swiss-Prot entries. Recently, an option to "download one protein sequence per gene (FASTA)" has been added. These FASTA files include Swiss-Prot and TrEMBL sequences to number about 20 000 protein sequences for a wide range of higher eukaryotic organisms.

The number of additional isoforms in a proteome varies considerably by species. In the human, mouse, and rat proteomes of the total number of entries, 25 %, 40 % and 48 % are canonical, respectively. The choice of including isoforms is related to the search algorithm and experimental goals. For instance, if differentiating isoforms is relevant, they should be included otherwise they will not be detected. In cases where isoforms are present in the FASTA (evident by shared protein names) but these cannot be removed prior to downloading (e.g., California sea lion, *Zalophus californianus*, proteome UP000515165, release 2023_04 has no options for downloading one protein sequence per gene), non-redundant FASTA files can be manually generated (i.e., "remove_duplicates.py" via [727]). If possible, retrieving canonical protein sequences via proteomes is the most straight forward approach and in general appropriate for most search algorithms, versus the method of searching and downloading Swiss-Prot and/or TrEMBL entries.

Though FASTA files are the typical input of many search algorithms, UniProt also offers an XML and GFF format download. In contrast to the flat FASTA file format, the XML format includes sequence information as well as associated information like PTMs, which is used in some search algorithms like MetaMorpheus [728].

Once a protein sequence collection has been selected and retrieved, there is the evergreen question of how to name and report this to others in a way that allows them to reproduce the retrieval. The minimum reporting information is the taxon identified and number of sequences used [722,723]. The following naming format (and those below) augments this and is suggested for UniProtKB FASTA files (the use of underscores or hyphens is not critical):

[common or scientific name]-[taxon id]-uniprot-[swiss-prot/trembl/proteome]-[UP# if used]-[canonical/canonical plus isoform]-[release]

example of a *Homo sapiens* (human) protein fasta from UniProtKB:

Human-9606-uniprot-proteome-UP000005640-canonical-2023_04.fasta

The importance of the taxon identifier has already been described above and is a consistent identifier across time and shared across resources. The choices of Swiss-Prot and TrEMBL in some combination was discussed above, and Proteome can be "proteome", "reference proteome" or "pan-proteome". The proteome identifier ('UP' followed by 9 digits) is conserved across releases, and release information should also be included. A confusing issue to newcomers is what the term "release" means. This is a year_month format (e.g., 2023_04), but it is not the date a FASTA file was downloaded or created, nor does it imply there are monthly updates. This release "date" is a traceable release identifier that is listed on UniProt's website. Including all this information ensures that the exact provenance of a FASTA file is known and allows the FASTA file to be regenerated.

## RefSeq

NCBI is a clearing house of numerous types of data and databases. Specific to protein sequence collections, NCBI Reference Sequence Database (RefSeq) provides annotated genomes across the Tree of Life. The newly developed NCBI Datasets portal [729] is the preferred method for accessing the myriad of NCBI data products, though protein sequence collections can also be retrieved from RefSeq directly[730,731]. Like UniProt described above, most of the additional functionality and information available through NCBI Datasets and RefSeq will not be described here, although the Eukaryotic RefSeq annotation dashboard [732] is a noteworthy resource to monitor the progress of new or re-annotations. We recommend exploring the resources available from NCBI [733], utilizing their tutorials and help requests.

RefSeq is akin to the "proteome" sequence collection from UniProtKB, where a release is based on a single genome assembly. If a more complete genome assembly is deposited or additional secondary evidence (e.g., RNA sequencing) is deposited, RefSeq can update the annotation with a new annotation release. Every annotation release will have an annotation report that contains information on the underlying genome assembly, the new genome annotation, secondary evidence used, and various statistics about what was updated. The current annotation release is referred to as the "reference annotation", but each annotation is numbered sequentially starting at 100 (the first release), though a recent naming change has abandoned the sequential release numbering and instead is the RefSeq assembly "-RS" and then the year month when it was annotated (e.g., the current human reference annotation is GCF_000001405.40-RS_2023_10). Certain species are on scheduled re-annotation, like human and mouse, while other species are updated as needed based on new data and community feedback (ex. release 100 of taxon 9704 was in 2018, but a more contiguous genome assembly resulted in re-annotation to release 101 in 2020). This general process for new and existing species is described in Heck and Neely [734].

Since RefSeq is genome assembly-centric, its protein sequence collections are retrieved for each species. This contrasts with being able to use a higher-level taxon identifier like 40674 (Mammalia) in UniProt to retrieve a single FASTA. To accomplish this same search in NCBI Datasets requires a Mammalia search, followed by browsing all 2847 genomes and then filtering the results to reference genomes with RefSeq annotations, and those resulting 223 could be bulk downloaded, though this will still be 223 individual FASTA files. It is possible to download a single FASTA from an upper-level taxon identifier using the NCBI Taxonomy Browser, though this service may be redundant with the new NCBI Datasets portal. Given the constant development of NCBI Datasets, these functionalities may change, but the general RefSeq philosophy of single species FASTA should be kept in mind. Likewise, when retrieving genome annotations there is no ability to specify canonical entries only, but it is possible to use computational tools to remove redundant entries ("remove_duplicates.py" from [727]).

Similar to the UniProtKB FASTA file naming suggestion, the following naming format is suggested for RefSeq protein sequence collection FASTA (the use of underscores or hyphens is not critical):

[common or scientific name]-[taxon id]-refseq-[release number]

Example of a *Equus caballus* (horse) protein FASTA from RefSeq:

Equus_caballus-9796-refseq-103.fasta

The release number starts at 100 and is consecutively numbered. Note, the human releases previously had a much longer number to be included (e.g., NCBI Release 109.20211119), then began following a consecutive numbering for Release 110, but have now switched to the new format related to assembly and annotation date. Also, in a few species (Human, Chinese hamster, and Dog, currently), there is a reference and an alternate assembly, both with an available annotation. In these cases, including the underlying assembly identifier would be needed. Note that when you retrieve the protein FASTA from NCBI it will include two more identifiers that aren't required in the file name since it can be determined from the taxon identifier and release number. These are the genome assembly used (this is generated by the depositor and follows no naming scheme) and the RefSeq identifier (GCF followed by a number string). These aren't essential for FASTA naming, but are for comparing between UniProt,

RefSeq and Ensembl when the same underlying assembly is used (or not, indicating how up to date one is versus the other).

## Ensembl

There are two main web portals for Ensembl sequence collections: the Ensembl genome browser [735] has vertebrate organisms and the Ensemble Genome project [736] has specific web portals for different non-vertebrate branches of the Tree of Life. This contrasts with NCBI and UniProt where all branches are centrally available. Recently, Ensembl has created a new portal "Rapid Release" focusing on quickly making annotations available (replacing the "Pre-Ensemble" portal), albeit without the full functionality of the primary Ensembl resources. Overall, Ensembl provides diverse comparative and genomic tools that should be explored, but, specific to this discussion, they provide species-specific genome annotation products similar to RefSeq.

To retrieve a protein sequence collection from Ensemble at any of the portals, a species can be searched using a name, which will then have taxon identifier displayed (but searching by identifier is not readily apparent). From the results you can select your species and follow links for genome annotation. Caution should be used when browsing the annotation products since the protein coding sequence (abbreviated "cds") annotations are nucleic acid sequences (a useable via 3-frame translation if using certain software), while actual translated peptide sequences are in the "pep" folders. The pep folders contain file names with "ab initio" and "all" in the FASTA file names (file extensions "fa" for FASTA and "gz" indicating gzip compression algorithm), while there may only be one pep product for certain species in the "Rapid Release" portal. The "ab initio" FASTA files contain mostly predicted gene products. The "all" FASTA files are the usable protein sequence collections. Ensembl FASTA files usually have some protein sequence redundancy.

Ensembl provides a release number for all the databases within each portal. Similar to the UniProt file naming suggestion, the following naming format is suggested for Ensembl protein sequence collection FASTA (the use of underscores or hyphens is not critical):

[common or scientific name]-[taxon id]-ensembl-[abinitio/all]-[rapid]-[release number]

Example of a *Sus scrofa* (pig) protein FASTA from Ensembl:

Pig-9823-ensembl-all-106.fasta

Similar to the FASTA download from RefSeq, the downloaded file name can include additional identifying information related to the underlying genome assembly. Again, this is not required for labeling, but is useful to easily compare assembly versions.

Since much of the data from Ensembl is also regularly processed into UniProt, using UniProt sequence collections instead may be preferred. That said, they are not on the same release schedule nor will the FASTA files contain the same proteins. Ensembl sequences still must go through the established protein sequence pipeline at UniProt to remove redundancy and conform to UniProt accession and FASTA header formats. Moreover, the gene-centric and comparative tools built into Ensembl may be more experimentally appropriate and using an Ensembl protein sequence collection can better leverage those tools.

## Other resources

There are other locations of protein sequence collections, and these will likewise have different FASTA file formatting; sequences may have unusual characters, and formats of accessions and FASTA header lines may need to be reformatted to be compatible with search software. These alternatives include institutes like the Joint Genome Institute's microbial genome clearing house, species-specific community resource (e.g., PomBase, FlyBase, WormBase, TryTrypDB, etc.), and one-off websites tenuously hosting in-house annotations. It is preferred to use protein sequence collection from the main three sources described here, since provenance can be tracked, and versions maintained. It is beyond the scope of this discussion to address other genome annotation resources, how they are versioned, or the best way to describe FASTA files retrieved from those sources. In these cases, defaulting to the minimum requirements of listing number of entries and supplying the FASTA along with data are necessary.

## Contaminants

Samples are rarely comprised of only proteins from the species of interest. There can be protein contamination during sample collection or processing. This may include proteins from human skin, wool from clothing, particles from latex, or even porcine trypsin itself, all of which contain proteins that can be digested along with the intended sample and analyzed in the mass spectrometer. Avoiding unwanted matching of mass spectra originating from contaminant proteins to the cellular proteins due to sequence similarities is important to the identification and quantitation of as many cellular proteins as possible. To avoid random matching, repositories of supplementary sequences for contaminant proteins have been added to a reference database for MS data searches. Appending a contaminants database to the reference database allows the identification of peptides that are not exclusive to one species. Peptides that are exclusive to the organism of interest are used to calculate abundance to avoid inflated quantitative results due to potential contaminant peptides.

As early as 2004, The Global Proteome Machine was providing a protein sequence collection of these common Repository of Adventitious Proteins (cRAP), while another contaminant list was published in 2008 [737]. The current cRAP version (v1.0) was described in 2012 [738] and is still widely in use today. cRAP is the contaminant protein list used in nearly all modern database searching software, though the documentation, versioning or updating of many of these "built-in" contaminant sequence collections is difficult to follow. There is also another contaminant sequence collection distributed with MaxQuant. Together, the cRAP and MaxQuant contaminant protein sequence collections are found in some form across most software, including MetaMorpheus and Philosopher (available in FragPipe) [739]. This list of known frequently contaminating proteins can either be automatically included by the software or can be retrieved as a FASTA to be used along with the primary search FASTA(s). Recently the Hao Lab has revisited these common contaminant sequences in an effort to update the protein sequences (ProtContLib), test their utility on experimental data, and add or remove entries [740].

In addition to these environmentally unintended contaminants, there are known contaminants that also have available protein sequence collections (or can be generated using the steps above) and should be included in the search space. These can include the media cells were grown in (e.g., fetal bovine serum [741,742], food fed to cells/animals (e.g., *Caenorhabditis elegans* grown on *Escherichia coli*) or known non-specific binders in affinity purification (i.e., CRAPome [743]). The common Repository of Fetal Bovine Serum Proteins (cRFP)[744] are protein lists of common protein contaminants and fetal serum bovine sequences used to reduced the number of falsely identified proteins in cell culture experiments. Cells washed or cultured in contaminant free media before harvest or the collection of secreted proteins depletes most high abundance contaminant proteins but the sequence similarity between contaminant and secreted proteins can cause false identifications and overestimation of the true protein abundance leading to wasted resources and time on validating false leads. As emphasized throughout this section, accurately defining the search space is essential for accurate results and, especially in the case of contaminants, requires knowledge of the experiment and sample processing to adequately define possible background proteins.

## Choosing the right database

Proteomics data analysis requires carefully matching the search space (defined by the database choice) with the expected proteins. A properly chosen database will minimize false positives and false negatives. Choosing a database that is too large will increase the number of false positives, or decoy hits, which in turn will reduce the total number of identifiable proteins. For this reason it is ill advised to search against all possible protein sequences ever predicted from any genomic sequence. On the other hand, choosing a database that is too small may increase false negatives, or missed protein identifications, because in order for a protein to be identified it must be present in the database. Some search algorithms can self-correct when a database is overly large such that higher identity thresholds are required for identification to minimize false positives (e.g., Mascot), while smaller experiment-specific search spaces (also referred to as "subsets") can have unintended effects on false positives if not managed appropriately [745,746,747] or may even improve protein identifications [748]. Whether to employ a search space that is sample-specific (i.e., subset), species-specific (with only canonical proteins, described below), exhaustive species-specific (including all isoforms), or even larger clade-level protein sequence set (e.g., the over 14 million protein sequences associated with Fungi, taxon identifier 4751) is a complex issue that is experiment and software dependent. Moreover, in cases where no species-specific protein sequence collection exists, homology-based searching can be used (as described in [734]). In each of these cases, proteomics practitioners must understand their specific experimental sample and search algorithm in order to know how to best define the search space, which is essential to yielding accurate results.

# 15. Protein Knowledge Bases

# PeptideAtlas and SRMAtlas

For most shared raw mass spectrometry data, none of the data was compared *in toto* to derive at a knowledge base of all detectable proteins in an organism. In addition, given the vast array of software for MS data analysis, the results are not directly comparable, nor combinable given the problem of false discovery rates (FDR) that pile up when dataset results are combined. For example, if we combine 3 datasets that were each filtered to 1% FDR, the maximum FDR of the combined dataset is now 3% because it is unlikely that the random decoy hits are shared across each dataset. To address this, in 2005 the PeptideAtlas concept was started to ingest as many publically available datasets as possible per organism, search the data through a single pipeline in total and arrive at a datasets with a total controlled 1% FDR at the protein level [749,750]. The PeptideAtlas website (www.peptideatlas.org) is a therefore a multi-organism, publicly accessible compendium of peptides identified in large sets of tandem mass spectrometry proteomics experiments. Mass spectrometer output files are collected for human, mouse, yeast, and many other organisms of research interest, and searched using the latest search engines and genome derived protein sequences. All results of sequence and spectral library searching on PeptideAtlas are processed through the Trans Proteomic Pipeline to derive a probability of correct identification for all results in a uniform manner to insure a high quality database, along with false discovery rates at the whole atlas level.

The most recognizable MS data compendium is the Human PeptideAtlas which is produced yearly since 2005 to derive all the peptide sequence knowledge of the current human proteome (Figure 17A). As of 2023, the Human PeptideAtlas contains the knowledge of over 93% of the human proteome, with over 189K MS runs and 3.6B spectra searched resulting in 3.4m peptides identified and 17,245 proteins identified from the 19,600 total proteins possible. The number of proteins has been incrementally increases year over year as new public data becomes available (Figure 17B).

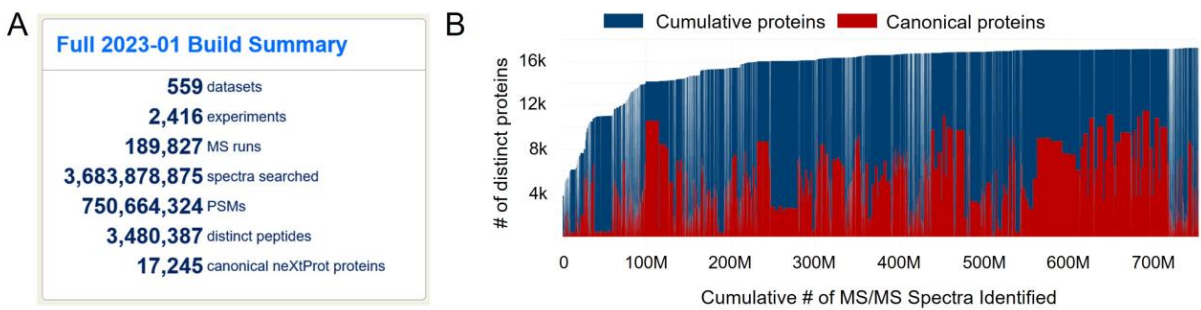

*Figure 17: **The Human PeptideAtlas as of 2023.** A) The current total search space and identified elements of the 2023 Human PeptideAtlas. B) Historical cumulative plot of the identified total proteins (blue vertical bars) and the unique proteins identified per dataset (red vertical bars) over the total period of 2005-2023.*

For the presentation of selected reaction monitoring (SRM) targeted peptide assays, there are two components of the PeptideAtlas ecosystem where the PeptideAtlas SRM Experiment Library (PASSEL) is presented to enable submission, dissemination, and reuse of SRM experimental results from analysis of biological samples [751,752]. The PASSEL system acts as a data repository by allowing researchers with SRM data to deposit their data in parallel with journal publication. and other users can search existing data to obtain the parameters for replication in their own laboratory. Another unique component for SRM data repositories is the SRMAtlas website, which provides definitive coordinates for all possible proteins within an organism to conduct targeted SRM assays that conclusively identify the respective peptide in biological samples. As an example, the Human SRMAtlas provides data on 166,174 synthetic proteotypic human peptides, providing multiple, independent assays to quantify any human protein and numerous spliced variants, non-synonymous mutations, and post-translational modifications [753]. The data are freely accessible as a resource at http://www.srmatlas.org/.

## Other knowledge bases

There are many more proteomics knowledgebases, such as PRIDE [754,755], the proteomics standards initiative (PSI) [756], Massive [651], Proteometools [757], Panorama [758], Chorus [759], and iProX [760].

# 16. Biological Interpretation

The most common untargeted proteomics experiment will produce a list of proteins or peptides of interest which require further validation and biological interpretation. This list usually results from statistical data analysis; the typical output of differentially expressed proteins usually contains hundreds of hits. In this section, we aim to present a concise overview of how proteomic data can be effectively contextualized and used to generate new hypotheses.

The simplest approach is to start manual lookup of every protein in the list to uncover groups that function together. Starting with a list of hundreds of protein changes, a smaller list can be prioritized by considering the level of significance and effect size. For example, proteins with the smallest p-values (significance) and largest abundance fold-changes (effect size). It is tempting to focus on proteins with the most extreme fold changes. In this case, the assumption is that the more significant the fold change (in either direction, up- or down-regulation), the higher the impact of those proteins on cellular behavior. This assumption is not always valid because protein signal in MS depends on abundance. The manual data interpretation approach is typically infeasible due to the number of proteins that would need to be individually looked up one-by-one.

A better strategy is to use computational methods. These methods may consider the whole list of proteins including some ranking by significance or fold change. One common interpretation method is to construct a protein network, which then lends itself to network analyses. Another method is to consider functional enrichment through annotation databases. These databases offer insights by examining the enrichment of certain functional annotations amongst the interesting proteins. Secondly, one could consider other evolutionary, structurally or regulatory based methods to identify interpretation of the data. To fully interpret analysis, it may be required to perform or examine other data such as data from biophysical, biochemical and alternative proteomic approaches. Finally, the data can further be interpreted using multi-omic, native or clinical approaches. Below we summarize these approaches and point out potential pitfalls with these methods.

# Constructing a protein network

A network is a representation of the relations between objects. Nodes are the entities of the network (e.g., users of a social platform, train stations, proteins), while edges are the connections between them (e.g., friendship, routes, and protein interactions, respectively). In the case of protein-protein interactions, evidence for the functional associations between proteins can be obtained experimentally. For example, co-immunoprecipitation, crosslinking, and proximity labeling can be used to reveal physical interactions [761]. The data is presented in a table with nodes and edges (e.g., "protein A interacts with protein B") from which the network can be constructed. A considerable wealth of protein-protein association data is stored in free databases like IntAct, which contain interactions derived from literature curation or direct user submissions [762]. Protein interactions can also be predicted by classifiers that consider many features, like orthology and co-localization, to produce a posterior odds ratio of interaction [763,764]. Finally, large repositories like STRING (Search Tool for the Retrieval of Interacting Genes/Proteins) collect and integrate protein-protein interaction data from several databases [764]. STRING also provides a web-based interface to survey the data, and users only have to feed a search box with the identifiers of the protein(s) of interest. STRING will retrieve the network and show the evidence supporting each interaction. Importantly, these databases do not indicate the direction of the interaction, so they produce undirected networks. If edges have directions (e.g., A influences B and not vice versa), then the network is directed. Signaling pathways are examples of directed graphs.

# Network analysis

Network analysis is a group of techniques that explore and investigate the network, yielding valuable knowledge about its structure and unveiling key players regulating the flow of information. One of the first steps in network analysis relates to centrality measurements. Centralities are indicators of the relative importance of a node corresponding to its position in the network, and each centrality measure provides new insights to interpret the data in new ways [765,766].

## Degree centrality

The degree of a node measures the number of edges incident to that node. Nodes with a high degree interact with many other nodes, called first neighbors. In particular, the node degree distribution in protein networks is highly skewed, with most nodes having a low degree and a few having high degrees, known as hubs. Hubs are usually regulatory proteins, being notable examples oncogenes and transcription factors. Moreover, hubs are attractive targets for directed interventions, as their alteration has a profound effect on the stability of the network [767].

## Closeness centrality

The route from one node to another is a path, and the shortest path is the one connecting them in the least amount of steps. Closeness centrality is the inverse of the average length of a node's shortest paths to all other nodes in the network. Nodes with a high closeness score have the shortest distances to all the others, so closeness centrality calculations detect nodes that can spread information very efficiently, as they are in a better position in the network for this task [768,769].

## Betweenness centrality

This centrality index is related to the amount of shortest paths transversing a node. Nodes with a high betweenness centrality usually bridge different parts of the network and strongly influence the flow of information, as they lie in communication paths. These connector hubs (or bottlenecks) are also interesting for follow–up experiments because their removal can disconnect different regions of the network [770].

Centrality measurements add new layers of information and allow for ranking differentially expressed proteins apart from their fold-change in abundance. **Figure 18** depicts a simple network consisting of proteins A to L, with A having the highest fold change (10) and L the lowest (2). In Panel A, the fill color for the nodes indicates this metric, where it can be easily seen that A stands out. However, protein A is a peripheral protein, only interacting with B. In Panel B, nodes are colored according to node degree. Clearly, protein F has the highest number of interactions and is also the closest to all other nodes, which can be appreciated when nodes are colored according to closeness centrality (Panel C). On the other hand, protein G acts as a bridge between two regions of the network and thus, has the highest betweenness centrality (Panel D). Except for fold change, node A has the lowest indices, and it will be up to the researcher to decide whether this protein warrants further examination.

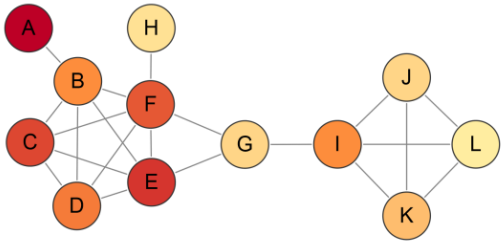

**A)** Fold change

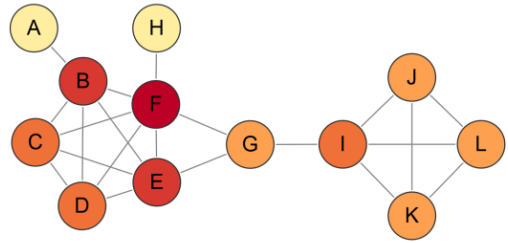

**B)** Degree

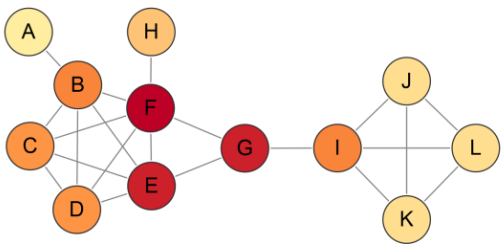

**C)** Closeness centrality

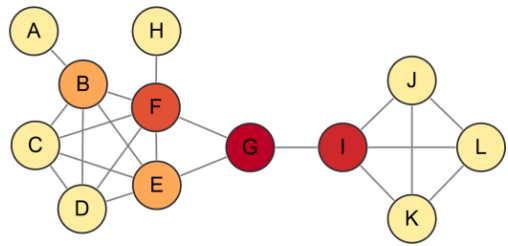

**D)** Betweenness centrality

*Figure 18:* ***Analysis of a simple network using different centrality measurements.*** *Nodes are colored according to each metric using a yellow-to-red gradient (yellow: lowest value, red: highest value). Network visualization and analysis were performed in Cytoscape.*

# Network clustering

In the small network presented in **Figure 18**, two groups of densely connected nodes exist. This topology suggests that these communities (or "clusters") work together or participate in a protein complex. Dividing a network into clusters helps identify underlying relationships among nodes, which is especially useful in large networks. In a broad sense, network clustering groups nodes according to a topological property, generally interconnectedness. There are many network clustering algorithms, each with its own merits and approaches [771,772]. The MCL (Markov CLustering) algorithm is suitable for protein networks in most situations . On the other hand, the Molecular COmplex DEtection (MCODE) algorithm helps detect very densely connected nodes, thus unveiling protein complexes [773]. In this regard, network clustering is useful for tentatively assigning the function of an uncharacterized protein. If the protein appears in a cluster, its function should be closely related to the cluster members, a principle known as "guilty by association." [774]

# Network visualization

A critical step in network analysis is to display the data in a structured and uncluttered graph. Networks can rapidly become a hairball unamenable to interpretation. Software platforms like Cytoscape can be used to visualize networks orderly by applying layout algorithms and format styles [775]. Since many of these platforms are open source, community-designed plugins enhance their capabilities. In Cytoscape, the stringApp adds a search bar to query the STRING database with accession numbers or protein names [776]. The network is directly retrieved into Cytoscape, where its built-in network analyzer can be used to calculate centralities. Moreover, user-defined information, like fold-change values, can be integrated and mapped into the network.

## Functional term enrichment analysis: KEGG, String, GO, GSEA, ORA, Reactome, and others

Term enrichment analysis is performed to assess whether particular 'functional terms' are over-represented in a list of proteins (e.g. from a proteomics experiment) [777,778,779]. For example, after a differential abundance analysis, we may wish to examine whether there is any shared function amongst the proteins which were determined to have significant changes. The simplest analysis to test whether this subset contains more of any particular functional terms than we would expect given the background of proteins. For example, the Gene Ontology is split into the classes: Cellular Component, Molecular Function and Biological Function and we might be interested as to whether our proteins may be more likely to localize to a particular subcellular niche [780]. The Cellular Component terms could give us a starting point if this might be the case, by examining if Cellular Component annotations are enriched.

There are a number of databases and tools to perform such analysis, which can even be extended to examine whole pathways, networks, post-translational modification and literature representation. For example, databases such as KEGG [781], String [764], Reactome [PMID:31691815] and PhosphoSitePlus [782] can be used to test or annotate a list of proteins. For example, proteomics analysis of human cardiac 3D microtissue exposed to anthracyclines (drugs used in cancer chemotherapy) unearthed several proteins with altered levels [783]. Many of these were specifically grouped under GO terms related to mitochondrial dysfunction, indicating the detrimental effects of these drugs on the organelle. GO terms [780] or descriptors from other annotation libraries (like KEGG [781] or REACTOME [784]) can be retrieved from STRING when constructing a network or from other freely available compendiums. We refer to a number of articles on the topics, including tools, reviews and best-practice [785,786,787]. The main points from such analysis is that we can obtain an insight about protein function by looking at whether our list of proteins have similar or the same annotations. A number of limitations should be taken into account for interpretation. The first is that proteins that are more abundant are more likely to be studied, measured and examined in the literature. Hence, abundant proteins will have more annotations than less abundant ones. One key part of the analysis is also to correctly select the background set; that is, the universe of protein which our list is being compared against. By including contaminates or proteins that are not expressed in our system within the list, the results may be unfaithful.

We may also have access to our own curated set of annotations derived either computational or experimental. One may be interested in seeing whether we have enrichment of these annotations amongst the differentially abundant proteins. Our list of proteins could be divided into two groups: differentially abundant or not. These groups could be divided into whether they have a particular annotation: yes or no. This information can be summarized in a two-by-two table, to which we can apply a statistical test to examine whether that annotation is enriched within our differentially abundant proteins. One test that could be used is the hypergeometric test, and another would be a Fisher Exact test.

There are many methods for performing functional enrichment analysis on the data, but they can mainly be classified into three categories (**Figure 19**), as follows.

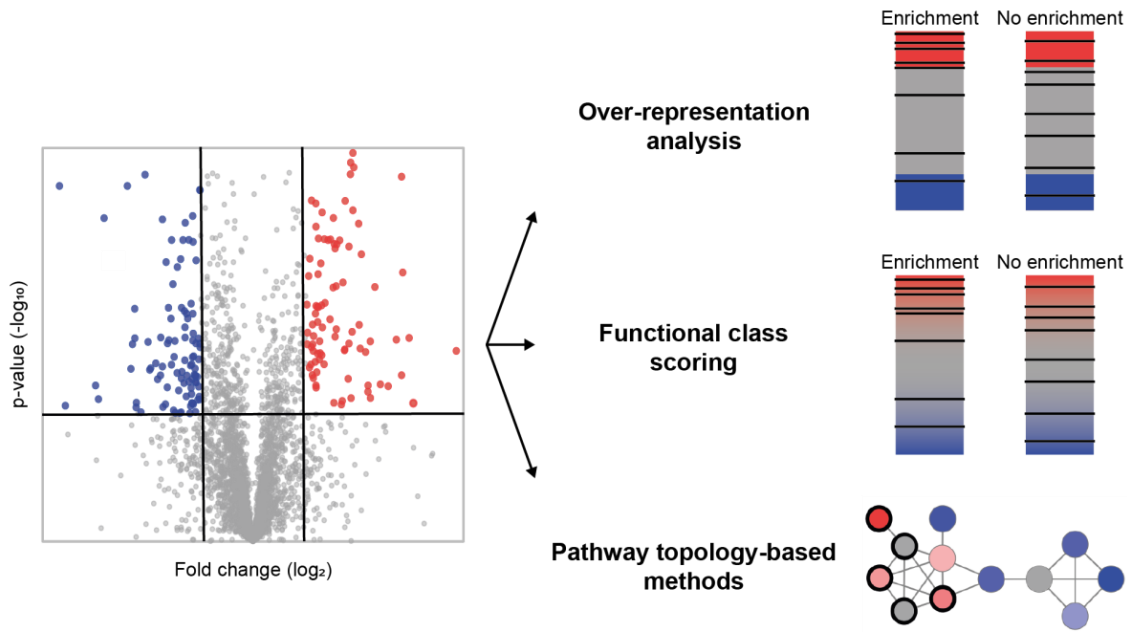

*Figure 19:* ***Types of functional enrichment methods.*** *In the volcano plot (left), proteins with altered values are colored blue or red according to arbitrarily chosen cut-off values for significance and fold change. Black bars or thick-bordered nodes indicate members of a GO category.*

## Over-representation analysis

In modern proteomics analysis, usually thousands of proteins are identified and quantified. Fold-change and significance thresholds are chosen (e.g., fold-change ≥ 2 and p ≤ 0.05) to obtain a list of proteins with altered levels among the tested conditions. In over-representation methods, a contingency table is created for every protein set to establish whether proteins with altered abundance show an enrichment or a depletion of the ontology term compared to the background observed proteome [788]. For example, suppose that 2000 proteins were quantified in a proteomics analysis, being 40 of these members of the set "tricarboxylic acid cycle (TCA)." Also, let us assume that 200 proteins showed altered abundance, with 15 belonging to the TCA set. Then, the contingency table can be constructed as follows:

| | Proteins with altered abundance | Proteins with unaltered abundance | Total |
|---|---|---|---|
| **Proteins in TCA set** | 15 | 25 | 40 |
| **Proteins not in TCA set** | 185 | 1775 | 1960 |
| **Total** | 200 | 1800 | 2000 |

Then, a suitable statistical test is conducted to ascertain if proteins with altered levels are enriched in members of the TCA cycle (in this case, they are; p < 0.00001). This is commonly achieved using Fisher's exact test [789]. The process is then repeated for every set as desired. Since multiple comparisons are tested, p values must be adjusted by a false discovery rate [790].

## Functional class scoring

The caveat of over-representation methods is that they rely on a list of differentially expressed genes or proteins with altered abundance, selected due to arbitrarily chosen cut-off values. For example, if we set a fold change cutoff of 2, a protein with a fold-change of 1.99 would not be included in the analysis. Moreover, several proteins belonging to the same set may have altered levels but are below the fold change threshold. However, moderate alterations of their abundance as a group could drive the observed phenotype, even more so than a single protein over the cut-off. Functional class scoring strategies aim at countering these limitations by disregarding thresholds altogether. GSEA (Gene Set Enrichment Analysis) is a widely used functional class scoring method in which all detected entities are first ranked according to a quantitative measurement (fold change, p-value, or their combination) [791]. Then, the distribution of members of a set is obtained. A scoring scheme based on the Kolmogorov – Smirnov test is used to assess whether there is an enrichment of the category towards the top or bottom of the ranked list.

## Pathway topology-based methods

Both methods mentioned above do not consider the functional relationships among proteins put forth by network analysis; i.e., they assume functional independence. Topology-based enrichment methods incorporate this information by, for example, assigning an importance value to a set when its members also participate in a pathway or cluster together in a network [792]. **Figure 19** shows how topology-based methods consider non-significant hits (grey nodes) that other strategies may not pick up, due to their position in a network.

# Other computational approaches: Network analysis, Isoform correlation analysis, AlphaFold, BLAST, protein language models

Additional computational analysis of a list of interesting proteins may uncover additional substructure, correlation or biologically useful hypothesis. Building a network between the proteins based on the experiments performed might be a useful approach to identify additional structure. For example, co-expression network analysis can be used to build a network from these proteins [793]. In these networks, proteins are nodes and edges describe relationships between those proteins. Network-specific methods can then be applied, such as community detection algorithms which could uncover clusters of proteins with shared functions [794,795].

One way the proteome generates complexity is through alternative-splicing, which results in protein isoforms [796].

Recently, a number of tools have been proposed to identify peptide isoforms that are quantitatively different across conditions by using a principle called peptide correlation analysis [797,798]. The idea is that the quantitative behavior of peptides should match each other. If there are subgroups that behave coherently within the group but not across groups suggest that peptide may have come from a different proteoform. These approaches can be used to identify specific proteoforms that are functional across different conditions.

For many, a protein's structure reveals important functional details [799]. There are a plethora of approaches to predict a protein's structure [800,801,802]. Recently, AlphaFold and RoseTTAFold have become dominant methods for predicting protein structures with high resolution [801,802]. If intrinsically disordered domains are of particular interest, methods explicitly designed for this task are recommended [803]. Once a structure is obtained more elaborate computational methods might be useful such as docking or molecular dynamics [804,805]. These approaches can give insight into how protein or molecules fit together and the dynamics of a protein's structure (conformational heterogeneity). A complete discussion of these topics is beyond the scope of this section.

Another way to obtain insights into a protein function is to look for protein with similar sequences or motifs. Using BLAST, a sequence alignment tool, one can align two or more protein sequences and determine their level of similarity [806]. For example, if a human protein of unknown function has a similar sequence to a yeast protein with known function this may be a starting place for the putative function of that protein.

Novel approaches to representing the similarity of proteins have proved successful at predicting the functional properties of proteins. Protein language models seek to learn "representation" of proteins, these are usually numerical vectors that represent a protein sequence [807,808]. Abstractly, these vectors preserve protein similarity or a notion of "proteinness". This usually means that two proteins that have a close vector may share similarities in protein function. These representations are also advantageous because they can easily become the inputs for machine learning algorithms to predict valuable protein properties; for example, thermal stability values [809], protein-protein binding affinities [810], secondary protein structure, and more.

# 17. Orthogonal experimental methods

## The importance of orthogonal experimental validation

The computational workflows to interpret mass spectrometry data are sophisticated, powerful tools, but also show important limitations and caveats due to their dependence on limited prior knowledge, specific experimental parameters or data quality restraints (see section 'Analysis of

Raw Data'). These inherent biases can give rise to ambiguous or spurious interpretation of the data even when these workflows are applied correctly and to the best of the experimenter's knowledge. Therefore, researchers will oftentimes be asked by scientific journals to provide independent orthogonal validation of their proteomics data and not performing such can be a major roadblock in the publication process.

The aim of validating data obtained by proteomics approaches should always be two-fold by demonstrating that the conclusions arrived at by proteomics data acquisition and analysis are, firstly, valid and, secondly, relevant. Depending on the question at hand, researchers can draw on an overabundance of techniques to validate MS-derived hypotheses in appropriate cellular, organismal or in vitro models. In the following paragraphs we aim to present only a high-level, stringent, non-exhaustive selection of orthogonal validation approaches and emphasize the importance of implementing assays that challenge assumptions gained from proteomics data analysis pipelines.

Before embarking on orthogonal validation of any hit, the success of the experiment should be established by assessing (internal) positive controls. Internal positive controls can be proteins whose behavior under the experimental conditions applied can be deduced from prior knowledge (i.e. the scientific literature or public databases). Once the expected changes in internal controls have been confirmed by computational analysis (see the above section), the orthogonal experimental validation of novel, perhaps unexpected findings can begin.

Orthogonal validation of new insights obtained from quantitative proteomics experiments can be a very time-consuming process and often requires familiarity with techniques not directly related to proteomics workflows. Given these challenges, the method(s) of choice warrant(s) careful consideration and is highly context-dependent. Importantly, proteomics experiments in one way or another generally yield comprehensive lists of potentially interesting candidate proteins or pathways, the researcher will have to shortlist candidates to be taken forward to the validation stage of the project. Which candidates should you validate by an orthogonal approach and which ones might not require further validation?

In general, candidates representing abundant proteins that show high sequence coverage and are detected with high confidence might not necessarily need extensive orthogonal validation when compared with proteins of intermediate to low abundance that might be more challenging to faithfully quantify by proteomics alone (i.e., many membrane proteins or transcription factors). Similarly, since the proteome is rarely comprehensively quantified in any single proteomics experiment, proteins of interest (POIs) that are critical for an observed biological change might not be part of the dataset. In these cases, additional, targeted analyses might help to support or discredit proteomics-based hypotheses.

Validation techniques are as manifold as biological questions and discussions thereof may easily fill multiple textbooks. The following sections are therefore merely meant to paint with a broad brush stroke a picture of useful methodologies with which to validate and follow up MS-data derived observations. As this is meant to orient the reader, wherever possible, we will explicitly point out useful literature reviews for a deeper dive into each of these techniques.

# General considerations

Once POIs have been selected based on prior agreed-upon selection criteria (i.e. (adjusted) p value and/or fold change thresholds), orthogonal validation experiments should ideally be conducted under physiologically relevant conditions to mitigate artificial and misleading outcomes. Therefore, in vitro experiments, while useful to isolate and dissect particular aspects of a biological system, can give highly artificial results as conditions are far removed from the POI's native environment. To investigate the biological function of a protein or pathway, direct genetic manipulation of the biological system at hand (e.g., modulating the expression of a POI by overexpression or knockout-/down experiments) can be minimally invasive when performed correctly. Should the POI be encoded by an essential gene, by definition, a complete and stable knockout might not be advisable [811,812]. In these extreme cases, attenuated expression (i.e., using RNA interference (RNAi) or controlled degradation, see below) rather than complete repression of a gene can be used to probe for protein function. Epitope tagging and/or exogenous expression of a gene of interest can be a powerful approach in assessing PPIs and investigating proteins of low abundance. However, overexpression artifacts are common [813].

It is not always possible to fully avoid the pleiotropic effects of protein (over-)expression or depletion, but a number of mitigation strategies (i.e., inducible expression, the use of multiple independent RNAi strategies) will be discussed below.

Extensive biochemical characterization of any overexpressed gene is critical to ensure it closely reflects the functions of its endogenous counterpart. These assays might involve assessing protein localization (i.e., by imaging techniques such as microscopy and flow cytometry), protein abundance (i.e., by mass spectrometry or immunoblot analysis) and phenotypic assays where applicable and practical.

# Functional genomics techniques in the validation of MS hits

Typical follow-up experiments to validate mass-spectrometry derived insights often involve the acute depletion or induction of a POI and assessing the impact on specific cellular phenotypes. Here we present a selection of methodologies to effectively modulate gene expression and discuss important considerations when planning functional genomics experiments for target validation.

Gene deletion or knockdown to prevent production of a functional protein is a powerful means to interrogate the role of one or more proteins in the phenotype(s) under investigation. To this end, well-established technologies deserving mention at this point are RNA interference (RNAi) in the form of siRNA/shRNA- or miRNA-mediated gene knockdown abd CRISPR/Cas9-or TALEN-mediated gene knockout [814]. Since each one of these technologies comes with its own unique advantages and caveats, the approach taken depends on the biological question at hand.

Clustered regularly interspaced short palindromic repeats (CRISPR)/Cas-based gene deletion technologies allow for the targeting of individual genes with relative ease, high efficiency and specificity [815]. When expressed in mammalian cells, the bacterially-derived Cas9 endonuclease can be guided with the help of a short guide RNA (gRNA) to a genomic location of interest, where it creates a DNA double strand break in a highly controlled manner (for a detailed discussion see [816]). The cell's DNA double-stand break repair machinery then introduces base pair insertions or deletions (indels) via non-homologous-end-joining (NHEJ), thus causing missense, and frameshift mutations (i.e. resulting in premature stop codons), leading to premature termination of gene expression or non-functional, aberrant gene products. Similarly, the concomitant provision of a complementary DNA donor template encoding a desired gene modification (i.e. insertion of a stretch of DNA or base pair modification) will trigger homology-directed repair (HDR), resulting in gene knock in or base editing [816]. Practical considerations of CRISPR/Cas9-mediated gene knock-in and base editing will not be addressed in detail but are expertly discussed in [817,818,819,820].

The relative ease-of-use and high efficiency of the CRISPR/Cas9 gene editing technology has rendered it the method of choice for gene manipulation in many fields of cell biology. However, it should be noted that CRISPR/Cas9-mediated gene deletion is not free from off-target effects ([821] for advice on how to minimize these off-target effects). Moreover, long-term depletion (or upregulation) of a POI itself can in some cases have dramatic systemic consequences and constitute an acute selection pressure leading to compensatory stress-induced adaptation that might obfuscate primary loss-of-function phenotypes and pose a substantial hurdle to the interpretability of biological data. As these compensatory mechanisms often manifest with time, controlled, transient genetic manipulation (gene depletion or transgene expression) is advised. Small interfering RNA (siRNA)-mediated knockdown by transient transfection is typically achieved at shorter time frames (24 – 96h), depending on the turnover of the POI. On an even shorter timescale, targeted, degron-based degradation systems enable depletion of a POI within minutes and further reduce off-target effects, but require the exogenous expression of a transgene and therefore some genetic manipulation. A more comprehensive discussion of a selection of these systems (anchor-away, deGradFP, auxin-inducible degron (AID), dTAG technologies) and their advantages and potential pitfalls is presented in [822].

Multiple eukaryotic and prokaryotic transcription-based systems have been developed that allow for the controlled biosynthesis or depletion of one or more POIs. Amongst these, a popular and dependable choice for mammalian cells are tetracycline-controlled operon systems, which allow up- or downregulation of a POI in the presence of the antibiotic tetracycline or its derivative doxycycline. These systems rely on the insertion of a bacteria-derived Tet operon (TetO) between the promoter and coding sequence of a GOI. In this configuration, the TetO binds a co-expressed Tet-repressor protein blocking transcription the of GOI. When tetracycline is added to the cells, the repressor then dissociates from the operon, thus de-repressing the GOI. Different variations of this potent system exist, allowing for more flexibility in experimental design. For instance, in the Tet-OFF system, the Tet repressor is fused to a eukaryotic transactivator (the chimeric fusion construct is termed tTA) and addition of tetracycline, or the related doxycycline, abolishes TetO binding and thus suppresses transcriptional activation [823]. Alternatively, a mutant form of tTA (rtTA) binds the TetO only in the presence of tetracycline, allowing for tetracycline-induced gene expression. For a detailed discussion of these systems, we refer the reader to an excellent review [824].

When generating stable expression cell lines, being able to precisely control the genomic integration site of the transgene reduces overall genetic heterogeneity in a cell population and thereby reduces potential off-target or pleiotropic effects. This ability is realised in the FlpIn-T-

REx technology which harnesses Flp-recombinase mediated DNA recombination at a strictly defined genomic locus (the FRT site) [825]. Site-directed isogenic integration of any GOI at the FRT site, which is under a tetracycline-inducible promoter and a hygromycin resistance gene, allows for facile generation of tetracycline/doxycycline-inducible isogeneic expression cell lines with minimal leaky expression (for an example, see [826]).

## Validation and interpretation of protein abundance changes

To validate protein abundance changes observed by quantitative bottom-up proteomics or simply assess the success of targeted genetic manipulation as part of an orthogonal follow-up experiment (see above), the experimenter typically resorts to antibody-based techniques such as immunoblotting analysis or immunofluorescence and immunohistological imaging of POIs. The latter also allows for validation of protein expression and localization in intact tissue or cells. However, these semi-quantitative methods are strongly influenced by the quality of the antibodies used and might not be sensitive enough to detect small changes in protein levels. In this case, more accurate orthogonal quantitation of proteins might be achieved by stable isotope labelling (SILAC/TMT/iTRAQ) and/or SRM/PRM (see section 'Experiment Types'). SDS-PAGE and immunoblot analysis are powerful and facile low-throughput tools to quickly validate protein abundance changes. However, short of introducing epitope tags to the endogenous POI, the success of immunoblotting is contingent on the availability of specific antibodies, which can present a formidable problem when investigating poorly characterized proteins or working with model organisms for which the commercial availability of specific antibodies is limited (this is particularly problematic for 'unconventional' or even well-established model organisms such as yeast). A detailed discussion of the strengths and pitfalls of immunoblotting for validation of semi-quantitative proteomics data can be found in an excellent review by Handler *et al.* [827].

Protein abundance changes detected in a proteomics experiment can be the result of a range of different cellular processes. The abundance of a protein in a complex sample (e.g. cell lysate or biological fluid) directly reflects a combination of the protein's intrinsic stability and the translational rate under the conditions of interest.

Both protein stability as well as gene expression activity can be quantified independently. Altered protein stability might be a direct consequence of specific or global changes in protein turnover. Radioisotope labelling is a well-established, accurate way to monitor protein synthesis, maturation and turnover [828,829]. This 'pulse-chase' methodology relies on the incorporation ('pulsing') of radioisotopes (typically 35S-labelled cysteine and methionine) into de-novo synthesized proteins. Upon withdrawal of the labeled amino acids from the culture medium, the decay of signal is monitored over time ('the chase') by SDS-PAGE and phosphoimaging, resulting in a temporal readout of protein abundances. The advantage of this technology is that a subpopulation (newly synthesized proteins) can be monitored directly, giving an accurate assessment of protein stability. Once a change in protein stability has been validated, the underlying mechanisms can be addressed by inhibiting protein degradation pathways; prominently proteasome-mediated degradation (using specific proteasome inhibitors such as bortemzomib/velcade or MG132), autophagy (pharmacologically inhibiting autophagic flux) or degradation by proteases (using protease inhibitors). The type of radiolabeling described above is relatively labor-intense, of low-throughput and has the obvious disadvantage of requiring radioactive material, which needs to be handled under strict safety precautions. Moreover, it critically depends on the presence of one or more methionines and/or cysteines in the POIs.

It is also possible to measure protein stability within complex protein mixtures (i.e. cell lysates or biological fluids) using an array of specialized mass spectrometry techniques as discussed in [830] and [114].

For purified proteins, well-established in vitro spectrometric and calorimetric methods such as circular dichroism, differential scanning calorimetry or differential scanning fluorometry can be used, but the relatively high sample amounts might be restrictive.

Finally, gene expression changes can also be determined with high fidelity using quantitative real-time PCR (qRT-PCR) or RNA-Seq can measure changes in gene transcription or mRNA turnover (for an extensive discussion of both technologies, please see [831] and [832], respectively).

## Validation of protein-protein interactions

The interaction of a protein with other proteins determines its function. Protein-protein interactions (PPIs) can be either mostly static (i.e. core subunits of a protein complex) or dynamic, varying with cellular state (i.e. cell cycle phase or cellular stress responses, posttranslational modifications) or environmental factors (i.e. availability of nutrients, presence of extracellular ligands of cell-surface receptors). Therefore, any given protein can typically bind a range of interaction partners in a spatially and temporally restricted manner, thus forming complex PPI networks (the interactome of a protein). The method of choice to experimentally examine altered PPI states depends on the model system and biological question (i.e. purified proteins vs complex protein mixtures, monitoring of PPIs in live cells or cell lysate etc). Popular methods for the validation of PPIs in vivo include protein fragment complementation (split protein systems), 2-hybrid assays (mammalian, yeast and bacterial), proximity ligation, proximity labelling and FRET / BRET. Protein fragment complementation assays rely on the principle that the two self-associating halves of reporter proteins can be expressed in an inactive form but when in spatial proximity bind one another to complement the functional, active reporter. When these split reporters are fused to two interacting proteins (so-called bait and prey proteins), the binding of bait to prey induces the spatial restriction needed to fully complement the reporter. Commonly used reporter complementation systems are split fluorescent proteins (i.e. GFP, YFP) [833], ubiquitin [834], luciferase [835], TEV protease [836], beta-lactamase [837], beta-galactosidase, Gal4, or DHFR [838]. The resulting functional readout of these complementation system depends on which split reporter is used. In general, the split luciferase system shows enhanced sensitivity over fluorescence-based systems as background luminescence is low.

Two-hybrid assays are based on a similar functional complementation strategy as fragment complementation systems. Conventionally, two self-complementing transcription factor fragments are fused to bait and prey proteins, respectively, leading to the restoration of a functional transcription factor only upon prey-bait interaction. The complemented transcription factor then induces the expression of a reporter gene that can be measured. Multiple variations of this system abound for different model organisms, but they almost always involve transcriptional activation or repression of a reporter gene ([839] for a detailed discussion).

The yeast-2-hybrid system (Y2H) is deserving of mention here as it had been the very first 2-hybrid system established [840] and has ever since proven to be extremely versatile (multiple auxotrophic reporters and markers of phenotypic sensitivity available), cheap, lends itself to functional high-throughput screening and variants have been developed that allow for the investigation of membrane-protein interactions (i.e. membrane Y2H) [839,841].

Despite the many advantages the Y2H offers, critical drawbacks include the potential of misfolding of bait and prey proteins when fused to a complementation reporter, expression at non-physiological levels, the lack of control over posttranslational modifications that might be important for the PPI under investigation, and the potential requirement of kingdom- or species-specific folding factors for the bait/prey under investigation (i.e. when probing PPI of mammalian proteins in Y2H). Principles of the Y2H technology have also been adapted to mammalian systems, which circumvent some of the aforementioned drawbacks of Y2H [842].

Perhaps the most commonly applied method of detecting and validating PPIs in vitro is affinity purification (AP, also known as affinity chromatography) of co-immunoprecipitation (Co-IP) either coupled with SDS-PAGE/immunoblotting or mass spectrometry to determine the identity of interacting proteins. AP typically relies on the isolation of a transgenic POI by an epitope tag (using epitope-specific matrix-conjugated proteins (antibodies or epitope-binding proteins)), while Co-IP harnesses specific antibodies directly targeting the POI. Specific interactors are expected to be enriched compared to the negative control (i.e an isotype control antibody, a knockout cell line or empty matrix). AP is not solely restricted to detecting PPIs, but can also be adapted to protein interactions with other biomolecules such as RNA [843]. It should be noted that AP and Co-IP can return multiple potential binding partners, many of which might be artefactual due to loss of cellular compartmentalization during sample preparation.

To reduce the probability of such artefacts and increase the confidence of a specific interaction, reciprocal affinity purification (by pulldown of each interaction partner) or in situ imaging might be performed (i.e. using fluorescence resonance energy transfer (FRET) [844], split-protein systems [845], proximity ligation assay [846] and immunofluorescence microscopy).

Forster and bioluminescence resonance energy transfer (FRET / BRET) can be used for in situ visualization of protein proximities and therefore PPIs. In FRET, non-radiative energy transfer between donor and receptor chromophores (each fused to prey and bait proteins, respectively), results in the emission of a characteristic fluorescence signal only when both prey and bait are in very close proximity (1-10 nm distance) and a suitable light source for donor excitation is provided [847].

The underlying principle of BRET is similar to that of FRET but with the exception of using a chemical substrate which activates bioluminescent donor, such as luciferase, resulting in energy transfer to a fluorescent acceptor molecule [848,849]. The main advantages of BRET over FRET are independence from an external light source (which can result in photobleaching), but requires at least one of the POIs to be fused to the donor (while in FRET, donor and acceptor can be chemically conjugated to POI-specific antibodies) [848]. FRET can be particularly useful in investigating cell surface protein interactions when using specific antibodies conjugated to donor and acceptor probes as antibodies are not cell-permeable and therefore restricted to targets presented on the cell surface in the absence of membrane permeabilization agents.

Other fluorescence-based PPI assays encompass Fluorescence correlation spectroscopy (FCS) and fluorescence cross-correlation spectroscopy (FCCS). These methods use small volumes of fluorescently labelled proteins and can determine their diffusion coefficients, which change in when proteins form a complex [850].

Proximity labelling methods (Proximity ligation and enzymatic proximity labelling (BirA, APEX2, HRP) can surveil labile or transient interaction in live cells in a high-throughput format when coupled with target identification by MS [851], [852]. These approaches harness a biotin ligase (i.e. BirA, BioID2, AirID, BASU, APEX2, HRP) fused to a POI whose interactome is to be determined. In the presence of biotin (for BirA, BioID2, AirID, BASU, APEX2 and HRP) or a biotin-phenol derivative (for APEX2), the biotin ligase will activate the biotin(-phenol) which then covalently biotinylates any protein in close proximity. The activated biotin has a short half-life, ensuring that the effective labelling radius is typically restricted to approximately 10 nm. Biotinylated proteins are isolated by affinity purification with streptavidin-conjugated beads and identified by mass spectrometry or SDS-PAGE/immunoblotting. TurboID, miniTurboID and ultraID, promiscuous biotin ligases faster than BirA, have been developed allowing for shorter treatment times and decreased background signal. The choice of a biotin ligase variant depends on the POI and experimental setup, but in general HRP does not work in cytoplasmic environments where conditions are chemically reducing, but is suitable for labelling proteins extracellular face of the plasma membrane or in the endoplasmic reticulum and golgi apparatus. While TurboID and similar variants have fast kinetics, they can cause depletion of endogenous biotin and therefore cytotoxicity.

A major drawback shared by all variants described above is that they necessitate fusion to the POI, which might alter its physiological behavior and give rise to false positives or false negatives. Moreover, detecting a biotin-labelled protein does not unequivocally designate it as an interaction partner as spatial proximity to the POI-biotin ligase fusion protein without direct binding can result in biotinylation. The inclusion of controls, such as expression of the biotinylating enzyme alone in the cellular compartment of interest, is therefore particularly important for enzymatic proximity labelling methods.

The *in situ* proximity ligation assay (PLA) combines the specificity of antibodies with the signal amplification capacity of a DNA polymerase reaction. Here, two antibodies, each conjugated to a short single-strand DNA (ssDNA) tag and each specific to one of the two proteins whose interaction is under investigation, are added to fixed cells or tissue. Once bound to their respective targets and only when in direct proximity, the addition of two connector oligonucleotides complementary to each tag ssDNA tag and phi29 DNA polymerase, triggers isothermal rolling circle amplification, eventually resulting in the generation of continuous stretches of repetitive DNA. These DNA products can then be visualized by in situ hybridization with fluorescently labelled oligonucleotides (see [853] for a detailed discussion). PLA has the advantage of visualizing the two interacting proteins in their native environment when high-resolution microscopy is used as a readout.

Chemical cross-linking (XL) of proteins can determine PPIs with amino-acid level resolution, and can thereby give valuable insights into the orientation of two or more proteins relative to one another [854]. Recent technical advances also enabled the visualization of protein-RNA interaction [855]. Various XL chemistries are available (amine-reactive, sulfhydryl and photoreactive crosslinkers; reversible vs irreversible) and cross-linked proteins detected by mass spectrometry [856]. In general, applying XL-MS to a mixture of interacting, purified proteins is preferable to in situ XL of complex protein mixtures (i.e. cell lysate) as detection and deconvolution of XL peptides is technically and computationally challenging.

Surface plasmon resonance can accurately measure several key kinetics of PPIs with high accuracy (e.g. association and dissociation kinetics, stoichiometry, affinity) [857]. It relies on the quantification of refractive index changes of polarized light shone onto a sensor chip containing a prey protein immobilized on a metal surface (typically gold). When prey and bait proteins interact, the mass concentration at the metal interface changes, altering the refractive index and SPR angle (intensity of the refracted light).

# Acknowledgements

The authors thank Phil Wilmarth for helpful input. Identification of certain commercial equipment, instruments, software, or materials does not imply recommendation or endorsement by the National Institute of Standards and Technology, nor does it imply that the products identified are necessarily the best available for the purpose. The authors thank Dasom Hwang for help with graphic design. The authors thank Anthony Gitter and Daniem Himmelstein for assistance using manubot. The authors thank Jordan Burton and Pierre-Alexander Mücke for minor edits to the text. This manuscript was written collaboratively using manubot [858]. The live and evolving version where anyone can contribute can be found here https://github.com/jessegmeyerlab/proteomics-tutorial/tree/main.

# Author contributions

JGM was assigned last author as the project initiator and leader. RLM was assigned second to last author based on their leadership role in curating all sections. All other authors were ordered by estimating their contributions using a quantitative score. The score for each author was a sum of the number of sentences added plus 33 lines for each figure added. Scores were adjusted for confounding factors, such as split contributions between multiple contributors. Authors with similar scores were assigned equal contributions.

Gundry *Current protocols* (2021-03) https://www.ncbi.nlm.nih.gov/pubmed/33750040 DOI: 10.1002/cpz1.85 · PMID: 33750040 · PMCID: PMC8011989

166. **Universal sample preparation method for proteome analysis.** Jacek R Wiśniewski, Alexandre Zougman, Nagarjuna Nagaraj, Matthias Mann *Nature methods* (2009-04-19) https://www.ncbi.nlm.nih.gov/pubmed/19377485 DOI: 10.1038/nmeth.1322 · PMID: 19377485

167. **Filter-Aided Sample Preparation for Proteome Analysis.** Jacek R Wiśniewski *Methods in molecular biology (Clifton, N.J.)* (2018) https://www.ncbi.nlm.nih.gov/pubmed/30259475 DOI: 10.1007/978-1-4939-8695-8_1 · PMID: 30259475

168. **[Group and type distribution of hemolytic streptococci isolated from clinical specimens--prevalence of group A type 3 isolates in 1985 in Toyama Prefecture].** H Kodama, N Tokuman, I Yasui, Y Gyobu, Y Kashiwagi *Kansenshogaku zasshi. The Journal of the Japanese Association for Infectious Diseases* (1987-04) https://www.ncbi.nlm.nih.gov/pubmed/3117935 DOI: 10.11150/kansenshogakuzasshi1970.61.482 · PMID: 3117935

169. **Evaluation of FASP, SP3, and iST Protocols for Proteomic Sample Preparation in the Low Microgram Range.** Malte Sielaff, Jörg Kuharev, Toszka Bohn, Jennifer Hahlbrock, Tobias Bopp, Stefan Tenzer, Ute Distler *Journal of proteome research* (2017-10-11) https://www.ncbi.nlm.nih.gov/pubmed/28948796 DOI: 10.1021/acs.jproteome.7b00433 · PMID: 28948796

170. **Protein Aggregation Capture on Microparticles Enables Multipurpose Proteomics Sample Preparation.** Tanveer S Batth, MaximA X Tollenaere, Patrick Rüther, Alba Gonzalez-Franquesa, Bhargav S Prabhakar, Simon Bekker-Jensen, Atul S Deshmukh, Jesper V Olsen *Molecular & cellular proteomics : MCP* (2019-03-04) https://www.ncbi.nlm.nih.gov/pubmed/30833379 DOI: 10.1074/mcp.tir118.001270 · PMID: 30833379 · PMCID: PMC6495262

171. **Comparison of In-Solution, FASP, and S-Trap Based Digestion Methods for Bottom-Up Proteomic Studies.** Katelyn R Ludwig, Monica M Schroll, Amanda B Hummon *Journal of proteome research* (2018-05-24) https://www.ncbi.nlm.nih.gov/pubmed/29754492 DOI: 10.1021/acs.jproteome.8b00235 · PMID: 29754492 · PMCID: PMC9319029

172. **Phase transfer surfactant-aided trypsin digestion for membrane proteome analysis.** Takeshi Masuda, Masaru Tomita, Yasushi Ishihama *Journal of proteome research* (2008-02) https://www.ncbi.nlm.nih.gov/pubmed/18183947 DOI: 10.1021/pr700658q · PMID: 18183947

173. **Colorimetric protein assay techniques.** CV Sapan, RL Lundblad, NC Price *Biotechnology and applied biochemistry* (1999-04) https://www.ncbi.nlm.nih.gov/pubmed/10075906 PMID: 10075906

*chemistry* (2004-01-01) https://www.ncbi.nlm.nih.gov/pubmed/14697044 DOI: 10.1021/ac030096q · PMID: 14697044

365. **Biomarkers discovery by peptide and protein profiling in biological fluids based on functionalized magnetic beads purification and mass spectrometry.** Fulvio Magni, Yuri EM Van Der Burgt, Clizia Chinello, Veronica Mainini, Erica Gianazza, Valeria Squeo, André M Deelder, Marzia Galli Kienle *Blood transfusion = Trasfusione del sangue* (2010-06) https://www.ncbi.nlm.nih.gov/pubmed/20606758 DOI: 10.2450/2010.015s · PMID: 20606758 · PMCID: PMC2897205

366. **Magnetic particles as powerful purification tool for high sensitive mass spectrometric screening procedures.** Jochen F Peter, Angela M Otto *Proteomics* (2010-02) https://www.ncbi.nlm.nih.gov/pubmed/20099258 DOI: 10.1002/pmic.200900535 · PMID: 20099258

367. **Robust and high-throughput sample preparation for (semi-)quantitative analysis of N-glycosylation profiles from plasma samples.** LRenee Ruhaak, Carolin Huhn, Carolien AM Koeleman, André M Deelder, Manfred Wuhrer *Methods in molecular biology (Clifton, N.J.)* (2012) https://www.ncbi.nlm.nih.gov/pubmed/22665312 DOI: 10.1007/978-1-61779-885-6_23 · PMID: 22665312

368. **Utilizing ion-pairing hydrophilic interaction chromatography solid phase extraction for efficient glycopeptide enrichment in glycoproteomics.** Simon Mysling, Giuseppe Palmisano, Peter Højrup, Morten Thaysen-Andersen *Analytical chemistry* (2010-07-01) https://www.ncbi.nlm.nih.gov/pubmed/20536156 DOI: 10.1021/ac100530w · PMID: 20536156

369. **On-line 1D and 2D porous layer open tubular/LC-ESI-MS using 10-microm-i.d. poly(styrene-divinylbenzene) columns for ultrasensitive proteomic analysis.** Quanzhou Luo, Guihua Yue, Gary A Valaskovic, Ye Gu, Shiaw-Lin Wu, Barry L Karger *Analytical chemistry* (2007-07-11) https://www.ncbi.nlm.nih.gov/pubmed/17625912 DOI: 10.1021/ac070583w · PMID: 17625912 · PMCID: PMC2570646

370. **Development of a monolithic silica extraction tip for the analysis of proteins.** Shota Miyazaki, Kei Morisato, Norio Ishizuka, Hiroyoshi Minakuchi, Yukihiro Shintani, Masahiro Furuno, Kazuki Nakanishi *Journal of chromatography. A* (2004-07-16) https://www.ncbi.nlm.nih.gov/pubmed/15317408 DOI: 10.1016/j.chroma.2004.03.025 · PMID: 15317408

371. **Solid-phase extraction strategies to surmount body fluid sample complexity in high-throughput mass spectrometry-based proteomics.** Marco R Bladergroen, Yuri EM van der Burgt *Journal of analytical methods in chemistry* (2015-01-27) https://www.ncbi.nlm.nih.gov/pubmed/25692071 DOI: 10.1155/2015/250131 · PMID: 25692071 · PMCID: PMC4322654

372. **Stop and Go Extraction Tips for Matrix-Assisted Laser Desorption/Ionization, Nanoelectrospray, and LC/MS Sample Pretreatment in Proteomics** Juri Rappsilber, Yasushi

*Cellular Proteomics* (2021) https://doi.org/gssrqj DOI: 10.1016/j.mcpro.2021.100138 · PMID: 34416385 · PMCID: PMC8453224

473.   **Characterization of Ion Dynamics in Structures for Lossless Ion Manipulations** Aleksey V Tolmachev, Ian K Webb, Yehia M Ibrahim, Sandilya VB Garimella, Xinyu Zhang, Gordon A Anderson, Richard D Smith *Analytical Chemistry* (2014-09-04) https://doi.org/f6jr4m DOI: 10.1021/ac502054p · PMID: 25152178 · PMCID: PMC4175726

474.   **Serpentine Ultralong Path with Extended Routing (SUPER) High Resolution Traveling Wave Ion Mobility-MS using Structures for Lossless Ion Manipulations** Liulin Deng, Ian K Webb, Sandilya VB Garimella, Ahmed M Hamid, Xueyun Zheng, Randolph V Norheim, Spencer A Prost, Gordon A Anderson, Jeremy A Sandoval, Erin S Baker, … Richard D Smith *Analytical Chemistry* (2017-04-05) https://doi.org/f9zgxj DOI: 10.1021/acs.analchem.7b00185 · PMID: 28332832 · PMCID: PMC5627996

475.   **High-Resolution Ion-Mobility-Enabled Peptide Mapping for High-Throughput Critical Quality Attribute Monitoring** James R Arndt, Kelly L Wormwood Moser, Gregory Van Aken, Rory M Doyle, Tatjana Talamantes, Daniel DeBord, Laura Maxon, George Stafford, John Fjeldsted, Bryan Miller, Melissa Sherman *Journal of the American Society for Mass Spectrometry* (2021-04-09) https://doi.org/gssrrb DOI: 10.1021/jasms.0c00434 · PMID: 33835810

476.   **An Introduction to Mass Spectrometry-Based Proteomics** Steven R Shuken *Journal of Proteome Research* (2023-06-01) https://doi.org/gsr77x DOI: 10.1021/acs.jproteome.2c00838 · PMID: 37260118

477.   **Protein Analysis by Shotgun/Bottom-up Proteomics** Yaoyang Zhang, Bryan R Fonslow, Bing Shan, Moon-Chang Baek, John R Yates III *Chemical Reviews* (2013-02-26) https://doi.org/ggkd3r DOI: 10.1021/cr3003533 · PMID: 23438204 · PMCID: PMC3751594

478.   **Periodicity of Monoisotopic Mass Isomers and Isobars in Proteomics** Long Yu, Yan-Mei Xiong, Nick C Polfer *Analytical Chemistry* (2011-09-20) https://doi.org/dbf94b DOI: 10.1021/ac201624t · PMID: 21932815

479.   **Protein sequencing by tandem mass spectrometry.** DF Hunt, JR Yates 3rd, J Shabanowitz, S Winston, CR Hauer *Proceedings of the National Academy of Sciences* (1986-09) https://doi.org/bhq2fw DOI: 10.1073/pnas.83.17.6233 · PMID: 3462691 · PMCID: PMC386476

480.   **DirectMS1: MS/MS-Free Identification of 1000 Proteins of Cellular Proteomes in 5 Minutes** Mark V Ivanov, Julia A Bubis, Vladimir Gorshkov, Irina A Tarasova, Lev I Levitsky, Anna A Lobas, Elizaveta M Solovyeva, Marina L Pridatchenko, Frank Kjeldsen, Mikhail V Gorshkov *Analytical Chemistry* (2020-02-20) https://doi.org/gsr77v DOI: 10.1021/acs.analchem.9b05095 · PMID: 32077687

481.   **Robust Algorithm for Alignment of Liquid Chromatography−Mass Spectrometry Analyses in an Accurate Mass and Time Tag Data Analysis Pipeline** Navdeep Jaitly,

MacCoss *Nature Methods* (2013-06-23) https://doi.org/gjqdct DOI: 10.1038/nmeth.2528 · PMID: 23793237 · PMCID: PMC3881977

Ranganathan *International journal of molecular sciences* (2023-01-05)
https://www.ncbi.nlm.nih.gov/pubmed/36674563 DOI: 10.3390/ijms24021050 · PMID: 36674563
· PMCID: PMC9865486

645. **Building consensus spectral libraries for peptide identification in proteomics.**
Henry Lam, Eric W Deutsch, James S Eddes, Jimmy K Eng, Stephen E Stein, Ruedi Aebersold
*Nature methods* (2008-09-21) https://www.ncbi.nlm.nih.gov/pubmed/18806791 DOI:
10.1038/nmeth.1254 · PMID: 18806791 · PMCID: PMC2637392

646. **mzSpecLib | HUPO Proteomics Standards Initiative** https://psidev.info/mzSpecLib

647. **Novor: real-time peptide de novo sequencing software.** Bin Ma *Journal of the
American Society for Mass Spectrometry* (2015-06-30)
https://www.ncbi.nlm.nih.gov/pubmed/26122521 DOI: 10.1007/s13361-015-1204-0 · PMID:
26122521 · PMCID: PMC4604512

648. **<i>De novo</i> mass spectrometry peptide sequencing with a transformer model**
Melih Yilmaz, William E Fondrie, Wout Bittremieux, Sewoong Oh, William Stafford Noble *Cold
Spring Harbor Laboratory* (2022-02-09) https://doi.org/gsthnz DOI: 10.1101/2022.02.07.479481

649. **Discovery and Visualization of Uncharacterized Drug-Protein Adducts Using Mass
Spectrometry.** Michael Riffle, Michael R Hoopmann, Daniel Jaschob, Guo Zhong, Robert L
Moritz, Michael J MacCoss, Trisha N Davis, Nina Isoherranen, Alex Zelter *Analytical chemistry*
(2022-02-20) https://www.ncbi.nlm.nih.gov/pubmed/35184559 DOI:
10.1021/acs.analchem.1c04101 · PMID: 35184559 · PMCID: PMC8892443

650. **Prosit: proteome-wide prediction of peptide tandem mass spectra by deep
learning** Siegfried Gessulat, Tobias Schmidt, Daniel Paul Zolg, Patroklos Samaras, Karsten
Schnatbaum, Johannes Zerweck, Tobias Knaute, Julia Rechenberger, Bernard Delanghe,
Andreas Huhmer, … Mathias Wilhelm *Nature Methods* (2019-05-27) https://doi.org/gjqdvt DOI:
10.1038/s41592-019-0426-7 · PMID: 31133760

651. **Assembling the Community-Scale Discoverable Human Proteome.** Mingxun Wang,
Jian Wang, Jeremy Carver, Benjamin S Pullman, Seong Won Cha, Nuno Bandeira *Cell systems*
(2018-08-29) https://www.ncbi.nlm.nih.gov/pubmed/30172843 DOI: 10.1016/j.cels.2018.08.004
· PMID: 30172843 · PMCID: PMC6279426

652. **OpenSWATH enables automated, targeted analysis of data-independent
acquisition MS data** Hannes L Röst, George Rosenberger, Pedro Navarro, Ludovic Gillet,
Saša M Miladinović, Olga T Schubert, Witold Wolski, Ben C Collins, Johan Malmström, Lars
Malmström, Ruedi Aebersold *Nature Biotechnology* (2014-03) https://doi.org/f2z4c3 DOI:
10.1038/nbt.2841 · PMID: 24727770

653. **CsoDIAq Software for Direct Infusion Shotgun Proteome Analysis** Caleb W
Cranney, Jesse G Meyer *Analytical Chemistry* (2021-09-01) https://doi.org/gs4qk3 DOI:
10.1021/acs.analchem.1c02021 · PMID: 34469131

Fallahi *Molecular Biotechnology* (2022-12-22) https://doi.org/gr5j2d DOI: 10.1007/s12033-022-00639-1 · PMID: 36547823

820. **Recent Advances in Genome-Engineering Strategies** Michaela A Boti, Konstantina Athanasopoulou, Panagiotis G Adamopoulos, Diamantis C Sideris, Andreas Scorilas *Genes* (2023-01-02) https://doi.org/gr5j2w DOI: 10.3390/genes14010129 · PMID: 36672870 · PMCID: PMC9859587

821. **Latest Developed Strategies to Minimize the Off-Target Effects in CRISPR-Cas-Mediated Genome Editing** Muhammad Naeem, Saman Majeed, Mubasher Zahir Hoque, Irshad Ahmad *Cells* (2020-07-02) https://doi.org/gmhjsc DOI: 10.3390/cells9071608 · PMID: 32630835 · PMCID: PMC7407193

822. **Targeted Protein Degradation Tools: Overview and Future Perspectives** Yuri Prozzillo, Gaia Fattorini, Maria Virginia Santopietro, Luigi Suglia, Alessandra Ruggiero, Diego Ferreri, Giovanni Messina *Biology* (2020-11-26) https://doi.org/gmfb42 DOI: 10.3390/biology9120421 · PMID: 33256092 · PMCID: PMC7761331

823. **Tet-On Systems For Doxycycline-inducible Gene Expression** Atze T. Das, Liliane Tenenbaum, Ben Berkhout *Current Gene Therapy* (2016-06-16) https://doi.org/f8sfzz DOI: 10.2174/1566523216666160524144041 · PMID: 27216914 · PMCID: PMC5070417

824. **How to Choose the Right Inducible Gene Expression System for Mammalian Studies?** Kallunki, Barisic, Jäättelä, Liu *Cells* (2019-07-30) https://doi.org/gr5j2v DOI: 10.3390/cells8080796 · PMID: 31366153 · PMCID: PMC6721553

825. **Rapid Creation of Stable Mammalian Cell Lines for Regulated Expression of Proteins Using the Gateway® Recombination Cloning Technology and Flp-In T-REx® Lines** Jessica Spitzer, Markus Landthaler, Thomas Tuschl *Methods in Enzymology* (2013) https://doi.org/f5fkw5 DOI: 10.1016/b978-0-12-418687-3.00008-2 · PMID: 24011039

826. **Using the Flp-In™ T-Rex™ System to Regulate GPCR Expression** Richard J Ward, Elisa Alvarez-Curto, Graeme Milligan *Methods in Molecular Biology* (2011) https://doi.org/dsvq7p DOI: 10.1007/978-1-61779-126-0_2 · PMID: 21607850

827. **The Art of Validating Quantitative Proteomics Data** David C Handler, Dana Pascovici, Mehdi Mirzaei, Vivek Gupta, Ghasem Hosseini Salekdeh, Paul A Haynes *PROTEOMICS* (2018-11-25) https://doi.org/gr5j2b DOI: 10.1002/pmic.201800222 · PMID: 30352137

828. **Pulse-Chase Analysis for Studying Protein Synthesis and Maturation** Susanne Fritzsche, Sebastian Springer *Current Protocols in Protein Science* (2014-11) https://doi.org/gr5jz9 DOI: 10.1002/0471140864.ps3003s78 · PMID: 25367008

829. **Degradation Parameters from Pulse-Chase Experiments** Celine Sin, Davide Chiarugi, Angelo Valleriani *PLOS ONE* (2016-05-16) https://doi.org/gbmt2f DOI: 10.1371/journal.pone.0155028 · PMID: 27182698 · PMCID: PMC4868333